\documentclass[aps,prb,showpacs,notitlepage,floatfix,twocolumn,superscriptaddress,eqsecnum]{revtex4-2}
\usepackage[utf8]{inputenc}
\usepackage{newunicodechar}
\newunicodechar{−}{-}
\usepackage[bookmarks=false,colorlinks=true,urlcolor=red3,citecolor=red3,linkcolor=red3]{hyperref}

\usepackage{graphicx,amsmath,amssymb,amsxtra}
\usepackage{xcolor}
\usepackage{tikz,rotating}
\usetikzlibrary{arrows.meta}
\usepackage{ellipse}
\usepackage{comment}

\def\bd{\begin{displaymath}}
\def\be{\begin{equation}}
\def\ed{\end{displaymath}}
\def\ee{\end{equation}}
\def\bsub{\begin{subequations}}
\def\esub{\end{subequations}}

\usepackage{tikz}
\usetikzlibrary{shapes}
\makeatletter
\newcommand*\dashline{\rotatebox[origin=c]{90}{$\dabar@\dabar@\dabar@$}}
\makeatother

\def\be{\begin{equation}}
\def\ee{\end{equation}}

\def\bea{\begin{eqnarray}}
\def\eea{\end{eqnarray}}

\usepackage{subfigure}
\usepackage{times}
\usepackage{graphicx}
\usepackage{xcolor,dsfont}
\usepackage{booktabs}

\graphicspath{{figures}}

\usepackage{autobreak}
\allowdisplaybreaks

\RequirePackage{color}

\definecolor{MyDarkGreen}{rgb}{0.02,0.60,0.06}

\definecolor{kellygreen}{rgb}{0.3, 0.73, 0.09}
\definecolor{deeppink}{rgb}{1.0, 0.08, 0.58}
\definecolor{myblue}{HTML}{0096FF}

\definecolor{red3}{HTML}{105C78}
\definecolor{red3}{HTML}{d60f0f}

\graphicspath{{xfigures999/}}

\begin{document}
\title{Breakdown of Quantum Chaos in the Staggered-Field XXZ Chain: Confinement and Meson Formation}

\author{Julia Wildeboer}
\affiliation{Division of Condensed Matter Physics and Materials Science, Brookhaven National Laboratory, Upton, NY 11973-5000, USA}
\affiliation{Perimeter Institute for Theoretical Physics, Waterloo, Ontario, N2L 2Y5, Canada}

\author{Marton Lajer}
\affiliation{Division of Condensed Matter Physics and Materials Science, Brookhaven National Laboratory, Upton, NY 11973-5000, USA}

\author{Robert M. Konik}
\affiliation{Division of Condensed Matter Physics and Materials Science, Brookhaven National Laboratory, Upton, NY 11973-5000, USA}
\begin{abstract}
Confinement produces mesonic bound states, but their existence alone does not determine the statistical organization of the many-body spectrum. We investigate this connection in the spin-\(\tfrac12\) XXZ chain subject to a longitudinal staggered field, using symmetry-resolved exact diagonalization for spin chains of up to \(N=22\) sites. As the exchange anisotropy is increased into the antiferromagnetic Ising regime, the finite-size spectra cross over from Gaussian-orthogonal-ensemble-like statistics to reduced level repulsion. This evolution occurs in both the zero-momentum and generic nonzero-momentum sectors examined. Direct measurements of the domain-wall-number operator and its eigenstate variance reveal increasingly sharp approximate domain-wall sectors, accompanied by banding of spin correlations and bipartite entanglement. Statistics within a selected domain-wall band further characterize the spectral organization beyond the full-sector gap-ratio analysis. We identify the one-meson branch and compare its energies and level spacings with established Airy and semiclassical descriptions and a finite-chain strong-anisotropy treatment, quantifying their accuracy and systematic deviations. These results connect the known confined two-spinon spectrum to the broader finite-anisotropy organization of many-body eigenstates and provide quantitative evidence for emergent domain-wall constraints in the full microscopic Hamiltonian.
\end{abstract}

\maketitle

\section{Introduction}\label{section_intro}
A central question in non-equilibrium many-body physics is how and when highly excited eigenstates depart from the eigenstate
thermalization hypothesis (ETH)~\cite{ETH1,ETH2}. While the ETH
explains why generic interacting systems thermalize, several
mechanisms can obstruct this behavior. The canonical example is
disorder-induced many-body localization~\cite{Basko2006,Nandkishore2015}. In clean translationally invariant systems, kinetic constraints, Hilbert-space fragmentation, and quantum many-body scars provide
distinct disorder-free routes to nonergodic behavior
~\cite{PhysRevB.101.174204,PhysRevX.15.011035,Sala2020Fragmentation,Moudgalya2022Review,Yang2020StrictConfinement,Kohlert2023Fragmentation,Chandran2023,PhysRevB.106.205142,PhysRevB.105.L060301,PhysRevB.102.235106,PhysRevB.104.L121103}.

Hilbert-space fragmentation occurs when exact or emergent conservation laws divide the Hilbert space into dynamically disconnected or weakly coupled subsectors, thereby preventing complete thermalization. In its strong form, as realized, for example, in dipole-conserving or fracton-like Hamiltonians, dynamics remain confined within independent fragments while in its weak form, approximate conservation laws produce long-lived subspaces that eventually thermalize but only on very long timescales~\cite{PhysRevB.101.174204,Sala2020Fragmentation,Moudgalya2022Review}. Quantum many-body scars, by contrast, correspond to special low-entanglement eigenstates embedded within a continuum of thermal ones, leading to coherent oscillations and slow relaxation from particular initial conditions~\cite{Bernien2017Simulator,Bluvstein2021Scars,Turner2018Scars,Turner2018PRB,Choi2019SU2Scars,Ho2019Periodic,Shiraishi2017ETH,Schecter2019Spin1Scar,Moudgalya2018Exact,PhysRevB.102.041118,PhysRevB.99.180302}. Together, these two ideas capture much of the current theoretical landscape of disorder-free nonergodicity~\cite{Chandran2023}.

A complementary, and physically transparent, route to ergodicity breaking arises from confinement. When local interactions generate a linear potential between fractionalized excitations, these excitations become bound into composite objects, i.e., ``mesons'' that can no longer freely propagate. This confinement suppresses the mixing of local degrees of freedom and thereby hinders thermalization~\cite{Rutkevich2010,James2013,PhysRevB.99.195108,James,PhysRevE.90.052105}. In this sense, confinement can be viewed as a natural, dynamical realization of {\it weak Hilbert-space fragmentation}, producing long-lived, structured eigenstates without fine-tuning or disorder~\cite{James}.

The paradigmatic realization of confinement in one dimension is the Ising chain in a longitudinal field, where spinons experience a linear potential and form bound states that manifest as a ``meson ladder.'' 
Recent theoretical work has shown that such confined systems display slow relaxation, nonthermal eigenstate structure, and characteristic entanglement features reflecting their composite excitations ~\cite{PhysRevB.99.195108,James}. 
The connection between confinement and anomalous thermalization has substantial precedents beyond Ising. Quasilocalized dynamics have also been demonstrated in the staggered-field XXZ chain~\cite{PhysRevB.102.041118}. Strict-confinement models provide a further connection to Hilbert-space fragmentation~\cite{Yang2020StrictConfinement}. Independently, the confined excitation spectrum of the staggered-field XXZ chain has been studied through numerical spectroscopy and analytic weak-field and strong-anisotropy approaches~\cite{Bera2017SrCo2VO8,PhysRevB.106.134405,Rutkevich2018}. Anomalous relaxation also occurs in the integrable XXZ chain without a staggered field. In Ref.~\cite{PhysRevLett.122.127202} high-temperature spin superdiffusion at the isotropic point and zero magnetization, with dynamical exponent $(z=3/2)$ was understood. In the easy-axis regime at nonzero magnetization, magnon bound states can yield anomalous local relaxation with continuously varying exponents~\cite{Gopalakrishnan2019}. These hydrodynamic effects provide complementary context for the distinct confinement-induced spectral reorganization studied here. 

Our focus is the organization of the finite-size many-body spectrum of the full staggered-field XXZ Hamiltonian as the exchange anisotropy is varied. Atypical meson eigenstates can coexist with a thermal continuum, and their existence does not by itself determine the level statistics of the surrounding spectrum. We therefore investigate how changes in symmetry-resolved level repulsion accompany the development of correlation and entanglement bands and increasingly well-defined domain-wall content. In particular, the eigenstate variance of the domain-wall-number operator measures the degree to which this organization survives at finite anisotropy, where domain-wall number is not exactly conserved.

We combine this spectral analysis with a quantitative study of the one-meson branch. Established meson theory supplies an analytic identification of these states, while comparisons of absolute energies and level spacings assess its performance for the finite chains studied. The contribution is thus a quantitative account of the relationship between the finite-anisotropy spectral crossover, approximate domain-wall sectorization, and meson spectroscopy. We describe the resulting band structure as fragmentation-like; the observed approximate sectorization does not by itself establish exact Hilbert-space fragmentation or a thermodynamic nonergodic phase.
\begin{figure}[b]
\includegraphics[width=1.00\columnwidth]{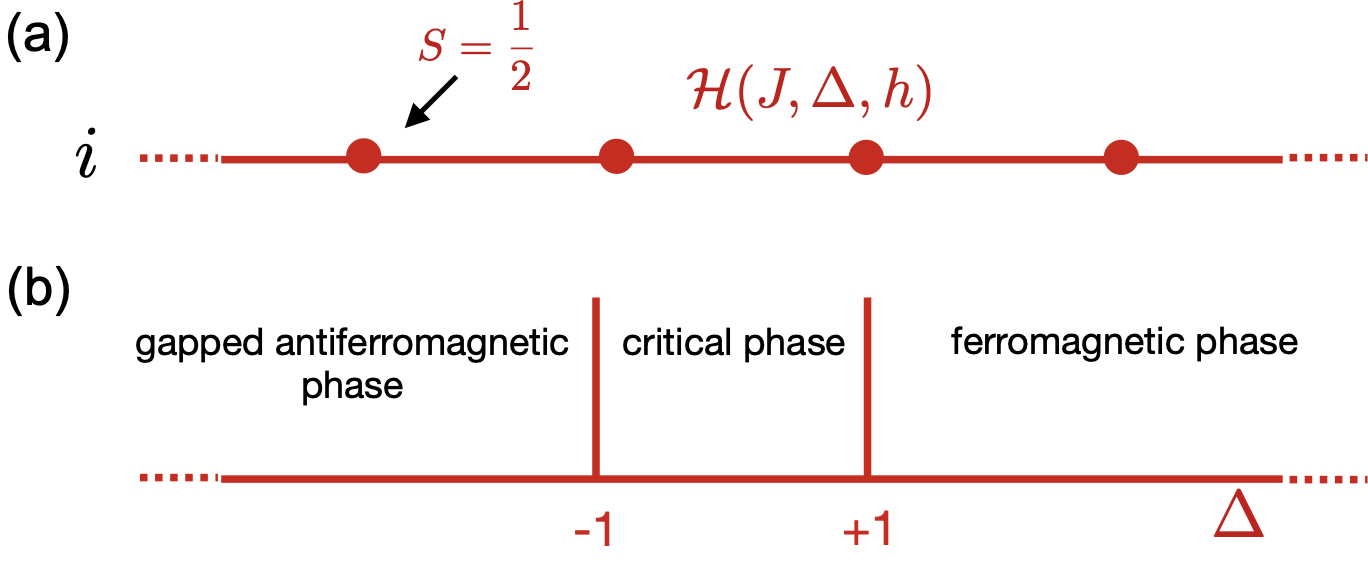}
\caption{ 
         {\it The spin-$\frac{1}{2}$ XXZ spin chain and its zero field phase diagram.} 
         (a) A one-dimensional chain of $S=\frac{1}{2}$ spins that interact via couplings $J$ and $\Delta$ in the presence of a staggered magnetic field. 
         (b) At zero staggered magnetic field $(h=0)$ the Hamiltonian $\mathcal{H}(J,\Delta,0)$ realizes three distinct phases depending on the anisotropy parameter $\Delta$, i.e. the ferromagnetic phase at $\Delta > 1$, the critical phase 
         at $-1 < \Delta < 1$, and the gapped antiferromagnetic (AF) phase at $\Delta < -1$. The gapped AF phase is the regime in which the confinement-induced meson states studied below occur.
        }
\label{figure1}
\end{figure}

The behavior of the XXZ chain connects directly to other disorder-free mechanisms of ETH violation. In constrained dipole-conserving models, strong versus weak fragmentation have been distinguished depending on whether exact higher-moment conservation strictly sectorizes dynamics or whether small symmetry-breaking perturbations eventually restore ergodicity while leaving long-lived subspaces~\cite{PhysRevB.101.174204,Sala2020Fragmentation}.  
Analogously, the approximate conservation of the number of domain walls $W$ in our model produces fragmentation-like spectral organization into weakly hybridizing, approximately $W$-resolved manifolds.

This picture naturally relates to the study of quantum many-body scars, where certain low-entanglement eigenstates coexist with an otherwise chaotic spectrum. Scarred dynamics have been observed in the PXP model relevant for Rydberg-atom chains, as well as in the Affleck–Kennedy–Lieb–Tasaki (AKLT) model and its deformations~\cite{Moudgalya2018AKLT,Bernien2017Simulator,Turner2018Scars,Turner2018PRB,Choi2019SU2Scars,Ho2019Periodic,Shiraishi2017ETH,Schecter2019Spin1Scar,Affleck1987,Burnell}. The AKLT chain provides a paradigmatic example since its exact matrix-product ground state and tower of quasiparticle excitations form a protected, nonthermal subspace within a local Hamiltonian~\cite{Affleck1987,Moudgalya2018AKLT}. Recent studies of deformed AKLT-type and spin-1 XY models have revealed families of scarred excitations and weakly fragmented dynamics~\cite{Schecter2019Spin1Scar,Burnell}.

Our staggered-field XXZ model complements these developments from a different angle. Instead of algebraically protected subspaces, confinement here provides an energetic mechanism that organizes the finite-size spectrum into meson bands and reduces entanglement within these bands. This organization is reminiscent of fragmentation and scar phenomenology, providing a bridge between exactly solvable and physically realistic models. 

Moreover, confinement is not merely a theoretical construct.
Signatures of confined bound states have been experimentally observed in several quasi-one-dimensional quantum magnets, including YbAlO$_3$ and members of the Ba(Sr)Co$_2$V$_2$O$_8$ family, where weak interchain couplings can generate an effective staggered molecular field  
and confine spinons into discrete bound-state ladders observed by inelastic neutron scattering and terahertz spectroscopy
~\cite{wu2019tomonaga,Grenier2015BaCo2VO8,
Bera2017SrCo2VO8,Wang2015SrCoConfine}. 
These materials thus provide realistic experimental platforms for exploring the interplay between confinement, fragmentation, and nonergodicity. 

A complementary realization of field-tuned XXZ-chain physics is
provided by Cs$_2$CoCl$_4$, which is well described as an
$S=1/2$ XXZ chain in a transverse magnetic field~\cite{Breunig2013_Cs2CoCl4}. Although its field geometry differs from the staggered longitudinal field studied here, it illustrates the broader experimental accessibility of nonintegrable, field-perturbed XXZ chains. 

Among the diagnostics studied below, the confined-meson spectrum has the most direct connection to experiment. The meson energies and their momentum dependence enter the dynamical spin structure factors and can therefore be probed by spectroscopic measurements. In particular, inelastic neutron scattering has directly resolved confined spinon bound states in quasi-one-dimensional magnets such as YbAlO$_3$ and
SrCo$_2$V$_2$O$_8$~\cite{wu2019tomonaga,Bera2017SrCo2VO8}, while
terahertz spectroscopy has observed the corresponding hierarchy of spinon-pair excitations in SrCo$_2$V$_2$O$_8$~\cite{Wang2015SrCoConfine}. For the staggered-field XXZ chain, both transverse and longitudinal
dynamical structure factors in the weak-confinement regime have been derived explicitly~\cite{PhysRevB.106.134405}, providing a direct link between the meson spectrum considered here and experimentally measured spectral response. By contrast, level statistics and eigenstate entanglement are primarily microscopic diagnostics in the present work. They reveal the many-body spectral reorganization accompanying the experimentally accessible confinement physics.

Real quasi-one-dimensional magnets are, of course, not exact
realizations of an isolated nearest-neighbor XXZ chain in an externally imposed staggered field. In fact, the weak interchain interactions present in materials such as YbAlO$_3$ and SrCo$_2$V$_2$O$_8$ are themselves responsible for generating an effective staggered molecular field and hence the confining potential between spinons~\cite{wu2019tomonaga,Bera2017SrCo2VO8,PhysRevB.106.134405}. The observation of well-defined confined modes in these weakly coupled systems indicates that the basic confinement mechanism does not require an idealized strictly one-dimensional Hamiltonian. Additional material-dependent interactions can shift or broaden individual meson levels and may weaken the approximate domain-wall-number conservation found below. Nevertheless, one expects the qualitative bound-state structure to persist provided such perturbations do not destroy the Ising-like antiferromagnetic regime or the effective confining potential. Establishing the quantitative robustness of the spectral-statistical and fragmentation-like signatures to such additional interactions is an important direction for future work. 

With these motivations, we use the staggered-field XXZ chain as a quantitative and conceptually transparent setting in which confinement can be related to fragmentation-like and scar-like spectral organization. 
We show that confinement produces a clear finite-size crossover from GOE-like level statistics at weak anisotropy toward reduced level repulsion deep in the antiferromagnetic regime, banded structures in spin correlations and entanglement entropy linked to $W$, and a confined meson spectrum that can be compared quantitatively with complementary Airy, WKB, and finite-$N$ strong-anisotropy descriptions. The model therefore brings together global and local diagnostics of confinement-driven spectral reorganization within a single, microscopically tractable Hamiltonian.

Our central findings on the XXZ model tie together four complementary diagnostics. 
{\it (i) Level statistics.} We observe a crossover from Gaussian
orthogonal ensemble (GOE) behavior at weak anisotropy toward reduced level repulsion deep in the antiferromagnetic (AF) regime, with the finite-size spectra remaining intermediate between the GOE and Poisson limits over the parameter range studied. {\it (ii) Correlation banding.} A simple nearest-neighbor correlator $C^{zz}_j$ (referenced to the ground state) reveals flat ``bands'' of eigenstates that are approximately labeled by a domain-wall number $W$. {\it (iii) Entanglement reorganization.} The bipartite von Neumann entropy $S^{\rm vN}$ exhibits the familiar ETH ``dome'' at small anisotropy $|\Delta|$, approaching the symmetry-matched random-state benchmark near mid-spectrum. As $|\Delta|$ increases and confinement strengthens, $S^{\rm vN}$ reorganizes into bands with entanglement well below this benchmark, mirroring the behavior seen in the correlation diagnostics. 
{\it (iv) Quantitative meson spectroscopy.} We compare the ED meson ladder with three complementary analytic descriptions and identify their distinct regimes of quantitative applicability.

Conceptually, this unifies global and local diagnostics in the form of level statistics and entanglement/correlations, respectively, with controlled analytics in a single lattice model. While our focus is on confinement, the connection to ``scar-like'' phenomenology is instructive. 
The meson bands furnish atypical, low-entanglement eigenstates embedded among more highly entangled states, and their structured organization is reminiscent of quantum many-body scar phenomenology~\cite{Bernien2017Simulator,Bluvstein2021Scars,Turner2018Scars,Turner2018PRB,Choi2019SU2Scars,Ho2019Periodic,Shiraishi2017ETH,Schecter2019Spin1Scar,Moudgalya2018Exact}.  
The key distinction is one of origin and scope: this organization follows from a generic binding mechanism rather than from an exact embedded tower, giving the resulting phenomenology a direct connection to realistic spin-chain materials.

The remainder of this work is organized as follows.
Section~\ref{section_spinchain} introduces the staggered-field XXZ model and its symmetries. 
Section~\ref{subsec:results_levelstats} presents the level-statistics analysis and the finite-size crossover toward reduced level repulsion. Sections~\ref{subsec:results_correlators} and~\ref{sec:results_entanglement} examine the accompanying
reorganization of local correlations and eigenstate entanglement, respectively, including the emergence of approximately domain-wall-resolved bands. Section~\ref{sec:meson_analytics} develops the analytic descriptions of the confined meson spectrum and compares them quantitatively with exact diagonalization. Section~\ref{section_conclusions} summarizes our conclusions and
discusses experimental implications and future directions. 
\begin{table*}[t] 
\small 
\centering 
\begin{ruledtabular}
\begin{tabular}{l|ccl} 
\textbf{symmetry} & \textbf{symbol} & \,\,\,\,\,\,\,\,\,\,\,\,\,\,\textbf{quantum number} & \,\,\,\,\,\,\,\,\,\,\,\,\,\,\textbf{action/description} \\ \hline 
U(1) spin (about $z$) & $S_{\rm tot}^{z}$ & \,\,\,\,\,\,\,\,\,\,\,\,\,\,$S_{\rm tot}^{z}$ & \,\,\,\,\,\,\,\,\,\,\,\,\,\,total magnetization $S_{\rm tot}^{z}=\sum_{i=1}^{N}\sigma_{i}^{z}$ is conserved \\ 
two-site translation & $\mathcal{T}_{2}$ &
\,\,\,\,\,\,\,\,\,\,\,\,\,\,$\kappa_{m}$ &
\,\,\,\,\,\,\,\,\,\,\,\,\,\,$i\mapsto i+2$; $\kappa_{m}=4\pi m/N$,
$m=0,\dots,N/2-1$ \\
inversion (parity) & $\mathcal{I}$ &
\,\,\,\,\,\,\,\,\,\,\,\,\,\,$I=\pm1$ at $\kappa=0,\pi$ &
\,\,\,\,\,\,\,\,\,\,\,\,\,\,sublattice-preserving site-centered reflection \\
& & & \,\,\,\,\,\,\,\,\,\,\,\,\,\,$i\mapsto 2-i\pmod N$ \\ 
translation--spin flip & $\mathcal{C}_{\rm flip} =\mathcal{T}_{1}\prod_{i}\sigma_{i}^{x}$ & \,\,\,\,\,\,\,\,\,\,\,\,\,\, 
$c_{\rm flip}=\zeta\, e^{i\kappa/2}$ $(\zeta=\pm1)$ &
\,\,\,\,\,\,\,\,\,\,\,\,\,\,one-site translation composed with a global spin flip \\[1pt] 
\end{tabular} 
\end{ruledtabular}
\caption{
         {\it Symmetries of the XXZ chain with a staggered field and periodic boundary conditions (PBCs).} 
         The staggered field breaks the one-site translation $\mathcal{T}_{1}$ while preserving the two-site translation $\mathcal{T}_{2}=\mathcal{T}_{1}^{2}$, leading to two-site translation phases $\kappa_{m}=4\pi m/N$. The inversion $\mathcal{I}$ is chosen as the site-centered reflection $i\mapsto 2-i\pmod N$, with site labels understood periodically. This reflection preserves the two staggered sublattices and hence the staggered-field term. It reverses the two-site momentum, $\mathcal{I}\,\mathcal{T}_{2}\,\mathcal{I}^{-1} =\mathcal{T}_{2}^{-1}$, so an independent inversion eigenvalue $I=\pm1$ can be assigned within a single momentum sector only for $\kappa=0$ and, when present, $\kappa=\pi$. The combined operation $\mathcal{C}_{\rm flip}$ compensates the sign change of the staggered field under $\mathcal{T}_{1}$ by a global spin flip and obeys $\mathcal{C}_{\rm flip}^{\,2}=\mathcal{T}_{2}$. Consequently, in a momentum-$\kappa$ sector its two eigenvalues are $c_{\rm flip}=\zeta\, e^{i\kappa/2}$ with $\zeta=\pm1$ and at $\kappa=0$ these reduce to $c_{\rm flip}=\pm 1$. The discrete-symmetry structure and its relation to the ED implementation are discussed in Appendix~\ref{app:Z2}. At $\kappa=\pi$, inversion exchanges the two $\zeta$ branches, so $I$ and $\zeta$ cannot be resolved simultaneously.
} 
\label{table1} 
\end{table*}

\section{XXZ spin chain models}\label{section_spinchain}
Motivated by the search for long-lived nonergodic excitations in realistic materials (e.g., rare-earth spin-chain compounds such as YbAlO$_3$), we study the spin-$\frac{1}{2}$ XXZ chain subject to a staggered magnetic field. Our focus throughout is the antiferromagnetic Ising regime where confined domain-wall bound states (``mesons'') emerge. 

\subsection{XXZ chain in a staggered magnetic field}
We consider the periodic spin chain described by the Hamiltonian  
\begin{align}
\begin{split}
\mathcal{H}(J,\Delta,h) = -&J \sum_{i=1}^{N} \left( \sigma_i^x \sigma_{i+1}^x
+ \sigma_i^y \sigma_{i+1}^y
+ \Delta \,\sigma_i^z \sigma_{i+1}^z \right)\\
-&h \sum_{i=1}^{N} (-1)^i \sigma_i^z 
\label{spinchain}
\end{split}
\end{align} 
with $J>0$, Pauli matrices $\sigma_i^{\alpha}$, $(\alpha=x,y,z)$ on site $i$, anisotropy $\Delta$, and staggered-field strength $h$. We assume periodic boundary conditions (PBC), i.e. $\sigma_{N+1}^{\alpha}\equiv\sigma_{1}^{\alpha}$.

At zero staggered field $h$ the model~\eqref{spinchain} exhibits three standard phases as $\Delta$ is varied:
(i) a ferromagnetic Ising phase for $\Delta>1$, 
(ii) a critical (Luttinger-liquid) phase for $-1<\Delta<+1$, and
(iii) a gapped antiferromagnetic Ising phase for $\Delta<-1$
[see Fig.~\ref{figure1}(a) and (b)]. 
Note that the overall minus sign in~\eqref{spinchain} is our convention. With $J>0$ this places the AF Ising regime at $\Delta<-1$. 
Turning on the staggered field $(h\neq0)$ breaks one-site translation symmetry and, in the Ising regime, confines spinons into mesonic bound states whose low-entanglement spectral organization is reminiscent of scar phenomenology.

\subsection{Symmetries of the spin chain Hamiltonian $\boldsymbol{\mathcal{H}(J,\Delta,h)}$}\label{subsec:symmetries} 

The model~\eqref{spinchain} possesses several useful symmetries, summarized in Table~\ref{table1}. First, the Hamiltonian is invariant under U(1) spin rotations about the $z$ axis, so that the total magnetization 
\begin{equation} 
S_{\rm total}^{z}=\sum_{i=1}^{N} \sigma_i^z 
\end{equation} 
is conserved. 

The staggered magnetic field breaks translation by one lattice site, but translation by two sites remains an exact symmetry. We denote the corresponding operator by $\mathcal{T}_{2}=\mathcal{T}_{1}^{2}$ and define the two-site translation phase $\kappa$ through
\begin{equation}
\mathcal{T}_{2}|\psi_{\kappa}\rangle
=e^{i\kappa}|\psi_{\kappa}\rangle 
\end{equation}
with
\begin{equation}
\kappa_m=\frac{4\pi m}{N},
\qquad
m=0,\ldots,\frac{N}{2}-1 \,.
\end{equation} 
For even $N$, the Hamiltonian also possesses a spatial inversion symmetry provided that the reflection center is chosen so that the two staggered sublattices are preserved. With the site convention of Eq.~\eqref{spinchain}, we define 
\begin{equation} 
\mathcal{I}:\qquad i\longmapsto 2-i\pmod N \,. 
\end{equation} 
Because $(-1)^{2-i}=(-1)^i$, this site-centered reflection leaves both the exchange interaction and the staggered-field term invariant. By contrast, the bond-centered reflection $i\mapsto N+1-i$ exchanges the two staggered sublattices and reverses the sign of the staggered-field term. It is therefore not a symmetry by itself for $h \neq 0$. The inversion reverses the two-site momentum, i.e. $\mathcal{I}\,\mathcal{T}_{2}\,\mathcal{I}^{-1} =\mathcal{T}_{2}^{-1}$, so an independent parity eigenvalue $I= \pm 1$ can be assigned within a single momentum sector only at the inversion-invariant momenta $\kappa=0$ and $\kappa=\pi$ (the latter being present when allowed by the finite-size momentum grid). Finally, although neither a one-site translation nor a global spin flip is separately a symmetry for $h \neq 0$, their product is. Defining 
\begin{equation} 
\mathcal{C}_{\rm flip} = \mathcal{T}_{1}\prod_{i=1}^{N}\sigma_i^x 
\end{equation} 
one finds 
\begin{equation} 
[\mathcal{C}_{\rm flip},\mathcal{H}]=0, \qquad \mathcal{C}_{\rm flip}^{\,2}=\mathcal{T}_{2}\,. 
\end{equation} 
Thus, in a momentum-$\kappa$ sector, $c_{\rm flip}^{\,2}=e^{i\kappa}$ and the two branches are $c_{\rm flip}=\zeta\, e^{i\kappa/2}$ with $\zeta=\pm1$. 
In particular, $c_{\rm flip}=\pm 1$ at $\kappa=0$. The inversion-center subtlety, the combined translation--spin-flip symmetry, and their relation to the symmetry operation used in our ED basis construction are derived explicitly in Appendix~\ref{app:Z2}. For reference, the size of the Hilbert-space block selected by $(S^{z}_{\rm total},\kappa,\mathcal{I},\mathcal{C}_{\rm flip}) =(0,0,+1,+1)$ as a function of $N$ is listed in Table~\ref{table2}, where the last column shows the fraction of the $S^{z}_{\rm total}=0$ sector retained after resolving $(\kappa,\mathcal{I},\mathcal{C}_{\rm flip})=(0,+1,+1)$. 
 
\begin{table}[b]
  \centering
  \begin{tabular}{@{}rrrr@{}}
    \toprule \toprule
    $N$ & $\dim\mathcal{H}(S^z_{\rm tot}{=}0)=\binom{N}{N/2}$ & $\dim\mathcal{H}_{\rm sub}$ & $\dim\mathcal{H}_{\rm sub}/\binom{N}{N/2}$ \\
    \midrule
     4 &      6   &      3   & 0.50  \\
     6 &     20   &      6   & 0.30  \\
     8 &     70   &     11   & 0.157 \\
    10 &    252   &     26   & 0.103 \\
    12 &    924   &     62   & 0.0671 \\
    14 &   3432   &    170   & 0.0495 \\
    16 &  12870   &    487   & 0.0379 \\
    18 &  48620   &   1530   & 0.0315 \\
    20 & 184756   &   4947   & 0.0268 \\
    22 & 705432   &  16718   & 0.0237 \\
    \bottomrule \bottomrule
  \end{tabular}
\caption{
         {\it Block dimensions for the XXZ chain (PBC) in the symmetry sector $(S^z_{\rm tot},\kappa,\mathcal{I},\mathcal{C}_{\rm flip})=(0,0,+1,+1)$.} The second column lists the size of the $S^z_{\rm tot}{=}0$ block, and the last column shows the fraction kept after projecting to $(\kappa,\mathcal{I},\mathcal{C}_{\rm flip}) = (0, +1, +1)$.
         }
\label{table2}
\end{table}

\begin{figure*}[t]
\includegraphics[width=1.00\textwidth]{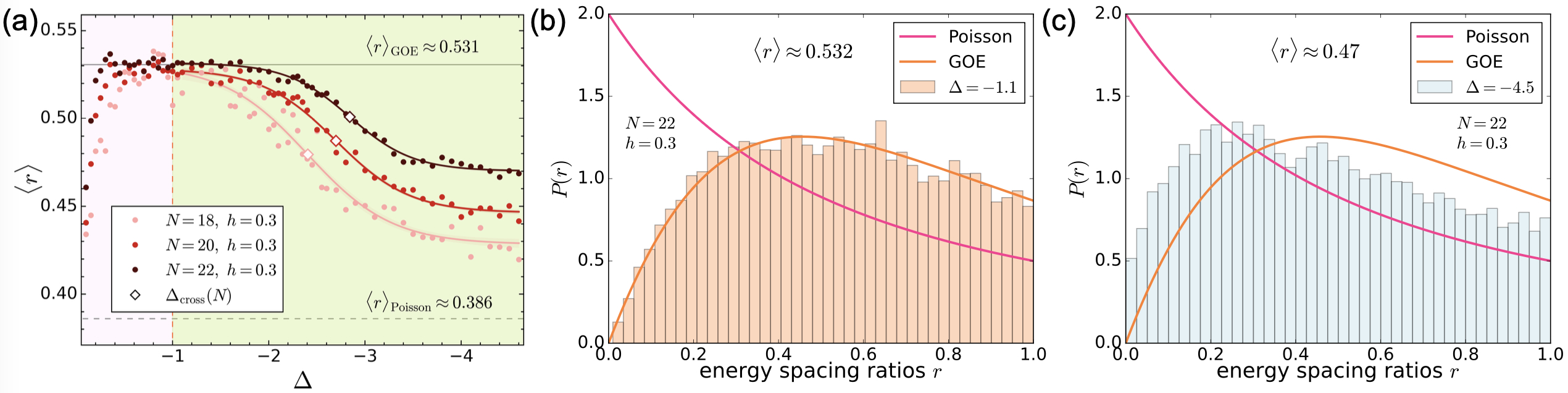}
\caption{
         {\it Level-statistics diagnostics for the spin-$\frac{1}{2}$ XXZ chain $\mathcal{H}(1,\Delta,h)$ in a staggered magnetic field, resolved in the symmetry sector $(S_{\mathrm{total}}^{z},\kappa,\mathcal{I},\mathcal{C}_{\mathrm{flip}})=(0,0,+1,+1)$.}
         (a) Mean adjacent-gap ratio $\langle r\rangle$ as a function of anisotropy $\Delta$ for chain lengths $N=18,20,22$ at $h=0.3$. The raw data are shown over the full calculated interval $-0.1\geq\Delta\geq-4.6$. The solid and dashed horizontal lines denote the large-matrix GOE reference value $\langle r\rangle_{\mathrm{GOE}}\simeq0.5307$ and the Poisson value $\langle r\rangle_{\mathrm{Poisson}} =2\ln2-1\simeq0.3863$, respectively. The solid colored curves are four-parameter sigmoidal fits performed only over $1.1\leq|\Delta|\leq4.6$. The points at $|\Delta|<1.1$ are displayed for context but are not included in the fits. The faint envelopes around the fitted curves indicate the 16th--84th percentile residual-bootstrap ranges. For each size, the open diamond marks the fitted midpoint $\Delta_{\mathrm{cross}}(N)$, at which the fitted curve has its maximum slope. The crossover shifts toward more negative anisotropy with increasing $N$. The vertical dashed line at $\Delta=-1$ and the background shading refer to the zero-field XXZ phase diagram: the region $-1<\Delta<0$ shown here belongs to the zero-field critical regime, whereas $\Delta<-1$ is the zero-field antiferromagnetic regime. They are included only as reference markers and do not represent a finite-$h$ phase boundary.
         (b) Distribution $P(r)$ of the adjacent-gap ratios for $N=22$, $\Delta=-1.1$, and $h=0.3$. The histogram is compared with the Poisson distribution (magenta) and the GOE Wigner--Dyson surmise (orange). The numerical mean $\langle r\rangle\simeq0.532$ is consistent with GOE-like level statistics.
         (c) Same as panel (b), but for $\Delta=-4.5$, deep in the antiferromagnetic regime. The distribution exhibits reduced level repulsion and shifts toward smaller spacing ratios relative to the GOE form, with $\langle r\rangle\simeq0.47$. The spectrum remains intermediate between the GOE and Poisson limits, consistent with a finite-size crossover away from GOE-like statistics and with the emergent domain-wall organization discussed in Sec.~\ref{subsec:results_correlators}. 
}
\label{figure2}
\end{figure*}

\section{Level statistics and finite-size crossover toward reduced level repulsion}\label{subsec:results_levelstats}
To characterize the global spectral properties of the
staggered-field XXZ chain in Eq.~\eqref{spinchain}, we begin
with the statistics of its many-body energy levels. Throughout
this section, the energy eigenvalues $E_{n}$ are ordered in ascending order and restricted to 
the fully resolved
symmetry block
\begin{equation}
(S^{z}_{\rm total},\kappa,I,\mathcal{C}_{\rm flip})=(0,0,+1,+1)\,.
\end{equation}
Resolving all available unitary symmetries is essential, since
combining independent symmetry sectors would artificially suppress
level repulsion. The corresponding block dimensions for
$N=4,\ldots,22$ are listed in Table~\ref{table2}. The $\kappa=0$ sector is inversion invariant and therefore admits the additional parity resolution $I=\pm1$ that is absent at generic $\kappa\neq0,\pi$.  To verify that the spectral crossover discussed below is not a consequence of this special residual symmetry structure, Appendix~\ref{app:momentum_robustness} repeats the level-statistics analysis in two generic nonzero-$\kappa$ sectors.

For each consecutive triplet of energy levels $(E_{n}, E_{n+1}, E_{n+2})$, we define the ratio
of adjacent spacings
\begin{equation}
r_{n} =
\frac{\min\!\left(E_{n+1}-E_n,E_{n+2}-E_{n+1}\right)}
{\max\!\left(E_{n+1}-E_n,E_{n+2}-E_{n+1}\right)}\,.
\label{eq:adjacent_gap_ratio}
\end{equation}
The corresponding spectrum-averaged quantity is
\begin{equation}
\langle r\rangle = \big\langle r_n\big\rangle 
\label{eq:mean_gap_ratio}
\end{equation}
where the average is taken over the complete spectrum within the
selected symmetry block. 

Random-matrix theory predicts distinct values of
$\langle r\rangle$ for chaotic and uncorrelated spectra~\cite{GOE,Poisson}. Time-reversal-invariant chaotic systems are
described by the Gaussian orthogonal ensemble (GOE), for which
$\langle r\rangle_{\rm GOE} \simeq 0.5307$, whereas an uncorrelated
Poisson spectrum of an integrable or a many-body localized system  gives $\langle r\rangle_{\rm Poisson} \simeq 0.386$.
In addition to the mean ratio, we examine the full probability
distribution $P(r)$ in order to resolve changes in the level
repulsion more directly.

Figure~\ref{figure2}(a) shows $\langle r\rangle$ as a function of the anisotropy $\Delta$ for $N=18,20,22$ at $h=0.3$. The anisotropy grid has been sampled more densely in the regions where the curves depart from their weak-anisotropy plateaus. Near the isotropic regime, the three system sizes give values close to the GOE benchmark, consistent with quantum-chaotic spectral correlations. As $\Delta$ becomes increasingly negative, $\langle r\rangle$ decreases continuously and the departure from the GOE regime shifts toward larger $|\Delta|$ with increasing system size. At the largest anisotropies studied, the values remain above the Poisson benchmark. The finite-size spectra therefore show a crossover toward reduced level repulsion rather than a complete
approach to the Poisson limit. This reduction in $\langle r\rangle$ therefore signals a finite-size crossover away from GOE-like statistics and is consistent with the emergent approximate domain-wall organization identified below.

The raw data in Fig.~\ref{figure2}(a) are displayed over the full 
calculated range $-0.1\geq\Delta\geq-4.6$, so that the approach to
the GOE-like weak-anisotropy regime is visible explicitly. The
sigmoidal crossover analysis, however, is restricted to the interval
$1.1\leq|\Delta|\leq4.6$, i.e. to $\Delta\leq-1.1$ on the
antiferromagnetic side of the zero-field XXZ phase diagram. The
additional points with $|\Delta|<1.1$ are shown for context but are
not included in the fits. The vertical dashed line at $\Delta=-1$
and the background shading in panel (a) therefore refer only to the
critical-to-antiferromagnetic boundary of the zero-field model and
should not be interpreted as a phase boundary at finite staggered
field.

The full distributions in Figs.~\ref{figure2}(b) and \ref{figure2}(c) illustrate the same evolution for $N=22$. At $\Delta=-1.1$, the distribution follows the Wigner--Dyson form characteristic of the GOE and gives $\langle r\rangle \simeq 0.532$. By contrast, at $\Delta=-4.5$ the probability weight is shifted toward smaller spacing ratios and the mean is reduced to $\langle r\rangle \simeq 0.47$. The latter distribution remains intermediate between the GOE and Poisson limits, but the suppression of level repulsion is consistent with the increasingly weak hybridization of the domain-wall-number bands identified below. 

We have also checked explicitly that this evolution is not specific
to the inversion-invariant $\kappa=0$ sector.  As detailed in
Appendix~\ref{app:momentum_robustness}, we repeated the calculation
for $N=18$ and $20$ in the first two generic nonzero two-site
translation sectors, $\kappa=4\pi/N$ and $8\pi/N$, resolving the
$\zeta=+1$ branch of $\mathcal{C}_{\rm flip}$.  In both sectors,
$\langle r\rangle$ shows the same qualitative crossover from GOE-like statistics at weak anisotropy to reduced level repulsion at large $|\Delta|$. The full adjacent-gap-ratio distributions likewise remain GOE-like at $\Delta=-1.1$ and shift toward reduced level repulsion at $\Delta=-4.5$, while remaining intermediate between the GOE and Poisson limits. Thus, while the precise finite-size values of $\langle r\rangle$ depend on the momentum sector, the crossover from GOE-like statistics toward reduced level repulsion is not a peculiarity of the special $\kappa=0$ block.

Since the evolution in Fig.~\ref{figure2}(a) is smooth, a crossover anisotropy cannot be identified uniquely by inspecting the first data point that falls below an arbitrarily chosen cutoff. Such a prescription would be sensitive both to local fluctuations of $\langle r\rangle$ and to the spacing of the sampled anisotropy values. We instead define an operational and reproducible finite-size crossover by fitting the complete curve for each system size. 

Introducing
\begin{equation}
x=|\Delta|
\end{equation}
we fit the data for each $N$ to the four-parameter sigmoidal form
\begin{equation}
\langle r \rangle_N(x) = r_{\rm low}^{(N)} +
\frac{r_{\rm high}^{(N)}-r_{\rm low}^{(N)}}
{1+\exp\!\left[\left(x-x_{\rm cross}^{(N)}\right)/w_N\right]}\,.
\label{eq:r_sigmoid}
\end{equation}
Here, $r_{\rm high}^{(N)}$ and $r_{\rm low}^{(N)}$ describe
the weak- and strong-anisotropy finite-size plateaus, respectively,
while $w_N$ characterizes the width of the crossover. The parameter
$x_{\rm cross}^{(N)}$ is the midpoint of the fitted curve, at which
\begin{equation}
\langle r \rangle_{N} =
\frac{r_{\rm high}^{(N)}+r_{\rm low}^{(N)}}{2} 
\end{equation}
and the magnitude of the fitted slope is maximal. We therefore
define
\begin{equation}
\Delta_{\rm cross}(N) = -x_{\rm cross}^{(N)}\,.
\label{eq:Delta_cross_definition}
\end{equation}
Equation~\eqref{eq:r_sigmoid} is used only as a smooth operational
parameterization of the finite-size crossover. It is not assumed to
be a universal scaling function or evidence for a thermodynamic
phase transition.
\begin{figure}[t]
\includegraphics[width=1.00\columnwidth]
{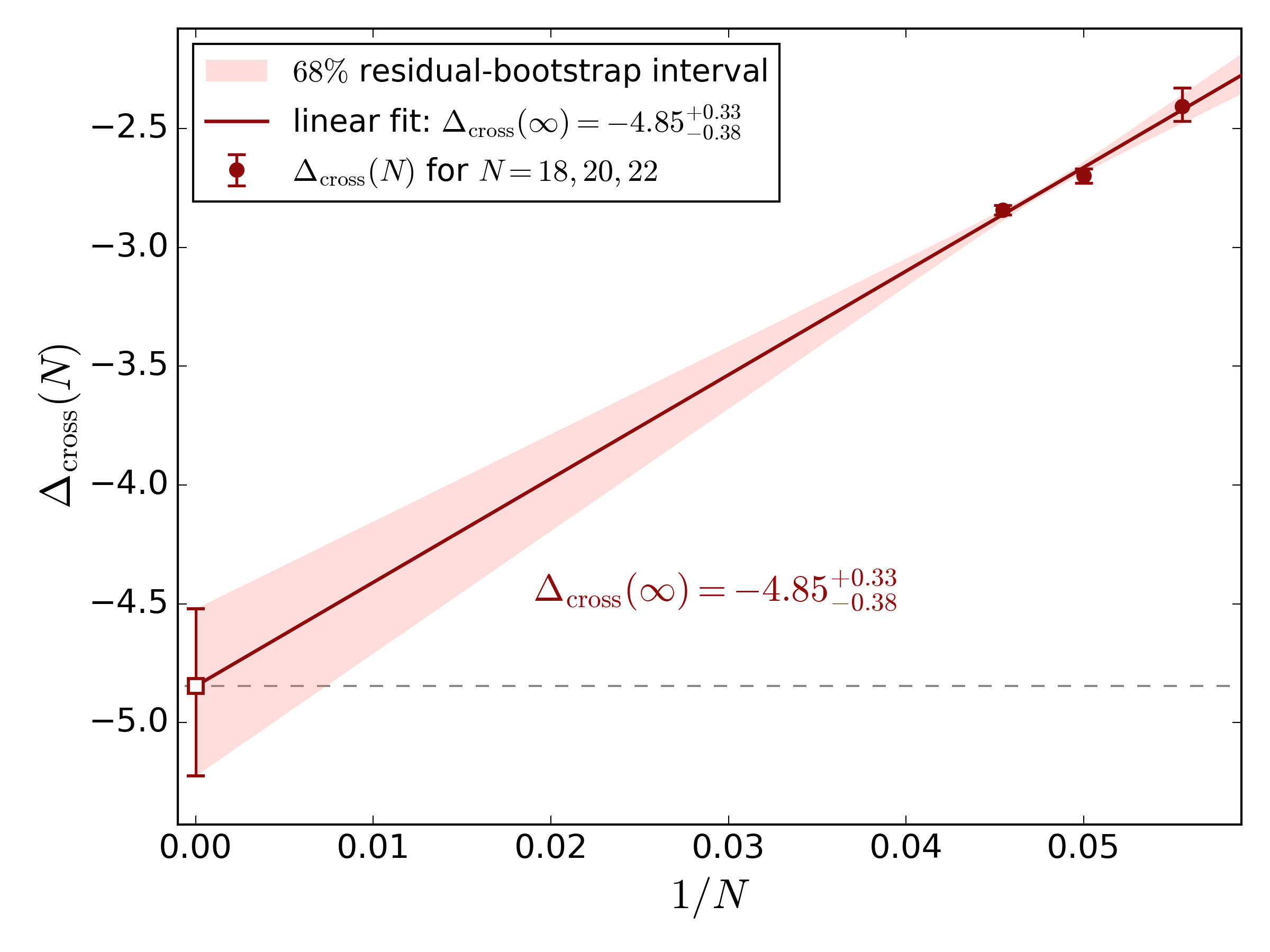}
\caption{
         {\it Finite-size scaling of the level-statistics crossover anisotropy $\Delta_{\rm cross}(N)$ for the spin-$\frac12$ XXZ chain $\mathcal{H}(1,\Delta,h)$ at $h=0.3$ in the symmetry sector $(S^{z}_{\rm total},\kappa,\mathcal{I},\mathcal{C}_{\rm flip}) = (0,0,+1,+1)$.}
         The points are the midpoints of the sigmoidal fits to $\langle r\rangle(\Delta)$ in Fig.~\ref{figure2}(a) for $N=18,20,22$. The vertical error bars denote the 16th--84th percentile intervals obtained from $5000$ residual-bootstrap realizations. The solid line is an unweighted fit to $\Delta_{\rm cross}(N) =\Delta_{\rm cross}(\infty)+a/N$. The shaded region is the corresponding 68\% interval obtained by propagating the matched bootstrap samples through the finite-size fit. The extrapolation gives $\Delta_{\rm cross}(\infty) =-4.85^{+0.33}_{-0.38}$. Because only three sizes are available, this result is interpreted as an indicative crossover scale rather than as a sharply determined thermodynamic transition point.
}
\label{figure3}
\end{figure}

The solid curves in Fig.~\ref{figure2}(a) show the resulting
sigmoidal fits, and the open diamonds mark their midpoints. The fits
describe the numerical data well, with coefficients of determination
$R^2\simeq0.954$, $0.988$, and $0.993$ for
$N=18,20$, and $22$, respectively. They yield
\begin{align}
\Delta_{\rm cross}(18)
&=
-2.406^{+0.077}_{-0.063}
\nonumber\\
\Delta_{\rm cross}(20)
&=
-2.699^{+0.030}_{-0.031}
\nonumber\\
\Delta_{\rm cross}(22)
&=
-2.843^{+0.020}_{-0.020}\,.
\label{eq:Delta_cross_values}
\end{align}
The quoted intervals correspond to the 16th and 84th percentiles of
a residual-bootstrap analysis described below. 
The background shading and the vertical line at $\Delta=-1$ in
Fig.~\ref{figure2}(a) refer to the zero-field XXZ phase diagram, as
described above, and are independent of the fitted crossover locations.

To estimate the uncertainty in $\Delta_{\rm cross}(N)$, we employ a residual bootstrap. For each system size, the residuals about the best-fit curve are centered and resampled with replacement. The resampled residuals are then added to the fitted curve, and the resulting synthetic data are refitted to Eq.~\eqref{eq:r_sigmoid}. We perform $5000$ matched bootstrap realizations, retaining one fitted crossover value for each of the three system sizes in every realization. The central intervals in Eq.~\eqref{eq:Delta_cross_values} are obtained from the 16th and 84th percentiles of the resulting distributions. This procedure accounts for the uncertainty associated with extracting a smooth crossover from the finite-size data and avoids assigning the 
crossover to a single fluctuating data point.

The fitted crossover anisotropies become progressively more negative
with increasing $N$. To characterize this drift, we use the linear
finite-size form
\begin{equation}
\Delta_{\rm cross}(N) =
\Delta_{\rm cross}(\infty) + \frac{a}{N}\,.
\label{eq:Delta_cross_scaling}
\end{equation}
Because only three system sizes are available and systematic finite-size corrections are likely more important than the modest differences between the bootstrap uncertainties of the individual points, we use an unweighted linear fit for the central values. For each matched bootstrap realization, we repeat the same linear fit, thereby propagating the uncertainty of the three sigmoid midpoints to the thermodynamic intercept.

The result is shown in Fig.~\ref{figure3}. The central extrapolation
gives
\begin{equation}
\Delta_{\rm cross}(\infty) = -4.85^{+0.33}_{-0.38}
\label{eq:Delta_cross_infinity}
\end{equation}
where the uncertainty is the propagated 68\% residual-bootstrap
interval. The corresponding 95\% interval is considerably broader,
approximately
\begin{equation}
-5.71 \lesssim \Delta_{\rm cross}(\infty) \lesssim -4.22\,.
\end{equation}
The extrapolated value should therefore be regarded as an indicative
thermodynamic crossover scale rather than as a precise critical
anisotropy. In particular, the quoted bootstrap interval propagates
the uncertainty of the sigmoid fits, but does not include systematic
uncertainties associated with the choice of fitting function, fitting
window, or linear $1/N$ scaling ansatz. Nevertheless, the consistent
finite-size drift demonstrates that the departure from GOE-like
statistics moves deeper into the antiferromagnetic Ising regime as
the chain length is increased. 
\begin{table}[b]
\begin{ruledtabular}
\begin{tabular}{ccc}
$h$ & $\langle r\rangle\ (N{=}20)$ & $\langle r\rangle\ (N{=}22)$ \\[2pt] \hline
    0.1 & 0.437191 & 0.443036 \\
    0.2 & 0.428236 & 0.472977 \\
    0.3 & 0.452605 & 0.496836 \\
    0.4 & 0.455710 & 0.500026 \\
    0.5 & 0.456062 & 0.505295 
\end{tabular}
\end{ruledtabular}
\caption{
         {\it Mean adjacent-gap ratio $\langle r\rangle$ for eigenstates belonging to the isolated spectral band associated with $W\simeq10$ at $\Delta=-4.5$ in the symmetry block $(S^z_{\rm total},\kappa,I,C_{\rm flip})=(0,0,+1,+1)$.}
        The states are selected from the full Hamiltonian spectrum using the well-separated correlation branch associated with $W\simeq10$, as in Fig.~\ref{figure5}(a). No projection onto a fixed-$W$ subspace is performed. The exact relation between $C_j^{zz}$ and $\langle\widehat W\rangle_j$ provides the equivalent direct domain-wall-number identification. Values are shown for $N=20$ and $N=22$ as a function of the staggered field $h$. For reference, $\langle r\rangle_{\rm GOE}\simeq0.5307$ and $\langle r\rangle_{\rm Poisson}=2\ln2-1\simeq0.3863$. The selected $W\simeq10$ band contains 1651 eigenstates for $N=20$ and 5419 eigenstates for $N=22$.
}
\label{table_r}
\end{table}
\begin{figure*}[t]
\includegraphics[width=1.00\textwidth]{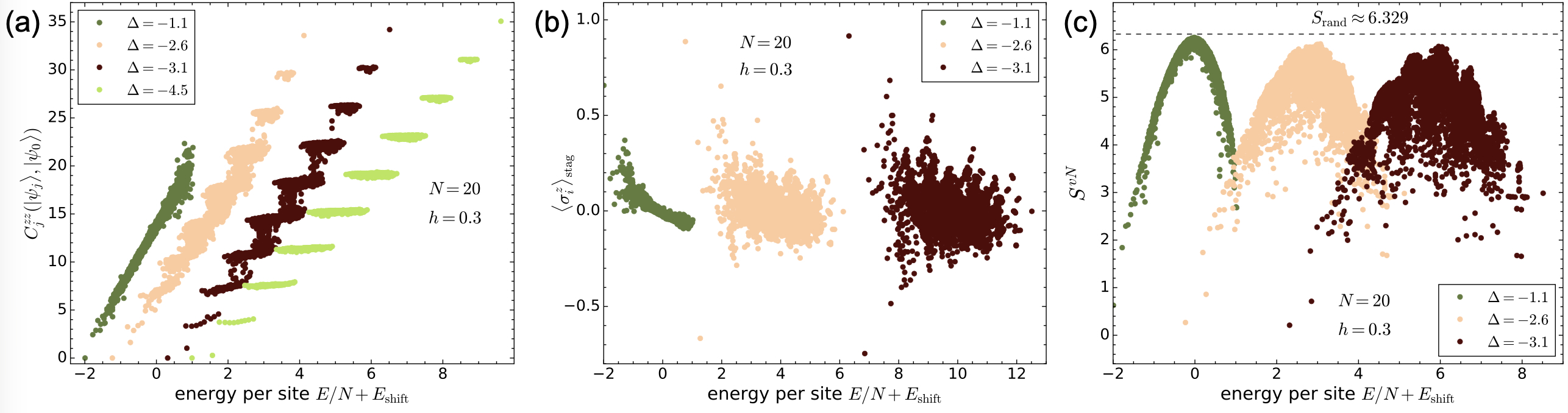}
\caption{
         {\it Correlation measure, staggered magnetization, and entanglement entropy across the spectrum of the staggered-field XXZ chain.}
        (a) Correlation measure $C^{zz}_j(\lvert\psi_j\rangle,\lvert\psi_0\rangle)$ versus  energy per site $E/N{+}E_{\rm shift}$ for $N=20$, $h=0.3$, and anisotropies $\Delta=-1.1,-2.6,-3.1,-4.5$ with respective energy shift $E_{\rm shift} = 0.0, 2.0, 4.0, 6.0$. For small $|\Delta|$, $C^{zz}_j$ forms a single broad cloud while for larger $|\Delta|$ the spectrum splits into distinct branches associated with different domain-wall numbers $W$ (banding). 
        (b) Staggered magnetization $\langle\sigma_i^z\rangle_{\rm stag}$ for the same system, which remains small and narrowly distributed around zero at weak anisotropy and develops a pronounced banded structure as $|\Delta|$ grows, consistent with strengthening antiferromagnetic correlations and domain-wall organization. The energy shift $E_{\rm shift} = 0.0, 4.0, 10.0$ was applied for $\Delta=-1.1,-2.6,-3.1$, respectively. 
        (c) Von Neumann entanglement $S^{\rm vN}$ for the same parameters with $E_{\rm shift} = 0.0, 3.0, 6.0$ for $\Delta=-1.1,-2.6,-3.1$. At small $|\Delta|$, $S^{\rm vN}$ exhibits a volume-law ``dome'' consistent with quantum chaos while for larger $|\Delta|$, clear bands of reduced entropy emerge, signaling confinement-induced banding and increasingly well-defined domain-wall organization [also see Fig.~\ref{figure5}(c)]. 
        A dashed horizontal line marks the symmetry-matched random-state benchmark $S_{\rm rand}=6.329$ for a half-chain cut ($N_A=N_B=10$), obtained from real random states in the same $(S_{\rm total}^{z},\kappa,I,C_{\rm flip})=(0,0,+1,+1)$ symmetry sector as the ED eigenstates.
}
\label{figure4}
\end{figure*}

To examine whether the reduction of level repulsion is tied to the domain-wall organization of the spectrum, we additionally evaluate the level statistics within the isolated spectral band associated with domain-wall number $W\simeq10$ at $\Delta=-4.5$. The states entering this analysis are eigenstates of the full Hamiltonian and are selected from the clearly separated $W\simeq10$ branch of the correlation-band
structure illustrated in Fig.~\ref{figure5}(a), i.e. no projection onto a fixed-$W$ subspace is performed. Through the exact relation Eq.~(\ref{eq:Czz_W_expectation_identity}) between $C_j^{zz}$ and $\langle\widehat W\rangle_j$, this branch assignment is equivalent to identifying the corresponding cluster of eigenstates centered near $\langle\widehat W\rangle_j\simeq10$. Table~\ref{table_r} lists the resulting mean adjacent-gap ratio $\langle r\rangle$ as a function of $h$ for $N=20$ and $N=22$.

For $N=22$, the mean ratio increases from approximately $0.443$ at $h=0.1$ to $0.505$ at $h=0.5$, indicating increased level repulsion within the selected $W\simeq10$ band as the staggered
field is increased. 
The variation is weaker and mildly nonmonotonic for $N=20$, where the smaller number of states makes finite-size effects more pronounced. These band-resolved results support the interpretation that the reduction of level repulsion in the complete symmetry block arises in part from the progressive organization of the spectrum into weakly coupled domain-wall manifolds. The direct expectation-value and variance diagnostics presented in Sec.~\ref{subsec:direct_domain_walls} make this emergent domain-wall structure explicit.

\section{Correlations and domain-wall diagnostics}
\label{subsec:results_correlators}
To complement the global level-statistics analysis of Sec.~\ref{subsec:results_levelstats}, we next examine observables that resolve the structure of individual many-body eigenstates. This section is organized in two parts. In Sec.~\ref{subsec:correlation_banding}, we introduce the staggered magnetization and nearest-neighbor correlation observables and show how their spectral distributions develop a pronounced banded structure as the system enters the antiferromagnetic Ising regime. In Sec.~\ref{subsec:direct_domain_walls}, we make the domain-wall
interpretation of this banding explicit by introducing the
domain-wall-number operator $\widehat{W}$ and analyzing both its
eigenstate expectation value and its variance. The corresponding
reorganization of the eigenstate entanglement is discussed separately in Sec.~\ref{sec:results_entanglement}. 

\subsection{Correlation observables and spectral banding}
\label{subsec:correlation_banding}
To further characterize the properties of the staggered-field XXZ chain~\eqref{spinchain}, we now turn to the study of correlation functions. These observables provide insight into the magnetic correlations of the system as well as the emergence of domain-wall excitations in different parameter regimes. For an eigenstate $|\psi\rangle$, we evaluate both local magnetization and nearest-neighbor spin correlations.

We first define the staggered longitudinal magnetization,
which quantifies the staggered spin polarization of an eigenstate, as 
\begin{equation}\label{expec1a}
\langle \sigma^{z} \rangle_{\mathrm{stag}}
= \frac{1}{N} \sum_{i=1}^{N} (-1)^i \langle \psi | \sigma_i^{z} | \psi \rangle\,.
\end{equation}
Here, $\sigma_i^z$ is the Pauli $z$-operator on site $i$, and
$N$ is the total number of sites~\footnote{The average longitudinal magnetization defined as Eq.~\eqref{expec1a} without the alternating factor $(-1)^{i}$ is zero since we work in the $S^{z}_{\rm total} = 0$ sector.}. Because the Hamiltonian contains an explicit staggered field, $\langle\sigma^z\rangle_{\rm stag}$ should be viewed here as a measure of the staggered magnetic response rather than as a spontaneous order parameter.

In addition, we analyze the average nearest-neighbor spin–spin correlation
\begin{equation}\label{expec2}
\langle \sigma_i^{z} \sigma_{i+1}^{z} \rangle_{\mathrm{av}}
= \frac{1}{N} \sum_{i=1}^{N} \langle \psi | \sigma_i^{z} \sigma_{i+1}^{z} | \psi \rangle\,.
\end{equation}  
This quantity is directly related to the density of domain walls and provides a convenient probe of the strength of nearest-neighbor antiferromagnetic correlations.

For large negative anisotropy $\Delta$ (deep in the antiferromagnetic regime), the number of domain walls is expected to become a quasi-conserved quantity. To test this hypothesis, we introduce the correlation measure 
\begin{align}
\label{test_corrSzSz}
\begin{split}
C^{zz}_{j}(|\psi_{j}\rangle,|\psi_{0}\rangle) &= 
\langle \psi_{j}| \sum_{i=1}^{N}\sigma^{z}_{i} \sigma^{z}_{i+1}|\psi_{j}\rangle \\
&- \langle \psi_{0}| \sum_{i=1}^{N}\sigma^{z}_{i} \sigma^{z}_{i+1}|\psi_{0}\rangle    
\end{split}
\end{align}
where $|\psi_{j}\rangle$ is an excited eigenstate and $|\psi_{0}\rangle$ is the ground state of the Hamiltonian~\eqref{spinchain}.
The quantity $C^{zz}_{j}(|\psi_{j}\rangle, |\psi_{0}\rangle)$ measures the deviation of nearest-neighbor spin correlations in 
$|\psi_{j}\rangle$ from those of the ground state and serves as an indicator of the domain-wall content of the excitation. 

\begin{figure*}[t]
\includegraphics[width=1.00\textwidth]{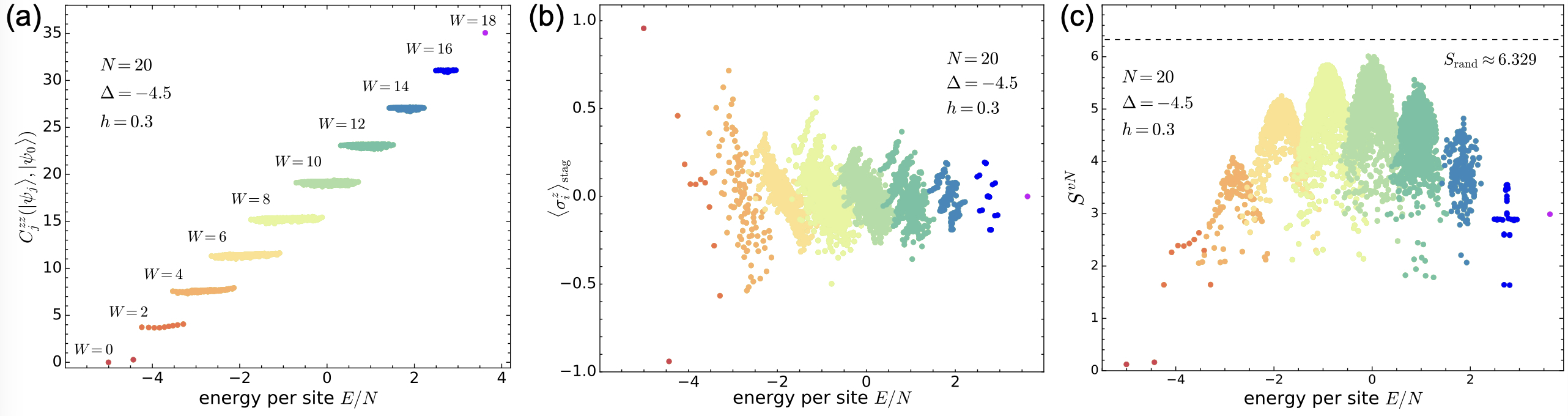}
\caption{
        {\it Banded structure at strong anisotropy ($\Delta=-4.5$) and its entanglement signature.}
        (a) Correlation measure $C^{zz}_{j}(|\psi_{j}\rangle, |\psi_{0}\rangle)$ versus energy per site $E/N$ for eigenstates of the $N=20$ chain at $h=0.3$ deep in the antiferromagnetic Ising regime at anisotropy $\Delta =-4.5$. 
        The eigenstates segregate cleanly into flat bands approximately labeled by the domain-wall number $W$, highlighting the quasi-conserved nature of $W$ in the deep antiferromagnetic Ising regime. Notably, the branch associated with the one-meson ($W\simeq2$) sector sits near $C^{zz}_{j}(|\psi\rangle,|\psi_{0}\rangle)\simeq+4$, corresponding to an average domain-wall content close to two aligned bonds (two kinks).
        (b) $\langle\sigma^z_i\rangle_{\rm stag}$ for the same parameters follows the same band pattern, further validating the domain-wall identification.  
        (c) The von Neumann entanglement entropy $S^{\rm vN}$ exhibits a corresponding banded structure: the low-lying $W\simeq2$ mesons have markedly reduced entanglement compared with the surrounding higher-entanglement states. The dashed horizontal line denotes the symmetry-matched random-state benchmark $S_{\rm rand}=6.329$ for a half-chain cut ($N_{A}=N_{B}=10$), obtained by averaging over $1000$ real random normalized states in the same $(S_{\rm total}^{z},\kappa,I,C_{\rm flip})=(0,0,+1,+1)$ symmetry sector as the ED eigenstates.
}
\label{figure5}
\end{figure*}

The behavior of these correlation functions across the many-body spectrum is shown in Figs.~\ref{figure4}(a) and ~\ref{figure5}(a) for a chain of size $N=20$ at several values of the anisotropy parameter $\Delta$.  
Figure~\ref{figure4}(a) displays the correlation measure $C^{zz}_{j}$ versus the energy per site $E/N + E_{\rm shift}$~\footnote{The energy shift $E_{\rm shift}$ is introduced to avoid the data belonging to the different values of anisotropy $\Delta$ (partially) lying on top of each other.} for different anisotropies $\Delta = -1.1, -2.6, -3.1, -4.5$. For weak anisotropy $(\Delta=-1.1)$, the values of $C^{zz}_{j}$ form a single broad distribution, indicating that domain-wall sectors are strongly mixed and not associated with well-defined quantum numbers. As $|\Delta|$ increases, the spectrum develops a clear banded structure, where states cluster into branches with nearly constant $C^{zz}_{j}$. This banding becomes more pronounced for large negative anisotropy $(\Delta=-4.5)$ as depicted in Figs.~\ref{figure4}(a) and ~\ref{figure5}(a), consistent with the picture that the number of domain walls $W$ becomes a quasi-conserved quantity in the deep antiferromagnetic Ising regime. 

Figure~\ref{figure4}(b) shows the corresponding staggered
magnetization $\langle\sigma_i^z\rangle_{\rm stag}$ for the same
parameters $\Delta=-1.1,-2.6,-3.1$. At weak anisotropy, the
distribution remains narrow around zero, indicating only a weak
staggered polarization across most of the many-body spectrum.  As $|\Delta|$ increases, the distribution broadens and develops a pronounced banded structure, reflecting stronger antiferromagnetic correlations and the progressive organization of the eigenstates by their domain-wall content.

Finally, Fig.~\ref{figure5} focuses on the case $\Delta = -4.5$, deep inside the antiferromagnetic regime.  
Figure~\ref{figure5}(a) shows that the eigenstates segregate into well-defined flat branches associated with the even domain-wall numbers $W=2n$, with $n=0,1,\dots,N/2$, confirming that $W$ acts as an emergent approximate quantum number.  Figure~\ref{figure5}(b) shows that the staggered magnetization exhibits the same band structure, with each branch associated with a distinct approximate domain-wall manifold. The low-energy branch centered near $W=2$ contains the one-meson excitations with strong antiferromagnetic correlations, while the higher-$W$ branches exhibit progressively smaller staggered magnetization.

In summary, the correlation analysis reveals a crossover from a
weak-anisotropy regime, where domain-wall sectors are strongly mixed and no clear band structure is visible, to a strongly confined regime at large negative $\Delta$, where the number of domain walls becomes approximately conserved, the spectrum organizes into bands, and meson-like bound states emerge. 

\subsection{Direct domain-wall-number diagnostics}
\label{subsec:direct_domain_walls}
The banding observed in the nearest-neighbor correlation measure
suggests that the eigenstates become approximately organized by their domain-wall content in the antiferromagnetic Ising regime. To make this interpretation explicit, we introduce the domain wall number operator 
\begin{eqnarray}
\widehat{W} =
\frac{1}{2}\sum_{i=1}^{N}
\left(\mathds{1}+\sigma_{i}^{z}\sigma_{i+1}^{z}\right)\,.
\label{eq:W_operator}
\end{eqnarray}
An antiferromagnetically aligned bond contributes zero to
$\widehat{W}$, whereas an aligned bond contributes one. Thus,
$\widehat{W}$ counts the aligned bonds, or equivalently the kinks relative to the N\'eel background. For an even periodic chain its eigenvalues are the even integers
\begin{eqnarray}
W=0,2,4,\ldots,N \,.
\end{eqnarray}

The operator in Eq.~\eqref{eq:W_operator} is directly related to the nearest-neighbor Ising correlation through
\begin{eqnarray}
\sum_{i=1}^{N} \sigma_{i}^{z}\sigma_{i+1}^{z} = 2\widehat W-N \,.
\label{eq:Czz_W_operator_identity}
\end{eqnarray}
Consequently, the ground-state-referenced correlation measure defined in Eq.~\eqref{test_corrSzSz} obeys the exact identity
\begin{eqnarray}
C_j^{zz}\left(|\psi_j\rangle,|\psi_0\rangle \right) =
2\left(\langle\widehat{W}\rangle_{j}-\langle\widehat{W}\rangle_{0}  \right)\,.
\label{eq:Czz_W_expectation_identity}
\end{eqnarray}
We have verified Eq.~\eqref{eq:Czz_W_expectation_identity}
numerically for all eigenstates used below, with deviations at the level of numerical roundoff. 
\begin{figure*}[t]
\includegraphics[width=1.00\textwidth]{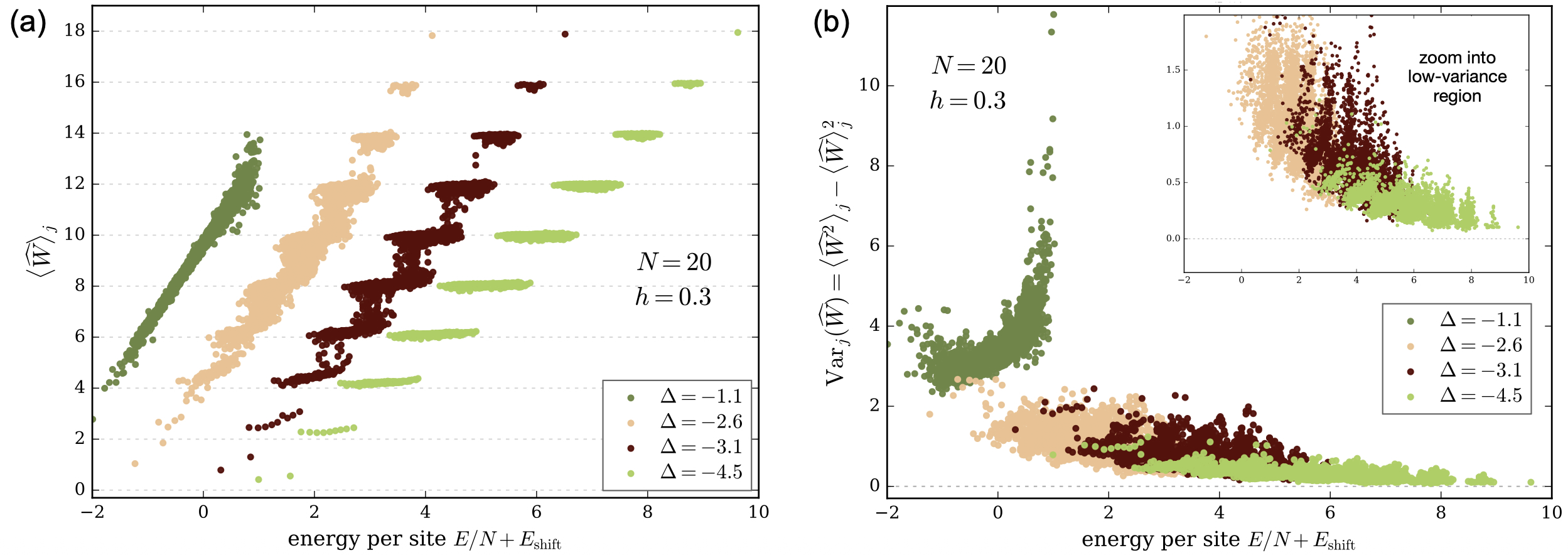}
\caption{
         {\it Direct domain-wall-number diagnostics for $N=20$, $h=0.3$, $S_{\mathrm{tot}}^{z}=0$, $\kappa=0$, $I=+1$, and $C_{\mathrm{flip}}=+1$.}
         (a) Eigenstate expectation value $\langle \widehat{W} \rangle_{j}$, where $\widehat{W}=\frac{1}{2}\sum_{i=1}^{N} (\mathds{1}+\sigma_{i}^{z} \sigma_{i+1}^{z})$. As the antiferromagnetic anisotropy is increased, the eigenstates organize into increasingly sharp bands near the even domain-wall numbers $W=0,2,4,\dots$.
         (b) Corresponding variance $\mathrm{Var}_{j}(\widehat{W})= \langle \widehat{W}^{2}\rangle_{j}-\langle \widehat{W}\rangle_{j}^{2}$. The progressive suppression of the variance shows that $W$ becomes an increasingly well-defined approximate quantum number, 
         while its remaining finite value is consistent with fragmentation-like organization rather than exact fragmentation. For visual clarity, the energy densities for $\Delta=-1.1,-2.6,-3.1$, and $-4.5$ are shifted by $E_{\mathrm{shift}}=0.0,2.0,4.0$, and $6.0$, respectively. The inset in panel (b) displays the same data with the vertical range restricted to $\operatorname{Var}_j(\widehat W)\lesssim 2$, resolving the structure in the low-variance regime. 
         }    
\label{fig:direct_W_diagnostics}
\end{figure*}

The Ising interaction and staggered-field terms are diagonal in
$\widehat{W}$, whereas the transverse exchange can change the
domain-wall number. Therefore $\widehat{W}$ is not an exactly conserved quantity of the full Hamiltonian. Deep in the antiferromagnetic regime, however, states belonging to different $W$ sectors are separated by the large Ising energy scale, suppressing their hybridization and allowing $W$ to emerge as an approximate quantum number.

Figure~\ref{figure5}(a)--(c) illustrates this organization at
$\Delta=-4.5$. 
The flat branches in the correlation measure can be associated
directly with the even domain-wall values $W=0,2,4,\ldots$, with the branch centered near $W=2$ describing the one-meson manifold. The staggered magnetization exhibits the same spectral organization. The corresponding entanglement bands in Fig.~\ref{figure5}(c) are discussed in Sec.~\ref{sec:results_entanglement}. 

Although the expectation value $\langle\widehat W\rangle_{j}$ identifies the average domain-wall content of an eigenstate, it does not by itself determine how sharply the state is concentrated in a single $W$ sector. We therefore also calculate the variance
\begin{equation}
\operatorname{Var}_j(\widehat W)
=
\langle\widehat{W}^{2}\rangle_{j} - \langle\widehat{W}\rangle_{j}^{2}\,.
\label{eq:W_variance}
\end{equation}
This quantity vanishes when $|\psi_j\rangle$ is an exact eigenstate of $\widehat{W}$ and remains finite when different domain-wall sectors are hybridized.

The resulting direct domain-wall diagnostics are shown in
Fig.~\ref{fig:direct_W_diagnostics}. At weak anisotropy, i.e. 
$\Delta=-1.1$, the values of $\langle\widehat{W}\rangle_{j}$ form a broad, nearly continuous distribution, consistent with strong mixing between different domain-wall sectors. As $|\Delta|$ is increased, the eigenstates separate progressively into bands centered near the even integers $W=0,2,4,\ldots$. At $\Delta=-4.5$, these bands are nearly horizontal and closely reproduce the domain-wall labeling inferred from the correlation measure.

The same crossover is visible more directly in
$\operatorname{Var}_{j}(\widehat{W})$. The domain-wall-number fluctuations are large throughout much of the spectrum at $\Delta=-1.1$, but are strongly suppressed as the system is taken deeper into the antiferromagnetic regime. In particular, the median variance over the spectrum decreases from approximately $3.29$ at $\Delta=-1.1$ to approximately $0.279$ at $\Delta=-4.5$, corresponding to a reduction by almost a factor of twelve. The inset of Fig.~\ref{fig:direct_W_diagnostics}(b) resolves the low-variance
region and makes the increasingly organized structure of the 
strong-anisotropy data visible.

The variances remain finite even at $\Delta=-4.5$, showing that
$\widehat{W}$ is not an exact integral of motion. 
The simultaneous formation of nearly integer-valued $\langle\widehat{W}\rangle_j$ bands and the strong, but incomplete, suppression of $\operatorname{Var}_j(\widehat W)$ therefore provide direct operator-level evidence for emergent approximate domain-wall sectorization rather than exact fragmentation. A spectrum-averaged summary of the variance, including the intermediate value $\Delta=-2.2$ and the corresponding interquartile ranges, is presented in Appendix~\ref{appC_domain_walls}.

\section{Eigenstate entanglement across the spectrum}\label{sec:results_entanglement}
Entanglement measures provide a powerful framework to characterize the structure of many-body eigenstates and to distinguish between different dynamical regimes. Unlike conventional correlation functions, which probe specific local observables, entanglement quantifies the global quantum correlations encoded in a wave function. 
This makes it an especially sensitive probe of the eigenstate
reorganization accompanying departures from GOE-like spectral
behavior in interacting quantum systems. 
We quantify it via the bipartite von Neumann entropy 
\begin{eqnarray}
S^{\mathrm{vN}} = -\mathrm{Tr}\!\left(\rho_A \ln \rho_A\right)
\end{eqnarray}
where the system is partitioned into contiguous regions $A$ and $B$ such that the Hilbert space factorizes as $\mathcal{H}=\mathcal{H}_A\otimes \mathcal{H}_B$ and the reduced density matrix of region $A$ is $\rho_A=\mathrm{Tr}_B\,|\psi\rangle\langle\psi|$ obtained by tracing out the degrees of freedom in region $B$. Throughout we use natural logarithms and evaluate $S^{\mathrm{vN}}$ for half-chain cuts ($N_A=N_B=N/2$) for all eigenstates $|\psi\rangle$ of the staggered-field XXZ Hamiltonian~\eqref{spinchain}, analyzing its dependence on anisotropy $\Delta$ and energy density $E/N$. 

As a benchmark for chaotic eigenstates, we use the average 
half-chain entanglement entropy of random real normalized states drawn from the same symmetry-resolved Hilbert-space sector as the ED eigenstates, i.e. $(S_{\rm total}^{z},\kappa,I,C_{\rm flip})=(0,0,+1,+1)$. 
We write a random state in an orthonormal basis 
$\{|\alpha\rangle\}$ of this sector as
\begin{equation}
|\psi_{\rm rand}\rangle
=
\sum_{\alpha=1}^{D_{\rm sym}} c_{\alpha}|\alpha\rangle 
\end{equation}
and then draw the coefficients $c_{\alpha}$ independently from a real Gaussian distribution and normalize the resulting state. The use of real coefficients is appropriate for the present symmetry sector, whose Hamiltonian belongs to the GOE symmetry class. Averaging the half-chain von Neumann entropy over $1000$ such random states for $N_A=N_B=10$ gives
\begin{equation}\label{eq:Srand}
S_{\rm rand}=6.3293 
\end{equation}
with a standard error of approximately $2\times10^{-4}$. For comparison, the unconstrained Page value for a $N_A+N_B=10+10$ bipartition is $10\ln(2)-1/2 \simeq 6.4315$. The latter corresponds to random states in the unrestricted tensor-product Hilbert space and does not incorporate the fixed-magnetization and additional discrete symmetry constraints imposed in our ED calculation. We therefore use $S_{\rm rand}$ as the symmetry-matched random-state benchmark for the chaotic eigenstates.

The behavior of the entanglement entropy across the many-body spectrum is shown in Figs.~\ref{figure4}(c) and \ref{figure5}(c) for $\Delta=-1.1,-2.6,-3.1$, and $-4.5$, respectively. For small $|\Delta|$, plotting $S^{\rm vN}$ versus the energy density $E/N+E_{\rm shift}$ produces the familiar broad, nearly symmetric ETH ``dome'', i.e. the entanglement is suppressed in the spectral tails and approaches the symmetry-matched random-state benchmark near mid-spectrum, as seen in Fig.~\ref{figure4}(c). By contrast, the confined regime contains eigenstates with substantially reduced entanglement and pronounced departures from this random-state benchmark.

Upon increasing $|\Delta|$ into the antiferromagnetic Ising regime, the distribution of $S^{\mathrm{vN}}$ is globally reduced and develops visible structure across the spectrum [Figs.~\ref{figure4}(c) and \ref{figure5}(c)]. The reorganization mirrors the band formation seen in correlation diagnostics. The correlation measure $C^{zz}_{j}(|\psi_{j}\rangle, | \psi_{0}\rangle)$~\eqref{test_corrSzSz} plotted versus the energy per site $E/N$ clusters eigenstates $|\psi \rangle$ into nearly flat branches that are approximately labeled by the domain-wall number $W$ [see Fig.~\ref{figure5}(a)]. Deep in the AF regime ($\Delta=-4.5$) this banding becomes sharp and is reflected one-to-one in $S^{\mathrm{vN}}$. 
The one-meson branch associated with $W\simeq2$ sits near
$C^{zz}_{j}(|\psi_{j}\rangle,|\psi_{0}\rangle)\simeq+4$ and has an average domain-wall content close to two aligned bonds (two kinks). It exhibits distinctly low entanglement compared with the surrounding higher-entanglement states, while higher-$W$ branches appear near $+8,+12,\dots$ with progressively larger entanglement that nevertheless remains well below the symmetry-matched random-state benchmark [Fig.~\ref{figure5}(c)]. 
This clear correspondence between the correlation measure and the entanglement entropy supports the interpretation that domain-wall number is a quasi-conserved quantity in the deep antiferromagnetic Ising regime. By contrast, as discussed previously, for small $|\Delta|$ (e.g. $\Delta=-1.1$), the spectrum forms a broad cloud without clear separation, consistent with GOE-like spectral statistics, ETH-like entanglement structure, and strongly mixed domain-wall content.

The staggered magnetization $\langle\sigma^z_i\rangle_{\rm stag}$ follows the same band pattern, reinforcing the domain-wall labeling and the emergence of an approximate conservation of $W$ in the confined regime.

Taken together, these observations establish the following
coherent picture. At small $|\Delta|$ the spectrum conforms to
ETH expectations, with a broad entanglement dome whose
mid-spectrum values approach the symmetry-matched random-state
benchmark. 
As $|\Delta|$ is made more negative, confinement suppresses
entanglement and reorganizes the spectrum into correlation and
entanglement bands approximately labeled by $W$, consistent with
increasingly constrained, approximately $W$-resolved spectral
organization. These observations are consistent with the level-statistics and correlation-function results, establishing a coherent picture of confinement and the formation of meson-like excitations in the staggered-field XXZ chain and setting the stage for the quantitative meson spectroscopy in Sec.~\ref{sec:meson_analytics}. 

\section{Continuum threshold, confinement, and analytic meson spectrum}\label{sec:meson_analytics}

In this section we analyze the confined meson spectrum of the staggered-field XXZ chain from both analytic and numerical perspectives. We begin in
Sec.~\ref{sec:analytic_meson_spectrum} by summarizing the analytic descriptions of the meson energies based on Rutkevich's theory~\cite{PhysRevB.106.134405} and the subsequent extensions~\footnote{Private Communication with S. Rutkevich.}.  
In particular, we collect the low-energy Airy expansion near the continuum
threshold, the semiclassical/Wentzel-Kramers-Brillouin (WKB) quantization formula for mesons lying well
inside the confinement window, and the strong-anisotropy expansion (SAE), including its finite-$N$ form for periodic chains. 
We then turn in Sec.~\ref{sec:numerical_comparison_mesons} to a
detailed comparison between these analytic predictions and our ED results for the one-meson branch centered near $\langle\widehat W\rangle\simeq2$. In the following, we refer to
this branch for brevity as the ``$W=2$ meson band''. As established in Sec.~\ref{subsec:direct_domain_walls}, this notation denotes an approximately $W$-resolved manifold and does not imply exact conservation of $\widehat W$.

\subsection{Analytic meson spectrum}\label{sec:analytic_meson_spectrum}

We now summarize the analytic descriptions of the confined meson spectrum
that will be compared to ED in Sec.~\ref{sec:numerical_comparison_mesons}. Our starting point is Rutkevich's theory of spinon confinement in the
antiferromagnetic XXZ chain in a staggered field~\cite{PhysRevB.106.134405}. 
That theory is formulated primarily in the thermodynamic limit, although a finite-$N$ periodic-chain version is also available for the
strong-anisotropy expansion. 

In the present work we confront these analytic results with ED data obtained in the symmetry sector $(S^z_{\rm total},\kappa,I,C_{\rm flip})=(0,0,+1,+1)$, where $\kappa$ denotes the phase of the two-site translation, $\mathcal{T}_2|\psi_\kappa\rangle=e^{i\kappa}|\psi_\kappa\rangle$. In the analytic formulas below, following Rutkevich Ref.~\cite{PhysRevB.106.134405}, we instead use $P$ for the total two-kink momentum, $P=p_{1}+p_{2}$, which is defined modulo $\pi$ in the staggered problem. With the translation-operator convention used in our ED construction, the two quantities are related by
\begin{equation}\label{eq:kappa_P_dictionary}
\kappa=-2P\pmod{2\pi}\,.
\end{equation}
Thus the zero-translation-phase sector $\kappa=0$ used in our ED calculation corresponds to the zero-total-momentum sector $P=0$ of the analytic meson theory, which is the case studied throughout this section. 

We also stress at the outset that Rutkevich's original formulas are written in a convention where the microscopic Hamiltonian carries a prefactor $J/2$ instead of $J$.
All expressions below are written consistently in the $J$-convention used
in our ED study and summarized in Appendix~\ref{appB}.
This affects the overall normalization of the kink dispersion,
the continuum threshold, and the confinement-induced meson energies.

At $h=0$, the neutral excitations of the antiferromagnetic XXZ chain may
be viewed as pairs of kinks (spinons). For fixed total momentum $P$, it is convenient to parameterize the two kink momenta as
\begin{equation}
p_1=\frac{P}{2}+p,
\qquad
p_2=\frac{P}{2}-p,
\label{6.1}
\end{equation}
where $p$ is the relative momentum.
The corresponding two-kink energy is then
\begin{equation}
\epsilon(p|P)
= \omega\!\left(\frac{P}{2}+p\right)
+ \omega\!\left(\frac{P}{2}-p\right)
\label{6.2}
\end{equation}
with $\omega(q)$ the single-kink dispersion.
In our convention, we have 
\begin{align}
\begin{split}
\omega(q) &= \Omega_{0}\sqrt{1-k^2\cos^2 q} \\
\Omega_{0} &= \frac{4JK(k)}{\pi}\sinh\eta \\
\Delta & =-\cosh\eta
\label{6.3}
\end{split}
\end{align}
where $K(k)$ is the complete elliptic integral of the first kind and the
elliptic modulus $k$ is fixed by
\begin{equation}
\frac{K(k')}{K(k)}=\frac{\eta}{\pi} \quad \mathrm{with}
\quad
k'=\sqrt{1-k^2}\,.
\label{6.4}
\end{equation}
Thus the anisotropy enters the dispersion through the parameter $\eta$ and the associated elliptic modulus.

For fixed total momentum $P$, the lower boundary of the deconfined
two-kink continuum is obtained by minimizing $\epsilon(p|P)$ with respect
to the relative momentum $p$.
In the regime of interest, this minimum occurs at $p=0$, giving
\begin{align}
\begin{split}
E_{\rm cont}(P) &= \epsilon(0|P) = 2\omega\!\left(\frac{P}{2}\right)\\
&= 2\Omega_{0}\sqrt{1-k^2\cos^2\left(\frac{P}{2}\right)}\,.
\label{6.5}
\end{split}
\end{align}
At the momentum relevant for our ED comparison, namely $P=0$, this reduces to
\begin{equation}
E_{\rm cont}(0)=2\omega(0)=2\Omega_{0}\sqrt{1-k^2}\,.
\label{6.6}
\end{equation}
This threshold plays a central role below since all three analytic
descriptions of the meson spectrum are naturally organized relative to the deconfined continuum edge.

Turning on a staggered field $h$ confines the two kinks through a linear
potential. The corresponding string tension is
\begin{equation}
f=2h\,\bar{\sigma}(\eta) 
\label{6.7}
\end{equation}
where $\bar{\sigma}(\eta)$ denotes the spontaneous staggered
magnetization. Therefore confinement converts the two-kink continuum into a discrete ladder of bound states. Depending on where these states lie within the confinement window, three analytic descriptions become useful. 
The first is the low-energy expansion, appropriate for mesons lying close
to the lower edge of the continuum.
In the spin-zero sector, the two channels are distinguished by the action of 
Rutkevich's modified one-site translation operator $\widetilde{T}_1^{(R)}$ and the
corresponding near-threshold ladder takes the form
\begin{equation}
E_{\pm,n}(P)
= E_{\rm cont}(P) + A(P)\,z_n + f\,a_{\pm}(P) + O(f^{4/3}) 
\label{6.8}
\end{equation}
where $z_{n}>0$ are the Airy magnitudes defined by 
$\mathrm{Ai}(-z_n)=0$. The coefficient
\begin{equation}
A(P) = f^{2/3} \left[\frac{\epsilon''(0|P)}{2} \right]^{1/3}
\label{6.9}
\end{equation}
sets the characteristic Airy scale, while the channel-dependent constants
$a_{\pm}(P)$ are fixed by the derivatives of the scattering phases.
The explicit evaluation of $a_{\pm}(0|\eta)$ in our convention is given in Appendix~\ref{appB}. 
For the even $+$ channel relevant to the
$(S^z_{\rm total},\kappa,I,C_{\rm flip})=(0,0,+1,+1)$ sector, we therefore write
\begin{align}
\begin{split}
E_n^{\mathrm{th}} &\equiv E_{+,n}(0)\\
& =
E_{\rm cont}(0) + A(0)\,z_n + f\,a_+(0|\eta) + O(f^{4/3})\,.
\label{6.10}
\end{split}
\end{align}
This is the low-energy, or Airy-ladder, theory used in our comparison to
the lowest ED meson levels. 

A useful consequence of the near-threshold Airy description is that it 
also provides an estimate for the stability window of the one-meson
states. At $h > 0$ the deconfined two-kink continuum is removed, and a
one-meson state remains stable only until it reaches the lightest
two-meson channel at the same total momentum. For fixed $P$, momentum
conservation assigns momenta $Q$ and $P-Q$ to the two mesons, so the
lowest two-meson threshold is
\begin{equation}
E_{\rm th}^{(2M)}(P) = \min_{Q}
\left\{ E_1^{(+)}(Q)+E_1^{(+)}(P-Q) \right\}\,.
\label{6.10a}
\end{equation}
Using the near-threshold form~\eqref{6.10}, the $n$-th one-meson level in
the even channel is stable provided that 
\begin{equation}
E_n^{(+)}(P) < E_{\rm th}^{(2M)}(P) 
\label{6.10b}
\end{equation}
or, equivalently,
\begin{equation}
z_{n} < \frac{E_{\rm th}^{(2M)}(P)-E_{\rm cont}(P)-f\,a_+(P)}{A(P)}\,.
\label{6.10c} 
\end{equation}
Since the Airy magnitudes $z_{n}$ grow monotonically with $n$, the number of
stable one-meson levels at fixed momentum $P$ is 
\begin{align}
\begin{split}
&n_{\rm th}^{\star}(P) = \\
&\qquad\max\left\{n: z_{n} < \frac{E_{\rm th}^{(2M)}(P)-E_{\rm cont}(P)-f\,a_+(P)}{A(P)}
\right\}\,.
\label{6.10d}
\end{split}
\end{align}

At $P=0$, for the lowest-meson dispersion relevant here, 
the minimum in Eq.~\eqref{6.10a} is attained at $Q=0$. 

If we work in the binding-energy reference
\begin{equation}
b_n \equiv E_n-E_{\rm cont}(0) 
\label{6.10e}
\end{equation}
then the corresponding two-meson threshold in the same binding reference is
\begin{equation}
b_{\rm th}^{(2M)}(0) =
E_{\rm cont}(0)+2\left[A(0)\,z_1+f\,a_+(0|\eta)\right]\,.
\label{6.10f}
\end{equation}
Accordingly, the number of stable mesons predicted by the low-energy
theory at $P=0$ is
\begin{equation}
n_{\rm th}^{\star}(0) = \max\left\{n:b_n < b_{\rm th}^{(2M)}(0)
\right\}\,.
\label{6.10g}
\end{equation}
Since $A(P)\propto f^{2/3}$, the right-hand side of
Eq.~\eqref{6.10c} scales as $f^{-2/3}$ for weak confinement, implying that $n_{\rm th}^{\star}(P)$ increases as $h \to 0$ and many shallow meson levels crowd near the deconfined threshold. 

For comparison with ED, it is useful to define an analogous threshold
directly from the finite-size spectrum by using the lowest level in the branch associated with $W\simeq4$
in
the same ED symmetry sector at $\kappa=0$: 
\begin{equation}
b_{\rm ED}^{(2M)}(0) =
E_{\min}^{(W\simeq4)}-E_{\rm cont}(0) 
\label{6.10h}
\end{equation}
and the associated ED stability count
\begin{equation}
n_{\rm ED}^{\star}(0) = \max\left\{n:
b_n<b_{\rm ED}^{(2M)}(0) 
\right\}\,.
\label{6.10i}
\end{equation}
In Sec.~\ref{sec:numerical_comparison_mesons} these two thresholds are displayed explicitly in the binding plot of Fig.~\ref{figure8}(a), allowing the theoretical and ED stability windows to be read off directly. 

The second description is the semiclassical, or
Wentzel--Kramers--Brillouin (WKB), approximation. 
Unlike the Airy ladder, which is designed specifically for the threshold
region, the WKB description is appropriate for mesons whose energies lie
well inside the continuum window, i.e. 
\begin{equation}
2\omega(0) < E < 2\omega\!\left(\frac{\pi}{2}\right) 
\label{6.11}
\end{equation}
that is, sufficiently far from both the lower and upper edges. At $P=0$, the semiclassical quantization condition determines the allowed 
relative momenta $p_{n}$ through 
\begin{align}
\begin{split}
&4\left[
\omega(p_n)\,p_n
-
\int_0^{p_n}dp'\,\omega(p')
\right]\\
&\qquad\qquad\,=
f\left[
2\pi\left(n-\frac14\right)
-
\Theta_\chi(2\alpha_n-\pi|\eta)
\right]
+
O(f^2),
\label{6.12}
\end{split}
\end{align}
where $\alpha_{n}$ is the rapidity corresponding to the momentum $p_{n}$ and
$\Theta_{\chi}$ is the channel-dependent scattering phase. Once the quantized momentum is known, the meson energy is simply
\begin{equation}
E_{\chi,n}(0)=2\omega(p_n)\,.
\label{6.13}
\end{equation}
Restricting again to the even $+$ channel, we denote the resulting
semiclassical energies by 
\begin{equation}
E_n^{\mathrm{WKB}} \equiv E_{+,n}(0)\,.
\label{6.14}
\end{equation}
The WKB formula should not be used for the very lowest mesons immediately above $E_{\rm cont}(0)$, where the Airy expansion~\eqref{6.10} is the correct asymptotic description. Rather, it is the natural approximation for the interior part of the meson ladder.

The third description is the strong-anisotropy expansion, which is
organized not around the continuum threshold but around the large Ising
scale $|\Delta|$. 
In the limit $|\Delta|\to\infty$ at fixed $h > 0$, one obtains for the
spin-zero channels at $P=0$ 
\begin{equation}
E_{\pm,n}^{\mathrm{SAE}} = 4J|\Delta|-4h\,\nu_n+O(|\Delta|^{-1}) 
\label{6.15}
\end{equation}
where the numbers $\nu_{n}$ are determined, in the infinite chain, by the
Bessel-function condition
\begin{equation}
J_{\nu_n}\!\left(\frac{2J}{h}\right)=0 
\label{6.16}
\end{equation}
with $\nu_{n+1}<\nu_{n}$. 

For the periodic chain with even length $N=2L$, the leading energy formula~\eqref{6.15} remains unchanged, but the quantization condition for
$\nu_n$ is replaced by
\begin{equation}
J_{\nu_n}\!\left(\frac{2J}{h}\right)
Y_{\nu_n+L}\!\left(\frac{2J}{h}\right)
=
Y_{\nu_n}\!\left(\frac{2J}{h}\right)
J_{\nu_n+L}\!\left(\frac{2J}{h}\right) 
\label{6.17}
\end{equation}
which has exactly $L-1$ real solutions.
This finite-size form is particularly relevant for the present work,
because our ED comparison is carried out for a periodic chain of length
$N=22$. After restricting to the even $+$ channel, we write
\begin{equation}
E_n^{\mathrm{SAE}} \equiv E_{+,n}^{\mathrm{SAE}}\,.
\label{6.18}
\end{equation}
Thus the strong-anisotropy expansion provides a third analytic benchmark,
complementary to the Airy and WKB results and incorporating the finite
system size explicitly when Eq.~\eqref{6.17} is used.

It is useful to emphasize that these three descriptions should not be
viewed as redundant formulas for the same regime. Rather, they are complementary approximations organized around different limits of the problem. The Airy ladder~\eqref{6.10} captures the mesons nearest the deconfined threshold.
The WKB quantization~\eqref{6.12}--\eqref{6.14} describes mesons lying
well inside the confinement window. The strong-anisotropy expansion~\eqref{6.15}--\eqref{6.18} is organized in powers of $|\Delta|^{-1}$ and is particularly useful in the regime where the large Ising scale dominates and finite-$N$ effects are important. 
In Sec.~\ref{sec:numerical_comparison_mesons} we compare all three analytic descriptions to the ED spectrum of the one-meson band at $\kappa=0$ for $N=22$, $\Delta=-4.5$, and $h=0.1$. Additional field-dependent comparisons at $h=0.05$ and $h=0.2$ are presented in Appendix~\ref{appC}. 
\begin{figure}[t]
\includegraphics[width=1.00\columnwidth]{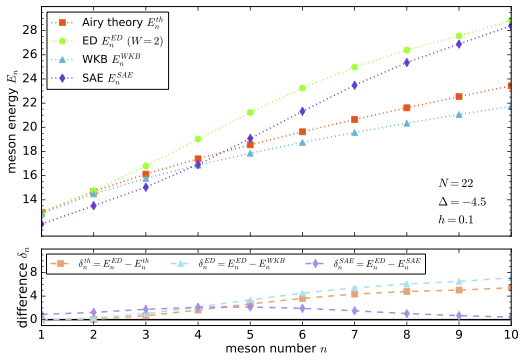}
\caption{
          {\it Comparison of $W=2$ meson energies at $P=0$ for the periodic chain with $N=22$, $\Delta=-4.5$, and $h=0.1$.} 
          The upper panel shows the ED energies $E_n^{\mathrm{ED}}$ together with three analytic descriptions, i.e. the low-energy Airy-ladder theory $E_n^{\mathrm{th}}$, the semiclassical/WKB energies $E_n^{\mathrm{WKB}}$, and the finite-$N$ strong-anisotropy expansion energies $E_n^{\mathrm{SAE}}$. The lower panel displays the corresponding residuals $\delta_n^\gamma=E_n^{\mathrm{ED}}-E_n^\gamma$, with $\gamma=\mathrm{th},\mathrm{WKB},\mathrm{SAE}$. The Airy-ladder theory provides the best description of the lowest mesons, while the finite-$N$ strong-anisotropy expansion tracks the higher members of the $W=2$ sequence most closely. 
          The WKB energies remain systematically below the ED values over the range shown.
}
\label{figure6}
\end{figure}
\begin{figure}[t]
\includegraphics[width=1.00\columnwidth]{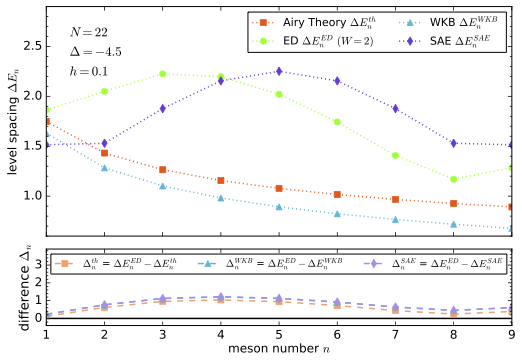}
\caption{
         {\it Level spacings for the $W=2$ meson ladder at $P=0$ for $N=22$, $\Delta=-4.5$, and $h=0.1$.} 
         The upper panel shows the spacings $\Delta E_{n}^{\gamma} = E_{n+1}^{\gamma}-E_{n}^{\gamma}$ for ED and for the three analytic descriptions $\gamma=\mathrm{th},\mathrm{WKB},\mathrm{SAE}$. The lower panel displays the spacing residuals $\Delta_{n}^{\gamma} = \Delta E_{n}^{\mathrm{ED}}-\Delta E_{n}^{\gamma}$. Since constant offsets cancel in the differences, this comparison isolates how well each analytic approximation reproduces the variation of the meson ladder with meson index $n$. The Airy theory captures the spacing pattern best for the lowest levels, whereas the finite-$N$ strong-anisotropy expansion provides the closest agreement at larger meson number.
}
\label{figure7}
\end{figure}
\begin{figure*}[t]
\includegraphics[width=1.00\textwidth]{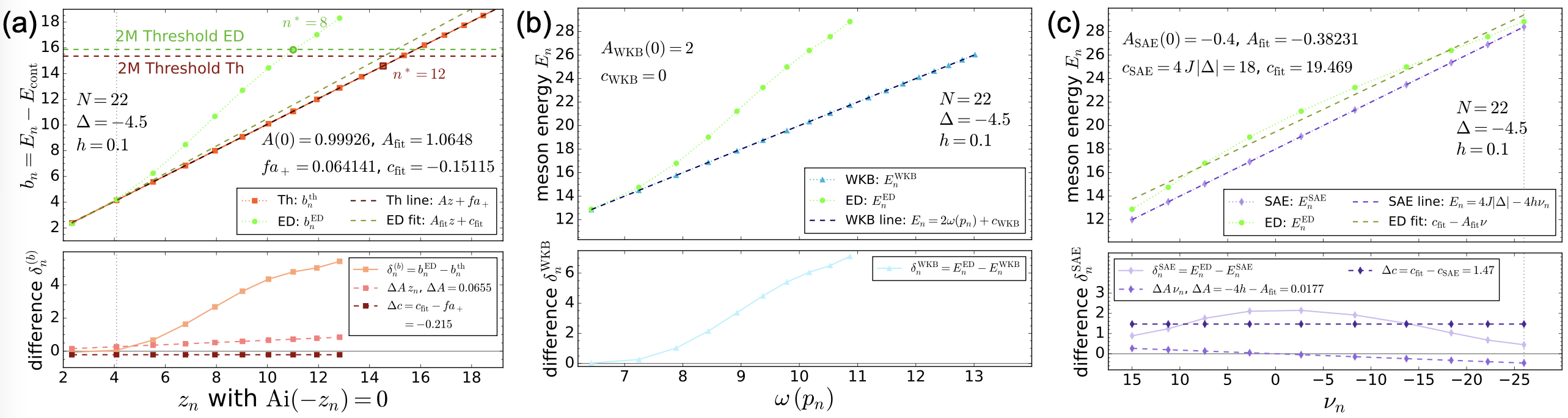}
\caption{
         {\it Theory-specific diagnostics for the $W=2$ meson spectrum at $P=0$ for $N=22$, $\Delta=-4.5$, and $h=0.1$.} 
         (a) Airy-ladder comparison: continuum-relative binding energies $b_{n}=E_{n}-E_{\rm cont}(0)$ plotted versus the Airy magnitudes $z_{n}$. The parameter-free theory line is $b_{n}^{\mathrm{th}}=A(0)z_n+f\,a_+(0|\eta)$, while the ED points are shown together with a linear fit performed over the lowest two mesons. The two horizontal lines mark the analytic two-meson threshold $b_{\rm th}^{(2M)}(0)$ and the ED-anchored threshold $b_{\rm ED}^{(2M)}(0)$ in the same binding reference, allowing the corresponding one-meson stability windows to be read off directly. The lower panel shows the residual $\delta_{n}^{(b)}=b_{n}^{\mathrm{ED}}-b_{n}^{\mathrm{th}}$ and its decomposition into slope and offset contributions. 
         (b) WKB comparison: ED energies plotted against the corresponding $\omega(p_{n})$ values entering the semiclassical relation $E_{n}^{\mathrm{WKB}}=2\omega(p_{n})$. The lower panel displays the residual 
         $\delta_n^{\mathrm{WKB}}=E_n^{\mathrm{ED}}-E_n^{\mathrm{WKB}}$. 
         (c) Strong-anisotropy comparison: ED energies plotted against the finite-$N$ strong-anisotropy variables $\nu_{n}$ entering $E_{n}^{\mathrm{SAE}} = 4J|\Delta|-4h\nu_{n}$ for the periodic chain of length $N=22$. The lower panel shows the residual 
         $\delta_n^{\mathrm{SAE}}=E_n^{\mathrm{ED}}-E_n^{\mathrm{SAE}}$. Together, panels (a)–(c) show that the Airy description is most accurate near threshold, while the finite-$N$ strong-anisotropy expansion gives the closest account of the higher mesons.
}
\label{figure8}
\end{figure*}

\subsection{Numerical comparison to the analytic meson spectrum at zero momentum $\boldsymbol{P=0}$ ($\boldsymbol{\kappa=0}$ in ED)}\label{sec:numerical_comparison_mesons}

We now compare the analytic meson spectra summarized in Sec.~\ref{sec:analytic_meson_spectrum} with our exact diagonalization results in the $(S^z_{\rm total},\kappa,I,C_{\rm flip})=(0,0,+1,+1)$ sector, restricting throughout to the $W=2$ one-meson band. Our primary focus is the parameter set $N=22$, $\Delta=-4.5$, and staggered magnetic field $h=0.1$ 
for which all three analytic descriptions --- the low-energy Airy ladder,
the semiclassical/WKB quantization, and the finite-$N$ strong-anisotropy
expansion --- can be confronted directly with the ED spectrum. 
The corresponding theory energies will be denoted by
$E_n^{\mathrm{th}}$, $E_n^{\mathrm{WKB}}$, and
$E_n^{\mathrm{SAE}}$, respectively, while the numerical values extracted
from the $W=2$ ED band are denoted by $E_n^{\mathrm{ED}}$.

The choice $h=0.1$ for the spectroscopy analysis should be distinguished from the choice $h=0.3$ used for the principal level-statistics finite-size analysis in Sec.~\ref{subsec:results_levelstats}. The two calculations probe different aspects of the problem and therefore have different practical requirements. For the level statistics, $h=0.3$ provides a well-developed, approximately monotonic crossover profile over the available system sizes, which permits a stable finite-size extraction of $\Delta_{\rm cross}(N)$. The analytic meson formulas, by contrast, are organized as weak-confinement and/or strong-anisotropy expansions. We therefore use $h=0.1$ as the principal spectroscopy example in order to compare the three complementary analytic descriptions with the same finite-size $W=2$ ED ladder. This field is not chosen so that the three approximations have comparable
accuracy. Rather, their different deviations from ED are themselves part of the comparison below.

\paragraph*{Absolute meson energies and residuals.} 
We begin with the most direct comparison, namely the meson energies
themselves, shown in Fig.~\ref{figure6}. The upper panel displays the ED energies $E_n^{\mathrm{ED}}$ together with the three analytic ladders
$E_n^{\mathrm{th}}$, $E_n^{\mathrm{WKB}}$, and $E_n^{\mathrm{SAE}}$ as a function of the meson index $n$. The lower panel shows the corresponding residuals
\begin{equation}
\delta_n^\gamma = E_n^{\mathrm{ED}}-E_n^\gamma 
\label{6.20}
\end{equation}
with $\gamma=\mathrm{th},\mathrm{WKB},\mathrm{SAE}$. Several clear trends emerge from this comparison. First, all three analytic descriptions reproduce the overall monotone growth of the meson ladder with increasing meson index $n$. Second, the low-energy Airy theory is closest to ED for the lowest few mesons, as expected from its threshold character.
Third, the residuals associated with the Airy ladder increase steadily
with $n$, indicating that the near-threshold expansion gradually loses
accuracy deeper in the spectrum. 
The WKB construction provides the appropriate semiclassical benchmark for the interior of the confinement window, but its energies remain systematically below the ED values over the range shown.

Finally, the finite-$N$ strong-anisotropy expansion provides the closest
agreement for the higher members of the $W=2$ sequence, consistent with
the fact that it incorporates the finite system size explicitly.
Thus Fig.~\ref{figure6} already shows that no single analytic description is
uniformly optimal across the entire ladder, i.e. the best approximation
depends on the meson number and, therefore, on where the state lies
within the confinement window. 

\paragraph*{Level spacings.}
A complementary diagnostic is provided by the level spacings
\begin{equation}
\Delta E_n^\gamma = E_{n+1}^\gamma-E_n^\gamma
\label{6.21}
\end{equation}
with $\gamma=\mathrm{ED},\mathrm{th},\mathrm{WKB},\mathrm{SAE}$ shown in Fig.~\ref{figure7}. The upper panel compares the spacings directly, while the lower panel displays the spacing residuals
\begin{equation}
\Delta_n^\gamma = \Delta E_n^{\mathrm{ED}}-\Delta E_n^\gamma 
\label{6.22}
\end{equation}
with $\gamma=\mathrm{th},\mathrm{WKB},\mathrm{SAE}$. Because the continuum edge and any channel-dependent constant offset cancel in the differences, this comparison isolates how well each analytic description reproduces the actual variation of the meson ladder with $n$. The Airy theory predicts spacings proportional to $A(0)(z_{n+1}-z_n)$ and therefore captures the gradual compression of the lowest part of the spectrum.
This description is indeed accurate for the first few levels, but its
residuals grow systematically as $n$ increases. 
The WKB construction provides the semiclassical benchmark appropriate to the interior of the confinement window. At the present finite size, however, the WKB spacings remain systematically displaced from the ED values, so that the expected interior regime is not realized with uniformly quantitative accuracy.
The SAE spacings, by contrast, track the ED spacings most closely toward
the upper half of the $W=2$ ladder.
This reinforces the conclusion suggested already by Fig.~\ref{figure6}. Near
threshold the Airy theory performs best, whereas for higher meson number
the finite-$N$ strong-anisotropy expansion gives the most accurate
description of the spectrum at the present system size. 
A complementary dimensionless comparison based on the relative spacings $\Delta E_n^\gamma/E_n^\gamma$ is presented in Appendix~\ref{appC_mesons_h01}.

\paragraph*{Theory-specific diagnostics.}
To sharpen this comparison further, Fig.~\ref{figure8} presents three
theory-specific diagnostics, each isolating one of the analytic
descriptions against the ED data.

Figure~\ref{figure8}(a) focuses on the low-energy Airy ladder by plotting the continuum-relative binding energies
\begin{equation}
b_{n} \equiv E_{n}-E_{\rm cont}(0)
\label{6.23}
\end{equation}
against the Airy magnitudes $z_{n}$. The parameter-free theory line is
\begin{equation}
b_n^{\mathrm{th}} = A(0)\,z_n+f\,a_+(0|\eta) 
\label{6.24}
\end{equation}
while the ED points are shown together with a linear fit performed over
the lowest two mesons. In addition, Fig.~\ref{figure8}(a) displays two horizontal two-meson threshold lines in the same binding reference, i.e.  the analytic threshold $b_{\rm th}^{(2M)}(0)$ from Eq.~\eqref{6.10f} and the ED-anchored threshold $b_{\rm ED}^{(2M)}(0)$ from Eq.~\eqref{6.10h}.
These lines delimit the approximate one-meson stability window and
therefore provide a direct graphical estimate of the maximal stable meson
number in the analytic and ED spectra, respectively.

This figure makes the threshold character of the Airy description
especially transparent.
The lowest ED levels lie close to a straight line in $z_{n}$, with a fitted
slope $A_{\rm fit}$ that remains close to the parameter-free value
$A(0)$, but the mismatch in both slope and intercept becomes
increasingly visible as one moves to higher $n$.
At the same time, the threshold lines show that the low-energy theory
predicts a broader one-meson stability window than is observed directly
in the finite-size ED data.
This is consistent with the expectation that finite-size effects and
hybridization with higher-$W$ sectors become increasingly important near
the top of the $W=2$ ladder.
The lower panel of Fig.~\ref{figure8}(a) shows the residual
\begin{equation}
\delta_n^{(b)} = b_n^{\mathrm{ED}}-b_n^{\mathrm{th}}
\label{6.25}
\end{equation}
together with its decomposition into slope and offset contributions.
In this representation, the Airy ladder is seen to capture the first few
mesons well, but to drift progressively away from the ED spectrum as one
moves farther from threshold.

Figure~\ref{figure8}(b) isolates the WKB description. Here the
ED meson energies are compared directly with the semiclassical
relation
\begin{equation}
E_n^{\rm WKB} = 2\omega(p_n)
\end{equation}
with the abscissa given by the corresponding values of
$\omega(p_n)$. Since the WKB construction is intended as an
interior semiclassical approximation rather than a threshold
expansion, we do not use the lowest mesons to define an
independent fit.  Instead, the parameter-free semiclassical
reference line is compared directly with the ED data over the
displayed sequence. The lower panel shows the residual
\begin{equation}
\delta_n^{\mathrm{WKB}} = E_n^{\mathrm{ED}}-E_n^{\mathrm{WKB}}
\end{equation}
which remains systematic across the spectrum. Thus the WKB
construction provides the appropriate semiclassical benchmark
for interior states, but does not give uniformly quantitative
agreement with the finite-size spectrum at $N=22$.

Finally, Fig.~\ref{figure8}(c) performs the analogous comparison for the finite-$N$ strong-anisotropy expansion. In this case the theory energies are
\begin{equation}
E_{n}^{\mathrm{SAE}}=4J|\Delta|-4h\,\nu_{n} 
\label{6.28}
\end{equation}
where the $\nu_{n}$ are the ten real solutions of the finite-chain
quantization condition for $N=22$. Plotting the ED energies against the corresponding $\nu_{n}$ shows that the data are well described by a nearly linear dependence of the same form, with fitted slope and intercept both close to the strong-anisotropy prediction. The lower panel shows the residual
\begin{equation}
\delta_n^{\mathrm{SAE}} = E_{n}^{\mathrm{ED}}-E_n^{\mathrm{SAE}}
\label{6.29}
\end{equation}
which is correspondingly smaller and less strongly drifting than in the
Airy and WKB cases over the same range of meson number. This provides a particularly direct explanation for why the SAE ladder tracks the upper part of the $W=2$ spectrum so well in Figs.~\ref{figure6} and \ref{figure7}.

Taken together, Figs.~\ref{figure6}--\ref{figure8} show that the three analytic descriptions are best viewed as complementary approximations organized around different limits of the problem. The Airy ladder provides the best account of the first few mesons nearest the deconfined threshold.  
The WKB approximation provides the appropriate semiclassical benchmark for the interior part of the confinement window, but remains systematically offset from the ED values at the present finite size. 
The finite-$N$ strong-anisotropy expansion gives the closest description
of the higher members of the $W=2$ sequence and therefore plays an
important role in understanding the finite-$N$ ED spectrum at large
anisotropy. This combination of comparisons clarifies both the strengths and the limitations of each analytic approach and will be complemented further by the additional field-dependent data presented in the Appendix~\ref{appC}. 

The field dependence of this hierarchy is examined explicitly in
Appendix~\ref{appC_mesons_additional}, where we repeat the spectroscopy
analysis at $h=0.05$ and $h=0.2$. These additional calculations show
that the precise quantitative accuracy of the three approximations
depends appreciably on the confinement strength: at $h=0.05$ the
finite-size ladder is considerably denser and the Airy and WKB
descriptions depart more rapidly from ED, whereas at $h=0.2$ both
comparisons improve over a broader portion of the resolved $W=2$
sequence. Thus the $h=0.1$ results should be viewed as one representative weak-confinement comparison rather than as an optimization of any one analytic description.

\section{Conclusions}\label{section_conclusions}

We have studied the spin-$\tfrac12$ XXZ chain in a staggered magnetic field as a minimal setting in which confinement reorganizes the many-body spectrum and drives a finite-size crossover away from fully chaotic spectral behavior. Using exact diagonalization in symmetry-resolved sectors, together with controlled analytic descriptions of the confined two-kink spectrum, we have established a coherent picture linking level statistics, correlation structure, eigenstate entanglement, and quantitative meson spectroscopy within a single microscopic model. 

At the global spectral level, we found a clear finite-size crossover from GOE-like statistics at weak anisotropy toward reduced level repulsion deep in the antiferromagnetic Ising regime. At the accessible finite sizes and anisotropies, the strong-anisotropy spectra remain intermediate between the GOE and Poisson limits rather than reaching the Poisson benchmark. This spectral evolution is robust against the choice of momentum sector, i.e. repeating the level-statistics analysis in the first two generic nonzero two-site translation sectors, $\kappa=4\pi/N$ and $\kappa=8\pi/N$, yields the same qualitative crossover as in the inversion-invariant $\kappa=0$
sector. The full adjacent-gap-ratio distributions likewise remain GOE-like at weak anisotropy and exhibit reduced level repulsion at strong anisotropy, while remaining intermediate between the GOE and Poisson limits. Thus, although the precise finite-size values depend on the symmetry sector, the reduction of level repulsion is not a peculiarity of the special
$\kappa=0$ block.

This crossover is accompanied by a pronounced reorganization of the eigenstates in local observables. A simple nearest-neighbor correlation measure $C^{zz}_{j}$ reveals flat bands across the spectrum that are approximately labeled by the domain-wall number $W$, while the staggered magnetization exhibits the same structure. The von Neumann entanglement entropy mirrors this reorganization, i.e.\ the broad high-entanglement dome characteristic of chaotic spectra at small $|\Delta|$, whose mid-spectrum values approach the symmetry-matched random-state benchmark, gives way in the confined regime to well-separated low-entanglement bands aligned with the same $W$-resolved structure. 
Taken together, these results show that confinement generates an
emergent quasi-conservation of domain-wall number and thereby organizes the many-body spectrum into weakly hybridizing, approximately $W$-resolved manifolds.

A central result of the paper is the quantitative analysis of the $W=2$ meson band at zero total momentum. Here the analytic description is not exhausted by a single approximation. Instead, three complementary regimes emerge naturally. Near the lower two-spinon threshold, the spectrum is captured by the
low-energy Airy ladder, including the channel-dependent offset and the associated estimate of the one-meson stability window.  
For mesons lying well inside the confinement window, the semiclassical WKB quantization provides the theoretically appropriate asymptotic description. For the higher part of the finite-size ladder, the finite-$N$ strong-anisotropy expansion supplies the most accurate analytic benchmark. The comparison with ED illustrates both the usefulness and the limitations of this hierarchy: the Airy theory performs best for the lowest mesons nearest threshold, the WKB construction provides the natural semiclassical benchmark for interior states but remains systematically offset at the accessible finite size, and the finite-$N$ strong-anisotropy expansion gives the closest account
of the upper part of the $W=2$ sequence. 
The additional field-dependent data in Appendix~\ref{appC} further show how the relative quality of these three descriptions evolves with the confinement strength.
 
These results place the staggered-field XXZ chain in a broader
conceptual context. In the confined regime, the emergent
quasi-conservation of the domain-wall number $W$ provides a natural mechanism for fragmentation-like spectral organization: although the Hilbert space is not exactly divided into disconnected components, the spectrum is organized into weakly hybridizing mesonic bands that remain visible over a broad parameter range. At the same time, the low-entanglement meson states are atypical relative to the surrounding higher-entanglement spectrum and exhibit scar-like features, including
suppressed entanglement and simple analytic organization. In this sense, the model connects confinement with fragmentation-like and scar-like phenomenology in a disorder-free and microscopically realistic setting.

The physical relevance of this picture is strengthened by the fact that the same ingredients that operate here --- Ising-like anisotropy and an effective staggered field --- arise naturally in several quasi-one-dimensional magnetic materials. 
Our results therefore suggest a set of portable diagnostics for
confinement-driven spectral reorganization: 
(i) level-statistics crossovers toward reduced level repulsion,
(ii) correlation and entanglement banding approximately labeled by domain-wall number, and 
(iii) near-threshold meson spectroscopy with an Airy scale controlled by the confinement strength. 
Among these diagnostics, meson spectroscopy has the most direct
experimental counterpart: the discrete bound-state energies and their momentum dependence enter the dynamical spin structure factors and can be accessed through inelastic neutron scattering and terahertz spectroscopy, particularly in the vicinity of the two-spinon threshold~\cite{PhysRevB.106.134405}. By contrast, the level-statistics, domain-wall, and entanglement diagnostics characterize the underlying many-body spectral reorganization and motivate future nonequilibrium tests using suitably prepared domain-wall-rich initial states. 

Confinement is expected to survive weak departures from the XXZ Hamiltonian, provided the gapped N\'eel regime and the energy-density difference between the two N\'eel orientations remain intact. A segment of the unfavored orientation between two domain walls then retains an energy cost proportional to its length. This mechanism does not require the precise XXZ exchange form or conservation of total $S^{z}$. Weak additional exchange anisotropies, such as unequal transverse couplings $J_x\ne J_y$, and further-neighbor interactions can therefore modify domain-wall dispersions and meson spectra without eliminating confinement. Weak uniform magnetic fields likewise preserve confinement under these conditions. For the XXZ model, a uniform longitudinal field produces Zeeman shifts without changing the eigenstates, while a weak uniform transverse field can hybridize meson modes~\cite{PhysRevB.109.014411}. Additional interactions may also open decay channels for higher mesons, so broadening of individual modes need not imply loss of confinement. 

Several natural extensions follow from the present work. On the spectroscopy side, it would be valuable to study systematically
the momentum dependence of the meson ladders and two-meson thresholds, as well as the onset of hybridization in higher-$W$ sectors. On the dynamical side, an important next step is to connect the eigenstate-based diagnostics developed here to real-time signatures such as revivals, slow relaxation, and state-dependent thermalization. More broadly, it will be important to determine how the confinement-driven organization identified here evolves beyond the minimal single-chain Hamiltonian. Natural extensions include additional exchange and anisotropy terms relevant to real quasi-one-dimensional magnets, as well as coupled-chain and ladder geometries. The latter provide a particularly interesting setting in which fragmentation-like spectral organization may arise from local constraints and interchain couplings without requiring an explicit staggered-field term. Such extensions would help distinguish which 
aspects of the confinement-induced spectral organization 
found here are specific to the present confinement mechanism and which persist more generally in interacting spin systems.

In summary, the staggered-field XXZ chain provides a simple and
microscopically transparent platform in which confinement simultaneously produces reduced level repulsion, banded correlation and entanglement structure, and a quantitatively accessible meson spectrum.  
Viewing this confined regime through the combined lens of meson
formation, approximate domain-wall sectorization, and scar-like
phenomenology unifies the diverse diagnostics studied here and
connects the resulting spectral organization directly to
experimentally relevant confinement physics.

\begin{acknowledgements}
The authors were supported by the U.S. Department of Energy,
Office of Science, Office of Basic Energy Sciences, Materials
Sciences and Engineering Division under Contract No.~DE-SC0012704.
We are grateful to S.~B.~Rutkevich for sharing private notes containing data relevant to our calculations. 
JW acknowledges the hospitality of the Perimeter Institute for Theoretical Physics as a Simons Emmy Noether Fellow, during which part of this work was completed. Research at Perimeter Institute is supported by the Government of Canada through the Department of Innovation, Science and Economic Development Canada and by the Province of Ontario through the Ministry of Colleges and Universities. 
\end{acknowledgements}
\appendix
\section{Discrete symmetries of the XXZ Hamiltonian in a staggered field}\label{app:Z2} 

In this Appendix we give a detailed account of the discrete symmetries used in the exact-diagonalization analysis. In particular, we distinguish the sublattice-preserving spatial inversion from the bond-centered reflection that exchanges the two staggered sublattices, and we explain the relation between the physical inversion quantum number and the equivalent reflection--spin-flip operation used in the numerical basis construction. We use periodic boundary conditions and the same one-based site convention as in Eq.~\eqref{spinchain}, $i=1,\dots,N$, with all site labels understood modulo $N$. The staggered pattern requires $N$ to be even. We write 
\begin{equation} 
\mathcal{H} = \mathcal{H}_{J}+\mathcal{H}_{h} 
\label{eq:H_total} 
\end{equation} 
where 
\begin{align} 
\mathcal{H}_{J} &= -J\sum_{i=1}^{N} \left( \sigma_i^x\sigma_{i+1}^x + \sigma_i^y\sigma_{i+1}^y + \Delta\,\sigma_i^z\sigma_{i+1}^z \right), \\ \mathcal{H}_{h} &= -h\sum_{i=1}^{N}(-1)^i\sigma_i^z \,. 
\end{align} 

\subsection{Combined translation--spin-flip symmetry} 
Let $\mathcal{T}_{1}$ denote translation by one lattice site, $\mathcal{T}_{1}\sigma_i^\alpha\mathcal{T}_{1}^{-1} =\sigma_{i+1}^\alpha$, and define $\mathcal{T}_{2}\equiv\mathcal{T}_{1}^{2}$. The staggered field breaks $\mathcal{T}_{1}$ but preserves $\mathcal{T}_{2}$. We also define the global spin flip 
\begin{equation} 
\mathcal{X} \equiv \prod_{j=1}^{N}\sigma_j^x \,. 
\end{equation} 
Its action on the Pauli matrices is 
\begin{align} 
\mathcal{X}\sigma_j^x\mathcal{X} &=\sigma_j^x,  \\ \mathcal{X}\sigma_j^y\mathcal{X} &=-\sigma_j^y, \\ \mathcal{X}\sigma_j^z\mathcal{X} &=-\sigma_j^z\,. 
\end{align} 
The exchange Hamiltonian is invariant under both $\mathcal{T}_{1}$ and $\mathcal{X}$ separately. By contrast, each of these operations reverses the staggered-field term: 
\begin{align}
\begin{split}
\mathcal{T}_{1}\mathcal{H}_{h}\mathcal{T}_{1}^{-1} &= -\mathcal{H}_{h}\\ 
\mathcal{X}\mathcal{H}_{h}\mathcal{X} &= -\mathcal{H}_{h}\,. 
\end{split}
\end{align}
Their product is therefore an exact symmetry. We choose the operator ordering 
\begin{equation} 
\mathcal{C}_{\rm flip} \equiv \mathcal{T}_{1}\mathcal{X} \label{eq:Cflip_def} 
\end{equation} 
as in the main text. Explicitly, 
\begin{align} 
\mathcal{C}_{\rm flip}\, \sigma_i^z\, \mathcal{C}_{\rm flip}^{-1} 
&= \mathcal{T}_{1} \mathcal{X}\sigma_i^z\mathcal{X} \mathcal{T}_{1}^{-1} \nonumber\\ 
&= -\sigma_{i+1}^z \,. 
\end{align} 
It follows that 
\begin{align} 
\mathcal{C}_{\rm flip} \mathcal{H}_{h} \mathcal{C}_{\rm flip}^{-1} 
&= -h\sum_i(-1)^i \left(-\sigma_{i+1}^z\right) \nonumber\\ 
&= +h\sum_j(-1)^{j-1}\sigma_j^z \nonumber\\ 
&= -h\sum_j(-1)^j\sigma_j^z \nonumber\\ 
&= \mathcal{H}_{h}\,. 
\end{align} 
In the second line we set $j=i+1$, and in the third line we used $(-1)^{j-1}=-(-1)^j$. This also makes explicit the two sign changes whose cancellation restores the symmetry. For the exchange term, $\mathcal{T}_{1}$ only permutes bonds, while $\mathcal{X}$ changes the sign of both factors in the $yy$ and $zz$ interactions and leaves the $xx$ interaction unchanged. Thus 
\begin{equation} 
\mathcal{C}_{\rm flip} \mathcal{H}_{J} \mathcal{C}_{\rm flip}^{-1} = \mathcal{H}_{J}\,. 
\end{equation} 
Together, 
\begin{equation} 
\mathcal{C}_{\rm flip} \mathcal{H} \mathcal{C}_{\rm flip}^{-1} = \mathcal{H}\,. 
\label{eq:claims} 
\end{equation} 
Since the global spin flip is itself translationally invariant, $\mathcal{T}_{1}\mathcal{X}\mathcal{T}_{1}^{-1} =\mathcal{X}$, and $\mathcal{X}^{2}=\mathbf{1}$. Consequently, 
\begin{align} \mathcal{C}_{\rm flip}^{\,2} &= \mathcal{T}_{1}\mathcal{X} \mathcal{T}_{1}\mathcal{X} \nonumber\\ 
&= \mathcal{T}_{1}^{2}\mathcal{X}^{2} \nonumber\\ 
&= \mathcal{T}_{2}\,. 
\label{eq:Cflip_square} 
\end{align} 
Thus $\mathcal{C}_{\rm flip}$ is a square root of the two-site translation. If 
\begin{equation} 
\mathcal{T}_{2}|\psi_{\kappa}\rangle = e^{i \kappa}|\psi_{\kappa}\rangle 
\end{equation} with 
\begin{equation} \kappa_{n}=\frac{4\pi n}{N}, \qquad n=0,1,\dots,\frac{N}{2}-1
\end{equation} 
then an eigenvalue $c_{\rm flip}$ of $\mathcal{C}_{\rm flip}$ in this momentum sector must satisfy 
\begin{equation} 
c_{\rm flip}^{\,2}=e^{i \kappa}\,. 
\end{equation} 
The two possible branches are therefore 
\begin{equation} 
c_{\rm flip} = \zeta\,e^{i\kappa/2}, \qquad \zeta=\pm 1\,. \label{eq:Cflip_eigs_appA} 
\end{equation} 
At $\kappa=0$ these quantum numbers reduce to $c_{\rm flip}=\pm1$. The principal ED calculations considered in the main text use the $c_{\rm flip}=+1$ branch.

We distinguish the two-site translation phase $\kappa$ used in the ED construction from the total two-kink momentum $P$ used in the analytic meson theory presented in  Ref.~\cite{PhysRevB.106.134405} where the one-site translation 
$T_1^{(R)}$ is defined by 
\begin{eqnarray}
(T_1^{(R)})^{-1}\sigma_i^\alpha T_1^{(R)} =\sigma_{i+1}^\alpha 
\end{eqnarray}
whereas our convention is 
\begin{eqnarray}
\mathcal T_1\sigma_i^\alpha\mathcal T_1^{-1}
=\sigma_{i+1}^\alpha\,.
\end{eqnarray}
Hence
$\mathcal T_1=(T_1^{(R)})^{-1}$ and
$\mathcal T_2=(T_1^{(R)})^{-2}$.
Since a two-kink state of total momentum $P$ has eigenvalue
$e^{2iP}$ under $(T_1^{(R)})^2$, comparison with
$\mathcal T_2|\psi_\kappa\rangle=e^{i\kappa}|\psi_\kappa\rangle$
gives
\begin{equation}
\kappa=-2P\pmod{2\pi}\,.
\end{equation}
At zero momentum, i.e. $\kappa=P=0$, the distinction disappears. 

\subsection{Spatial inversion and the choice of reflection center} 
For an even periodic chain with a staggered field, the location of the reflection center matters. Consider first the bond-centered reflection 
\begin{equation} 
\mathcal{R}:\qquad i\longmapsto N+1-i\,. 
\label{eq:R_bond} 
\end{equation} 
The uniform exchange term is invariant under this operation. However, because $N$ is even, 
\begin{equation} 
(-1)^{N+1-i} = -(-1)^i 
\end{equation} 
so the reflection exchanges the two staggered sublattices. Consequently, 
\begin{align}
\begin{split}
\mathcal{R}\mathcal{H}_{J}\mathcal{R}^{-1} &= \mathcal{H}_{J}\\[2pt]
\mathcal{R}\mathcal{H}_{h}\mathcal{R}^{-1} &= -\mathcal{H}_{h}\,. \label{eq:R_action} 
\end{split}
\end{align}
Thus $\mathcal{R}$ by itself is not a symmetry for $h \neq 0$. A genuine spatial inversion symmetry is obtained by shifting the reflection center by one lattice site. We define 
\begin{equation} 
\mathcal{I} \equiv \mathcal{T}_{1}\mathcal{R}
\label{eq:I_physical} 
\end{equation} 
which acts on the site labels as 
\begin{equation} 
\mathcal{I}:\qquad i\longmapsto 2-i\pmod N\,. 
\label{eq:I_action} 
\end{equation}
This is a site-centered reflection. It preserves each staggered sublattice because 
\begin{equation} 
(-1)^{2-i}=(-1)^i\,. 
\end{equation} 
Hence 
\begin{equation} \mathcal{I}\mathcal{H}\mathcal{I}^{-1} = \mathcal{H}, \qquad \mathcal{I}^{2}=\mathbf{1}\,. 
\label{eq:I_symmetry} 
\end{equation} 
The inversion eigenvalues are therefore $I=\pm 1$ whenever $\mathcal{I}$ acts within the momentum sector under consideration. The inversion reverses the two-site translation, 
\begin{equation} 
\mathcal{I}\mathcal{T}_{2}\mathcal{I}^{-1} = \mathcal{T}_{2}^{-1}
\label{eq:I_T2} 
\end{equation} 
and therefore maps 
\begin{equation} 
\kappa\longmapsto -\kappa\,. 
\end{equation} 
An independent inversion quantum number can consequently be assigned within a single momentum sector only when $\kappa=-\kappa$ modulo $2\pi$, namely at $\kappa=0$ and $\kappa=\pi$. The $\kappa=\pi$ sector is present in the finite-size momentum grid when $N$ is a multiple of four. The principal calculations in the main text use $\kappa=0$, for which inversion parity is therefore a well-defined quantum number. 

\subsection{Relation to the inversion operation used in the ED basis} 

It is useful to make explicit the relation between the physical site-centered inversion $\mathcal{I}$ and the symmetry operation used to construct the even-$N$ ED basis. The bond-centered reflection $\mathcal{R}$ of Eq.~\eqref{eq:R_bond} is not itself a symmetry because it reverses the staggered field. The global spin flip $\mathcal{X}$ reverses the staggered field as well. Their product 
\begin{equation} 
\widetilde{\mathcal{I}} \equiv \mathcal{R}\mathcal{X} \label{eq:I_tilde} 
\end{equation} is therefore an exact symmetry: 
\begin{equation} \widetilde{\mathcal{I}} \mathcal{H} \widetilde{\mathcal{I}}^{-1} = \mathcal{H}, \qquad \widetilde{\mathcal{I}}^{\,2} = \mathbf{1}\,. 
\end{equation} 
This is the reflection--spin-flip operation used in our even-$N$ numerical basis construction. The two exact symmetry operations are not independent. Using $\mathcal{C}_{\rm flip}=\mathcal{T}_{1}\mathcal{X}$, $\widetilde{\mathcal{I}}=\mathcal{R}\mathcal{X}$, $\mathcal{X}^{2}=\mathbf{1}$, and the fact that $\mathcal{X}$ commutes with the spatial reflection, one obtains the operator identity \begin{align} 
\mathcal{C}_{\rm flip}\widetilde{\mathcal{I}} 
&= (\mathcal{T}_{1}\mathcal{X}) (\mathcal{R}\mathcal{X}) \nonumber\\ 
&= \mathcal{T}_{1}\mathcal{R} \nonumber\\ 
&= \mathcal{I}\,. 
\label{eq:I_relation} 
\end{align} 
This relation explains the symmetry labeling used in the main text. In the $\kappa=0$ sector, $\mathcal{T}_{2}=\mathbf{1}$ and $c_{\rm flip}=\pm 1$, so the relevant discrete symmetries can be resolved simultaneously. If a state satisfies  
\begin{align}
\begin{split}
\mathcal{C}_{\rm flip}|\psi\rangle &= c_{\rm flip}|\psi\rangle \\[2pt]
\widetilde{\mathcal{I}}|\psi\rangle &= \widetilde I|\psi\rangle
\end{split}
\end{align}
then Eq.~\eqref{eq:I_relation} gives 
\begin{equation} I = c_{\rm flip}\,\widetilde I\,. \label{eq:I_eigen_relation} 
\end{equation} 
All principal ED calculations in the main text use $c_{\rm flip}=+1$. In this sector, 
\begin{equation} I=\widetilde I\,. 
\end{equation} 
Therefore selecting the $+1$ eigenvalue of the reflection--spin-flip operation in the numerical construction is exactly equivalent to selecting the physical sublattice-preserving inversion sector $I=+1$ quoted throughout the main text.

\section{Analytic description of meson energies}\label{appB}

In this Appendix we collect the analytic formulas used in Sec.~\ref{sec:meson_analytics} for the comparison between exact diagonalization (ED) and the confined two-kink (``meson'') spectrum of the staggered-field XXZ chain. Throughout we work in the same convention as in the main text, namely
\begin{align}
\begin{split}
\mathcal{H}(J,\Delta,h) = -&J \sum_{i=1}^{N} \left( \sigma_i^x \sigma_{i+1}^x
+ \sigma_i^y \sigma_{i+1}^y
+ \Delta \,\sigma_i^z \sigma_{i+1}^z \right)\\
-&h \sum_{i=1}^{N} (-1)^i \sigma_i^z 
\label{B1}
\end{split}
\end{align}
with $J>0$ and $\Delta< -1$ in the antiferromagnetic Ising regime. 
We parameterize the anisotropy by
\begin{equation}
\Delta = -\cosh\eta
\quad \mathrm{with} \quad \eta>0\,.
\label{B2}
\end{equation}
We stress that Rutkevich's original formulas are written with a $J/2$
prefactor in the Hamiltonian~\cite{PhysRevB.106.134405}. In the formulas below we consistently use our $J$-convention, i.e.\ the one employed in the ED study.

The main purpose of this Appendix is threefold. First, we summarize the one-kink dispersion and the associated two-kink continuum threshold. Second, we record the near-threshold low-energy (Airy-ladder) expansion used in the comparison to low-lying ED mesons. 
Third, we collect the two complementary analytic descriptions that become
useful away from the immediate threshold region, namely the
semiclassical/WKB quantization formula, where WKB stands for the
Wentzel--Kramers--Brillouin approximation, and the strong-anisotropy
expansion.

\subsection{One-kink dispersion and two-kink kinematics}\label{app:B_dispersion}
In the antiferromagnetic Ising regime at $h=0$, neutral excitations can be viewed as pairs of kinks (spinons). In our $J$-convention the single-kink dispersion takes the standard elliptic form 
\begin{equation}
\omega(q) = \Omega_{0} \sqrt{1-k^2 \cos^2 q} 
\label{B3}
\end{equation}
with
\begin{equation}
\Omega_{0} = \frac{4J K(k)}{\pi}\sinh\eta
\label{B4}
\end{equation}
where $K(k)$ is the complete elliptic integral of the first kind and the modulus $k$ is fixed implicitly by
\begin{equation}
\frac{K(k')}{K(k)}=\frac{\eta}{\pi} 
\label{B5}
\end{equation}
with $k'=\sqrt{1-k^2}$. 
Equations~\eqref{B3}--\eqref{B5} are the $J/2\mapsto J$ version of the
dispersion formulas used by Rutkevich~\cite{PhysRevB.106.134405}. 

For a two-kink state with total momentum $P$ and relative momentum $p$,
we parameterize the individual kink momenta as
\begin{equation}
p_1=\frac{P}{2}+p,
\qquad
p_2=\frac{P}{2}-p,
\label{B6}
\end{equation}
so that the two-kink energy in the deconfined problem ($h=0$) is
\begin{equation}
\epsilon(p|P) = \omega\!\left(\frac{P}{2}+p\right)
+ \omega\!\left(\frac{P}{2}-p\right)\,.
\label{B7}
\end{equation}
The lower boundary of the two-kink continuum at fixed total momentum $P$ is obtained by minimizing $\epsilon(p|P)$ with respect to $p$. In the regime relevant here this minimum occurs at $p=0$, yielding
\begin{align}
\begin{split}
E_{\rm cont}(P) &= \epsilon(0|P)
= 2\omega\!\left(\frac{P}{2}\right)\\
&= 2\Omega_{0}\sqrt{1-k^2\cos^2\!\left(\frac{P}{2}\right)}\,.
\label{B8}
\end{split}
\end{align}
In particular, at $P=0$ one has
\begin{equation}
E_{\rm cont}(0)=2\Omega_{0}\sqrt{1-k^2}\,.
\label{B9}
\end{equation}

Turning on a staggered field confines the two kinks through a linear
potential with string tension
\begin{equation}
f=2h\,\bar{\sigma}(\eta) 
\label{B10} 
\end{equation}
where the spontaneous staggered magnetization is
\begin{equation}
\bar{\sigma}(\eta) =
\prod_{n=1}^{\infty}\left(\frac{1-e^{-2n\eta}}{1+e^{-2n\eta}}\right)^2
= \prod_{n=1}^{\infty}\tanh^2(n\eta)\,.
\label{B11}
\end{equation}
Equations~\eqref{B10}--\eqref{B11} set the confinement scale entering all three analytic descriptions discussed below. 

\subsection{Scattering phases and low-energy meson energies}\label{app:scattering_phases_mesons}
We now summarize the scattering phases that enter the low-energy 
description of confined two-kink states and state the corresponding
near-threshold meson energies. Throughout this subsection we focus on the spin sector $S^z_{\rm total}=0$.  
Following Ref.~\cite{PhysRevB.106.134405}, we denote Rutkevich's modified one-site translation by $\widetilde T_{1}^{(R)}$. In the spin-zero sector, we have 
\begin{equation}
\widetilde T_1^{(R)}
|\pi_{0,\iota,n}(P)\rangle
=
\iota e^{iP}
|\pi_{0,\iota,n}(P)\rangle,
\qquad
\iota=\pm1 
\label{B12}
\end{equation}
so that the two spin-zero meson channels are labeled by $\iota=\pm$. At momentum $P=0$, the $\iota=+$ channel corresponds to the $c_{\rm flip}=+1$ ($\zeta=+1$) ED sector used in the main-text meson comparison.

The exchange of the two kinks is encoded in the channel-dependent
scattering phases
\begin{equation}
\Theta_{\pm}(\alpha) = \Theta_0(\alpha)+\chi_{\pm}(\alpha)
\label{B13}
\end{equation}
where the ``bulk'' contribution is
\begin{equation}
\Theta_0(\alpha) = \alpha + \sum_{n=1}^{\infty}
\frac{e^{-n\eta}\sin(2n\alpha)}{n\cosh(n\eta)}
\label{B14}
\end{equation}
and the channel-resolved pieces are
\begin{align}
\chi_+(\alpha)
&=
-i\ln\!\left[
-\frac{\sin[(\alpha+i\eta)/2]}{\sin[(\alpha-i\eta)/2]}
\right],
\label{B15}
\\
\chi_-(\alpha)
&=
-i\ln\!\left[
\frac{\cos[(\alpha+i\eta)/2]}{\cos[(\alpha-i\eta)/2]}
\right].
\label{B16}
\end{align}
These phases determine the short-distance matching conditions for the
confined two-kink wave function and therefore fix the $O(f)$ channel
offsets in the low-energy meson spectrum.

At weak staggered field, the meson energies near the lower edge of the
two-kink continuum admit the Airy-ladder expansion
\begin{equation}
E_{\text{meson}}(n,\pm,P) = \epsilon(0|P) + A(P)\,z_n + f\,a_{\pm}(P)
+ O(f^{4/3})
\label{B17}
\end{equation}
where $-z_n<0$ is the $n$-th zero of the Airy function on the negative real axis, i.e.\ $\mathrm{Ai}(-z_n)=0$ with $z_n>0$.  
Here
\begin{equation}
\epsilon(0|P)=E_{\rm cont}(P)
\label{B19}
\end{equation}
is the lower boundary of the two-kink continuum at fixed total momentum
$P$, while the coefficient
\begin{equation}
A(P) = f^{2/3}
\left(
\frac{\partial_p^2 \epsilon(p|P)\vert_{p=0}}{2}
\right)^{1/3}
\label{B20}
\end{equation}
sets the characteristic Airy scale.

The $P=0$ sector is of particular importance for the comparison to ED.
In that case, the continuum edge reduces to
\begin{equation}
E_{\rm cont}(0)=2\omega(0)=2\Omega_{0}\sqrt{1-k^2}
\label{B21}
\end{equation}
and the low-energy meson ladder becomes
\begin{equation}
E_{\text{meson}}(n,\pm,0) = E_{\rm cont}(0) + A(0)\,z_n + f\,a_{\pm}(0)
+ O(f^{4/3})\,.
\label{B22}
\end{equation}
Thus the near-threshold spectrum consists of a sequence of bound states
whose spacings are controlled at leading order by the Airy magnitudes
$z_n$, while the channel dependence enters through the constants
$a_{\pm}(0)$.

In the main text we compare this low-energy description to ED data in the $(S^z_{\rm total},\kappa,I,C_{\rm flip})=(0,0,+1,+1)$ sector, for which the
relevant analytic branch is the even $+$ channel. The corresponding
continuum-relative binding energies are therefore
\begin{align}
\begin{split}
b_n \equiv E_n-E_{\rm cont}(0) 
= A(0)\,z_n + f\,a_+(0) + O(f^{4/3})\,.
\label{B23}
\end{split}
\end{align}
This form underlies the Airy-ladder comparisons presented in Sec.~\ref{sec:meson_analytics}. 
The coefficients $a_{\pm}(P)$ are determined by the derivatives of the
scattering phases and will be discussed in detail below. In particular, we will show that the $P=0$ constants retain a nontrivial dependence on the anisotropy parameter $\eta$.

\subsection{Curvature and Airy scale}\label{app:B_curvature}
We now evaluate the curvature entering Eq.~\eqref{B20}. 
Differentiating Eq.~\eqref{B7} gives 
\begin{align}
\begin{split}
\epsilon''(0|P) &=
\left.
\frac{\partial^2}{\partial p^2}
\left[\omega\!\left(\frac{P}{2}+p\right)
+ \omega\!\left(\frac{P}{2}-p\right)\right]\right|_{p=0}\\
&= 2\,\omega''\!\left(\frac{P}{2}\right).
\label{B24}
\end{split}
\end{align}
For the dispersion~\eqref{B3}, direct differentiation yields
\begin{equation}
\omega''(q) = \Omega_{0}\left[\frac{k^2 \cos(2q)}{\sqrt{1-k^2\cos^2 q}}
-\frac{k^4\cos^2 q\,\sin^2 q}{\left(1-k^2\cos^2 q\right)^{3/2}}
\right]\,.
\label{B25}
\end{equation}
Specializing to $P=0$, i.e.\ $q=0$, this simplifies to
\begin{equation}
\omega''(0) = \Omega_{0}\frac{k^2}{\sqrt{1-k^2}} 
\label{B26}
\end{equation}
and hence
\begin{equation}
\epsilon''(0|0) = 2\Omega_{0}\frac{k^2}{\sqrt{1-k^2}}\,.
\label{B27}
\end{equation}
Therefore the Airy scale at $P=0$ becomes
\begin{equation}
A(0) = f^{2/3}\left[\Omega_{0}\,\frac{k^2}{\sqrt{1-k^2}}\right]^{1/3}\,.
\label{B28}
\end{equation}
This is the expression used for the parameter-free Airy theory in the main text. 

\subsection{Evaluation of $\boldsymbol{a_{\pm}(P)}$ at $\boldsymbol{P=0}$}\label{app:evaluation_a_pm}
We now evaluate the channel-dependent constants $a_{\pm}(P)$ in the
zero-momentum sector $P=0$, which is the case relevant for the ED
comparison in Sec.~\ref{sec:meson_analytics}. As discussed above, these coefficients determine the $O(f)$ offsets in the
low-energy meson ladder and are fixed by the derivatives of the scattering phases at zero rapidity.

In our $J$-convention the channel shifts are
\begin{equation}
a_{\pm}(P) = -\frac{2J\sinh\eta}{\omega(P/2)}
\left.
\frac{\partial \Theta_{\pm}(\alpha)}{\partial \alpha}
\right|_{\alpha=0} 
\label{B29}
\end{equation}
or, equivalently, using Eq.~\eqref{B3},
\begin{equation}
a_{\pm}(P) = -\frac{\pi}{2K(k)\sqrt{1-k^2\cos^2(P/2)}}
\left.
\frac{\partial \Theta_{\pm}(\alpha)}{\partial \alpha}
\right|_{\alpha=0}.
\label{B30}
\end{equation}
Thus the problem reduces to evaluating the derivatives of the scattering phases at $\alpha=0$.

We begin from
\begin{equation}
\Theta_{\pm}(\alpha)=\Theta_0(\alpha)+\chi_{\pm}(\alpha) 
\label{B31}
\end{equation}
with
\begin{equation}
\Theta_0(\alpha) = \alpha + \sum_{n=1}^{\infty}
\frac{e^{-n\eta}\sin(2n\alpha)}{n\cosh(n\eta)} 
\label{B32}
\end{equation}
and
\begin{align}
\chi_+(\alpha)
&=-i\ln\!\left[
-\frac{\sin[(\alpha+i\eta)/2]}{\sin[(\alpha-i\eta)/2]}
\right]
\label{B33}
\\
\chi_-(\alpha)
&=-i\ln\!\left[
\frac{\cos[(\alpha+i\eta)/2]}{\cos[(\alpha-i\eta)/2]}
\right]\,.
\label{B34}
\end{align}
The derivative of the bulk phase is obtained directly from
Eq.~\eqref{B32}, i.e. 
\begin{equation}
\Theta_0'(\alpha) = 1 + 2\sum_{n=1}^{\infty}
\frac{e^{-n\eta}\cos(2n\alpha)}{\cosh(n\eta)}\,.
\label{B35}
\end{equation}
Setting $\alpha=0$ gives
\begin{equation}
\Theta_0'(0) = 1 + 2\sum_{n=1}^{\infty}
\frac{e^{-n\eta}}{\cosh(n\eta)}\,.
\label{B36}
\end{equation}
Introducing the nome
\begin{equation}
q=e^{-\eta}
\label{B37}
\end{equation}
and using
\begin{equation}
\frac{e^{-n\eta}}{\cosh(n\eta)} = \frac{2}{e^{2n\eta}+1}
= \frac{2q^{2n}}{1+q^{2n}} 
\label{B38}
\end{equation}
we may rewrite Eq.~\eqref{B36} in the convenient form
\begin{equation}
\Theta_0'(0) = 1 + 4\sum_{n=1}^{\infty}\frac{q^{2n}}{1+q^{2n}}\,.
\label{B39}
\end{equation}

We next differentiate the channel-dependent pieces. 
Utilizing 
\begin{align}
\begin{split}
\dfrac{d}{d\alpha}\ln\sin\!\dfrac{\alpha\pm i\eta}{2}&=\dfrac{1}{2}\cot\!\dfrac{\alpha\pm i\eta}{2} \\
\dfrac{d}{d\alpha}\ln\cos\!\dfrac{\alpha\pm i\eta}{2}&=-\dfrac{1}{2}\tan\!\dfrac{\alpha\pm i\eta}{2}
\end{split}
\end{align}
one finds 
\begin{equation}
\chi_+'(\alpha) = -\frac{i}{2}
\left[\cot\!\left(\frac{\alpha+i\eta}{2}\right)
-\cot\!\left(\frac{\alpha-i\eta}{2}\right)\right]\,.
\label{B40}
\end{equation}
Evaluating at $\alpha=0$ and using $\cot(ix)=-i\,\coth x$, we obtain
\begin{equation}
\chi_+'(0) = -\coth\!\left(\frac{\eta}{2}\right)\,.
\label{B41}
\end{equation}
Similarly, differentiation of $\chi_-(\alpha)$ gives
\begin{equation}
\chi_-'(\alpha) = -\frac{i}{2}
\left[-\tan\!\left(\frac{\alpha+i\eta}{2}\right)
+ \tan\!\left(\frac{\alpha-i\eta}{2}\right) \right]\,.
\label{B42}
\end{equation}
Using $\tan(ix)=i\,\tanh x$ at $\alpha=0$, we arrive at
\begin{equation}
\chi_-'(0) = -\tanh\!\left(\frac{\eta}{2}\right)\,.
\label{B43}
\end{equation}

Combining Eqs.~\eqref{B31}, \eqref{B39}, \eqref{B41}, and \eqref{B43},
we obtain
\begin{equation}
\Theta_+'(0) = -\coth\!\left(\frac{\eta}{2}\right) + 1 +
4\sum_{n=1}^{\infty}\frac{q^{2n}}{1+q^{2n}} 
\label{B44}
\end{equation}
and
\begin{equation}
\Theta_-'(0) = -\tanh\!\left(\frac{\eta}{2}\right) + 1 +
4\sum_{n=1}^{\infty}\frac{q^{2n}}{1+q^{2n}}\,.
\label{B45}
\end{equation}
For reference, it is also useful to introduce the corresponding
``spin'' or $\chi=0$ channel quantity
\begin{equation}
\Theta_0'(0) = 1 + 4\sum_{n=1}^{\infty}\frac{q^{2n}}{1+q^{2n}}
\label{B46}
\end{equation}
which leads to an auxiliary coefficient
\begin{equation}
a_0(0|\eta) = -\frac{\pi}{2K(k)\sqrt{1-k^2}}\, \Theta_0'(0)\,.
\label{B47}
\end{equation}

Specializing Eq.~\eqref{B30} to $P=0$, we therefore obtain 
\begin{eqnarray}
a_+(0|\eta) &=& -\frac{\pi}{\gamma_{k}}
\left[-\coth\!\left(\frac{\eta}{2}\right) + 1 +
4\sum_{n=1}^{\infty}\frac{q^{2n}}{1+q^{2n}}
\right]\nonumber \\
& & \\
\label{B48}
a_-(0|\eta) &=& -\frac{\pi}{\gamma_{k}}
\left[-\tanh\!\left(\frac{\eta}{2}\right) + 1 +
4\sum_{n=1}^{\infty}\frac{q^{2n}}{1+q^{2n}}
\right]\nonumber \\
& &
\label{B49}
\end{eqnarray}
with $\gamma_{k} = 2K(k)\sqrt{1-k^2}$. 
Together with Eq.~\eqref{B47}, these are precisely the $P=0$ low-energy
constants entering the confined meson ladder in our convention.

At this point it is worth emphasizing an important conceptual point. The expressions \eqref{B47}--\eqref{B49} retain an explicit dependence on the anisotropy parameter $\eta$ through both the factor $1/\gamma_k=[2K(k)\sqrt{1-k^2}]^{-1}$ and the nome-dependent series. Thus the zero-momentum constants are not universal numbers independent of $\eta$.

For the parameters used in the main text, $J=1$, $\Delta=-4.5$, 
Eqs.~\eqref{B47}--\eqref{B49} evaluate to
\begin{align}
\begin{split}
a_0(0|\eta)&\approx -1.7479 \\
a_+(0|\eta)&\approx 0.337582 \\
a_-(0|\eta)&\approx -0.420807\,.
\label{B51}
\end{split}
\end{align}
In the $(S^z_{\rm total},\kappa,I,C_{\rm flip})=(0,0,+1,+1)$ sector considered in Sec.~\ref{sec:meson_analytics}, the relevant low-energy branch is the even $+$ channel, and therefore the parameter entering the near-threshold comparison to ED is $a_+(0|\eta)$ from Eq.~\eqref{B48}.  

\subsection{Semiclassical/WKB quantization at $\boldsymbol{P=0}$}\label{app:WKB_quantization}
The low-energy Airy expansion discussed above describes the mesons closest to the lower edge of the two-kink continuum. For states lying well inside the confinement window, but not too close to either continuum boundary, one should instead use the semiclassical quantization condition derived by Rutkevich~\cite{PhysRevB.106.134405}. 
In the main text we use this semiclassical construction to obtain the energies denoted $E_{n}^{\mathrm{WKB}}$.

We specialize from the outset to the sector of principal interest for the ED comparison, namely $P=0$. In the deconfined problem the two-kink energy is
\begin{equation}
\epsilon(p|0)=\omega(p)+\omega(-p)=2\omega(p) 
\label{B52}
\end{equation}
since the single-kink dispersion is an even function of momentum. The
continuum window relevant for the semiclassical description is therefore bounded by
\begin{equation}
2\omega(0) < E < 2\omega\!\left(\frac{\pi}{2}\right)\,.
\label{B53}
\end{equation}
The WKB formula is intended for mesons whose energies lie well inside this interval, whereas the Airy expansion is the appropriate asymptotic description near the lower threshold. 

The semiclassical quantization condition may be written as
\begin{align}
\begin{split}
&4\left[\omega(p_n)\,p_n - \int_0^{p_n} dp'\,\omega(p')
\right]\\
&\qquad\qquad=
f\left[2\pi\left(n-\frac14\right) -\Theta_\chi(2\alpha_n-\pi|\eta)
\right] + O(f^2)
\label{B54}
\end{split}
\end{align}
where $\chi=0,\pm$ labels the channel and $\alpha_{n}$ is the rapidity
corresponding to the momentum $p_{n}$. Once $p_{n}$ is known, the meson energy is
\begin{equation}
E_{\chi,n}(0)=2\omega(p_{n})\,.
\label{B55}
\end{equation}
Equations~\eqref{B54} and \eqref{B55} are the working WKB formulas used in the numerical comparison.

It is useful to explain in more detail the structure of
Eq.~\eqref{B54}. The left-hand side depends only on the one-kink
dispersion and has the form of a semiclassical action integral for the
relative motion of the confined two-kink pair. Defining
\begin{equation}
S(p)= 4\left[\omega(p)\,p - \int_0^p dp'\,\omega(p')\right]
\label{B56}
\end{equation}
Eq.~\eqref{B54} takes the compact form
\begin{equation}
S(p_n) = f\left[
2\pi\left(n-\frac14\right)-\Theta_\chi(2\alpha_n-\pi|\eta)
\right]
+ O(f^2)\,.
\label{B57}
\end{equation}
Differentiating Eq.~\eqref{B56} gives
\begin{equation}
\frac{dS}{dp} = 4p\,\omega'(p)
\label{B58}
\end{equation}
which is positive on the interval $p\in(0,\pi/2)$ for the dispersion
relevant here. 
Thus $S(p)$ is monotone increasing on the physical domain, 
which facilitates the numerical solution of Eq.~\eqref{B56}
for the allowed momenta $p_n$.

The rapidity variable $\alpha$ enters through the scattering phase. It is related to the momentum by the elliptic map
\begin{equation}
p(\alpha) = -\frac{\pi}{2} +
\mathrm{am}\!\left(\frac{2K\alpha}{\pi},\,k\right) 
\label{B59}
\end{equation}
where $\mathrm{am}(u,k)$ is the Jacobi amplitude, defined implicitly by
$u=F(\phi|k^2)$ with $\phi=\mathrm{am}(u,k)$, where
$F(\phi|k^2)$ is the incomplete elliptic integral of the first kind. 

At $P=0$ the two kinks carry momenta $p$ and $-p$. If the positive-momentum kink is assigned rapidity $\alpha(p)$, then the negative-momentum kink corresponds to the rapidity $\pi-\alpha(p)$. Hence the rapidity difference entering the scattering phase is
\begin{align}
\begin{split}
\alpha_1-\alpha_2 &= \alpha(p)-(\pi-\alpha(p))\\
&= 2\alpha(p)-\pi\,.
\label{B60}
\end{split}
\end{align}
This is the origin of the argument $2\alpha_n-\pi$ in the quantization
condition~\eqref{B54}.

For numerical work it is convenient to define
\begin{equation}
\beta_n \equiv 2\alpha_n-\pi,
\label{B61}
\end{equation}
so that Eq.~\eqref{B54} may be rewritten as
\begin{align}
\begin{split}
&4\left[\omega(p_n)\,p_n - \int_0^{p_n} dp'\,\omega(p')\right]\\
&\qquad\qquad= f\left[2\pi\left(n-\frac14\right) - \Theta_\chi(\beta_n|\eta)\right]
+ O(f^2)\,.
\label{B62}
\end{split}
\end{align}
This is the form we use in the Python implementation described in the text.

The integral appearing on the left-hand side can be expressed in closed
form in terms of the incomplete elliptic integral of the second kind.
Using
\begin{equation}
\omega(p)=\Omega_{0}\sqrt{1-k^2\cos^2 p}\,,
\label{B63}
\end{equation}
one finds
\begin{equation}
\int_0^{p}\omega(q)\,dq = \Omega_{0}\int_0^{p}\sqrt{1-k^2\cos^2 q}\,dq.
\label{B64}
\end{equation}
Introducing the variable $\phi=\frac{\pi}{2}-q$ gives
\begin{equation} 
\int_0^{p}\omega(q)\,dq = \Omega_{0}\int_{\pi/2-p}^{\pi/2}\sqrt{1-k^2\sin^2\phi}\,d\phi 
\label{B65}
\end{equation}
and hence
\begin{equation}
\int_0^{p}\omega(q)\,dq = \Omega_{0}\left[
E(k)-E\!\left(\frac{\pi}{2}-p\,\middle|\,k^2\right)
\right]
\label{B66}
\end{equation}
where $E(k)$ is the complete elliptic integral of the second kind and
$E(\varphi|k^2)$ is the corresponding incomplete elliptic integral.
Equation~\eqref{B66} provides a convenient exact evaluation of the action integral in Eq.~\eqref{B62}.

Combining Eqs.~\eqref{B62} and \eqref{B55}, we obtain the semiclassical
meson ladder
\begin{equation}
E_n^{\mathrm{WKB}} \equiv E_{+,n}(0) = 2\omega(p_n)
\label{B67}
\end{equation}
for the even $+$ channel relevant to the
$(S^z_{\rm total},\kappa,I,C_{\rm flip})=(0,0,+1,+1)$ sector. These are the
$E_n^{\mathrm{WKB}}$ values compared to ED in Sec.~\ref{sec:meson_analytics}.

\subsection{Strong-anisotropy expansion}\label{app:strong_anisotropy}
A third analytic description of the meson spectrum is obtained in the strong-anisotropy limit $|\Delta|\to\infty$ at fixed staggered field $h>0$. In this regime the dominant energy scale is set by the Ising term, while the transverse exchange generates subleading corrections in powers of $|\Delta|^{-1}$. For the spin-zero channels at $P=0$, the leading strong-anisotropy result gives a ladder that is linear in a set of auxiliary numbers $\nu_n$.

For the infinite chain, the strong-anisotropy expansion takes the form
\begin{equation}
E_{\pm,n}^{\mathrm{SAE}}(J,\eta,h) = 4J|\Delta|-4h\,\nu_n+O(|\Delta|^{-1})
\label{B68}
\end{equation}
where the numbers $\nu_n$ are determined by
\begin{equation}
J_{\nu_n}\!\left(\frac{2J}{h}\right)=0 
\label{B69}
\end{equation}
with $\nu_{n+1}<\nu_{n}$. 
Here $J_\nu(x)$ denotes the Bessel function of the first kind.
Equation~\eqref{B68} shows that, to leading order in the
strong-anisotropy expansion, the meson energies are obtained by shifting the Ising value $4J|\Delta|$ by an amount proportional to $h\,\nu_n$.

For the periodic chain with even length $N=2L$, the leading energy formula~\eqref{B68} remains unchanged, but the quantization condition for the numbers $\nu_n$ is modified. Instead of Eq.~\eqref{B69}, one must solve
\begin{equation}
J_{\nu_n}\!\left(\frac{2J}{h}\right)
Y_{\nu_n+L}\!\left(\frac{2J}{h}\right)
=
Y_{\nu_n}\!\left(\frac{2J}{h}\right)
J_{\nu_n+L}\!\left(\frac{2J}{h}\right)
\label{B70}
\end{equation}
where $Y_\nu(x)$ is the Weber--Neumann function. For a chain of length $N=2L$, Eq.~\eqref{B70} has exactly $L-1$ real solutions $\{\nu_n\}_{n=1}^{L-1}$, which may be ordered as
\begin{equation}
\nu_1>\nu_2>\cdots>\nu_{L-1}.
\label{B71}
\end{equation}
Substituting these roots into Eq.~\eqref{B68} yields the finite-chain strong-anisotropy ladder
\begin{equation}
E_{\pm,n}^{\mathrm{SAE}} = 4J|\Delta|-4h\,\nu_{n} 
\label{B72}
\end{equation}
with $n=1,\dots,L-1$. 

In the main text we compare to ED in the even $+$ channel. Accordingly, once the branch has been selected, we suppress the explicit channel index and write simply
\begin{equation}
E_n^{\mathrm{SAE}} \equiv E_{+,n}^{\mathrm{SAE}}\,.
\label{B73}
\end{equation}
This is the notation used in the revised figures and in the discussion of the finite-$N$ strong-anisotropy data.

For the system size relevant to the main ED comparison, namely $N=22$ with $L = \frac{N}{2} = 11$, Eq.~\eqref{B70} has exactly ten real solutions, and therefore produces ten finite-size strong-anisotropy energies $E_n^{\mathrm{SAE}}$.
For the parameter set emphasized in the main text, i.e. $N=22$, $J=1$, $\Delta=-4.5$, and $h=0.1$ these ten values provide the finite-$N$ SAE benchmark used in the comparison to the $W=2$ ED meson energies.

The strong-anisotropy expansion should be viewed as complementary to the Airy and semiclassical descriptions discussed above. Near the lower continuum threshold, the Airy ladder provides the relevant
asymptotic form. For states well inside the confinement window, the semiclassical/WKB quantization is the appropriate description. By contrast, the strong-anisotropy expansion is organized in powers of
$|\Delta|^{-1}$ and captures the regime in which the large Ising scale is the dominant organizing principle. Its finite-$N$ version, Eq.~\eqref{B72}, is particularly useful in the present work because it incorporates the system size explicitly and thus provides a direct analytic benchmark for the $N=22$ periodic chain. 

\subsection{Airy zeros and useful asymptotics}\label{app:Airy_zeros}
For completeness, we collect here the basic properties of the Airy zeros used throughout Sec.~\ref{sec:meson_analytics} and Appendix~\ref{appB}. In the low-energy expansion, the meson ladder is expressed in terms of the positive magnitudes $z_n>0$ defined by
\begin{equation}
\mathrm{Ai}(-z_n) = 0\quad \mathrm{with} \quad  n=1,2,3,\dots\,.
\label{B75}
\end{equation}
Thus $-z_n<0$ are the zeros of the Airy function on the negative real axis, while $z_n$ themselves are positive numbers. We adopt this convention throughout because it makes the near-threshold
meson ladder 
\begin{equation}
E_n^{(+)} = E_{\rm cont}(0) + A(0)\,z_n + f\,a_+(0|\eta) + O(f^{4/3})
\label{B76}
\end{equation}
particularly transparent, i.e. the continuum edge $E_{\rm cont}(0)$ is shifted upward by a positive Airy contribution $A(0)z_n$ together with the channel offset $f a_+(0|\eta)$.

The first few values of $z_n$ are
\begin{align}
\begin{split}
&z_1 \approx 2.33811 \\
&z_2 \approx 4.08795 \\
&z_3 \approx 5.52056
\label{B77}
\end{split}
\end{align}
and they increase monotonically with $n$. For large $n$, the Airy magnitudes satisfy the well-known asymptotic form
\begin{equation}
z_n = \left[\frac{3\pi}{2}\left(n-\frac14\right)\right]^{2/3}
\left[1+O(n^{-2})\right]\,.
\label{B78}
\end{equation}
This expression is useful both for estimating the overall growth of the low-energy ladder and for initializing numerical procedures.

An immediate consequence of Eq.~\eqref{B76} is the asymptotic form of the level spacings. Since the continuum edge and the $O(f)$ offset do not depend on $n$, one finds
\begin{align}
\begin{split}
\Delta E_n^{\rm th} &\equiv E_{n+1}^{(+)}-E_n^{(+)} \\
& \simeq A(0)\left(z_{n+1}-z_n\right)\,.
\label{B79}
\end{split}
\end{align}
Thus the spacings are controlled entirely by the Airy scale $A(0)$ and the differences of successive Airy magnitudes. Because the Airy zeros spread sublinearly, the quantity $z_{n+1}-z_n$ decreases slowly with increasing $n$, implying a gradual compression of the low-energy ladder.

It is also convenient, when comparing to ED, to subtract the
$n$-independent offset
\begin{equation}
E_{\rm ref}(0) \equiv E_{\rm cont}(0)+f\,a_+(0|\eta) 
\label{B80}
\end{equation}
so that
\begin{equation}
E_n^{(+)}-E_{\rm ref}(0) \simeq A(0)\,z_{n}\,.
\label{B81}
\end{equation}
Equation~\eqref{B81} is the basis of the Airy-scaling plots used in the main text. After removing the continuum edge and the channel-dependent constant offset, the low-lying meson energies should fall on a straight line when plotted against $z_n$, with slope $A(0)$.

Finally, once the even $+$ branch has been selected, as is appropriate for the $(S^z_{\rm total},\kappa,I,C_{\rm flip})=(0,0,+1,+1)$ sector studied in the main text, we suppress the explicit channel label and write the low-energy theory energies simply as
\begin{equation}
E_n^{\mathrm{th}}
\equiv
E_n^{(+)} ,
\label{B82}
\end{equation}
and similarly for the corresponding continuum-relative binding energies and level spacings. This is the notation adopted in the revised comparison plots. 

\section{Additional numerical results}\label{appC} 

In this Appendix we provide numerical analyses that complement the main-text discussion of the level-statistics crossover, the emergent domain-wall-number organization of the spectrum, and the confined meson ladder. Appendix~\ref{appC_levelstats} gives further details of the sigmoidal crossover extraction and residual-bootstrap procedure used in Figs.~\ref{figure2} and \ref{figure3}, and examines the robustness of the level-statistics crossover at additional staggered fields. Appendix~\ref{appC_domain_walls} presents a spectrum-wide
summary of the domain-wall-number fluctuations, thereby connecting the global level-statistics crossover to the approximate $W$-resolved structure of the eigenstates. 
Appendix~\ref{app:momentum_robustness} tests the momentum-sector
robustness of the level statistics at $h=0.3$ by comparing the
inversion-invariant $\kappa=0$ block with the first two generic nonzero two-site translation sectors.  Appendix~\ref{appC_mesons_h01} collects supplementary diagnostics for the main-text $W=2$ meson comparison at $N=22$, $\Delta=-4.5$, and $h=0.1$, while Appendix~\ref{appC_mesons_additional} extends the
meson-spectroscopy analysis to $h=0.05$ and $h=0.2$. 

\subsection{Level-statistics crossover: fitting procedure and field dependence}\label{appC_levelstats}

In this subsection we provide additional details of the procedure used in Sec.~\ref{subsec:results_levelstats} to define and extract the finite-size crossover anisotropy from the mean adjacent-gap ratio $\langle r\rangle$. We first describe the sigmoidal fitting and residual-bootstrap analysis underlying Figs.~\ref{figure2}(a) and \ref{figure3}. We then examine the field dependence of the same level-statistics diagnostic at $h=0.1$, $0.2$, and $0.4$ and discuss the circumstances under which a unique finite-size crossover
$\Delta_{\rm cross}(N)$ can, or cannot, be assigned. Throughout this subsection we work in the symmetry-resolved sector
\begin{equation}
\left(S_{\rm total}^{z},\kappa,\mathcal{I},\mathcal{C}_{\rm flip}\right)
= (0,0,+1,+1)
\end{equation}
for periodic chains of length $N=18,20$, and $22$.

\paragraph*{Sigmoidal extraction and bootstrap uncertainties.} 
The level-statistics curves in Fig.~\ref{figure2}(a) evolve smoothly with anisotropy and do not contain a sharply defined finite-size transition point. In particular, identifying the crossover with the first computed value of $\Delta$ for which $\langle r\rangle$ falls below a prescribed threshold would make the result depend explicitly on the spacing of the sampled anisotropy grid and would also make it sensitive to the residual finite-size fluctuations of $\langle r\rangle$. We therefore use the full crossover profile to define an operational finite-size scale.

Introducing
\begin{equation}
x=|\Delta|\,
\end{equation}
we parameterize the crossover for each system size by the four-parameter sigmoid
\begin{equation}\label{eq:app_sigmoid}
f_N(x) = r_{\rm low}^{(N)}
+
\frac{r_{\rm high}^{(N)}-r_{\rm low}^{(N)}}{1+\exp\!\left[
\left(x-x_{\rm cross}^{(N)}\right)/w_N \right]}\,. 
\end{equation}
Here $r_{\rm high}^{(N)}$ and $r_{\rm low}^{(N)}$ are the fitted values on the weak- and strong-anisotropy sides of the crossover, respectively, and $w_N>0$ controls its width. The parameter $x_{\rm cross}^{(N)}$ has a particularly simple interpretation, i.e. 
substitution into Eq.~\eqref{eq:app_sigmoid} gives
\begin{equation}\label{eq:app_sigmoid_midpoint}
f_N\!\left(x_{\rm cross}^{(N)}\right) =
\frac{r_{\rm high}^{(N)}+r_{\rm low}^{(N)}}{2}\,
\end{equation}
so that $x_{\rm cross}^{(N)}$ is the midpoint between the two fitted plateau values. Moreover,
\begin{equation}
\left.
\frac{d f_N}{d x}
\right|_{x=x_{\rm cross}^{(N)}} =
-\frac{r_{\rm high}^{(N)}-r_{\rm low}^{(N)}}{4w_N}\,
\label{eq:app_sigmoid_slope}
\end{equation}
and the magnitude of the derivative is maximal at this point. The midpoint therefore also identifies the maximum-slope position of the fitted crossover. We define
\begin{equation}
\Delta_{\rm cross}(N) = -x_{\rm cross}^{(N)}\,.
\label{eq:app_delta_cross}
\end{equation}

We emphasize that Eq.~\eqref{eq:app_sigmoid} is used only as a
smooth operational parameterization of the finite-size spectral
crossover. We do not assume that the sigmoid is a universal scaling function, nor do we identify its midpoint with a thermodynamic critical point. The fitting procedure is intended instead to replace a visually selected crossover point by a reproducible quantity that depends on the entire crossover profile. The fits are constrained to be monotonically decreasing with $x$, with $r_{\rm high}^{(N)}>r_{\rm low}^{(N)}$ and $w_N>0$. In the numerical optimization we use
\begin{align}
\begin{split}
0.34 &\leq r_{\rm low}^{(N)} \leq 0.52 \\
0.005 &\leq r_{\rm high}^{(N)}-r_{\rm low}^{(N)} \leq 0.22 \\
1.2 &\leq x_{\rm cross}^{(N)} \leq 4.5 \\
0.03 &\leq w_N \leq 2\,.
\label{eq:app_sigmoid_bounds}
\end{split}
\end{align}
For the main-text $h=0.3$ analysis, all three optimized midpoints lie well inside these bounds.

The fitted parameters for $h=0.3$ are collected in
Table~\ref{tab:sigmoid_parameters}. The weak-anisotropy values are
close to the large-matrix GOE benchmark
$\langle r\rangle_{\rm GOE}\simeq0.5307$, whereas the
strong-anisotropy values remain above the Poisson result
$\langle r\rangle_{\rm Poisson}=2\ln 2-1\simeq0.3863$ at the
available system sizes. Most importantly, the fitted midpoint moves
systematically toward more negative $\Delta$ as the chain length is
increased. 

As an additional check on the fitting interval for the main-text
$h=0.3$ data, we repeat the analysis with $|\Delta|_{\rm max}=4.6$ fixed and $|\Delta|_{\rm min}=1.0,1.1,1.2$, and $1.3$. For $N=20$ and $22$, the extracted crossover positions vary by less than approximately $0.006$ and $0.002$, respectively. The smaller $N=18$ system shows a larger spread of approximately $0.105$, comparable to its broader residual-bootstrap interval. Thus the finite-size drift of $\Delta_{\rm cross}(N)$ is not tied to the particular choice $|\Delta|_{\rm min}=1.1$, which we retain as the common fitting boundary used in the main analysis.
\begin{table}[t]
\begin{ruledtabular}
\begin{tabular}{cccccc}
$N$ & $r_{\rm high}^{(N)}$ & $r_{\rm low}^{(N)}$ & $w_N$ & $\Delta_{\rm cross}(N)$ & $R^2$ \\[2pt] \hline
18 & 0.5306 & 0.4284 & 0.4377 & $-2.406^{+0.077}_{-0.063}$ & 0.954 \\
20 & 0.5283 & 0.4463 & 0.3843 & $-2.699^{+0.030}_{-0.031}$ & 0.988 \\
22 & 0.5316 & 0.4701 & 0.3292 & $-2.843^{+0.020}_{-0.020}$ & 0.993
\end{tabular}
\end{ruledtabular}
\caption{
         {\it Parameters of the sigmoidal fits in Eq.~\eqref{eq:app_sigmoid} for the main-text data at $h=0.3$.}  The quoted uncertainties of $\Delta_{\rm cross}(N)$ are the 16th--84th percentile intervals of the residual-bootstrap distributions described below. The final column gives the coefficient of determination of the central fit.
        }
\label{tab:sigmoid_parameters}
\end{table}

We next estimate how precisely the midpoint can be extracted from a given finite-size curve. Since the ED values of $\langle r\rangle$ are deterministic, there is no statistical sampling error associated with the individual data points. The small nonmonotonic variations of the data about the smooth sigmoid instead reflect finite-size structure that is not represented explicitly by Eq.~\eqref{eq:app_sigmoid}. We use a residual bootstrap to quantify the sensitivity of the extracted midpoint to this unresolved structure.

For a given size $N$, let
\begin{equation}\label{eq:app_residuals}
\epsilon_i^{(N)} = \langle r\rangle_i^{(N)} - f_N(x_i)
\end{equation}
be the residual of the $i$th data point about the best-fit sigmoid. We first center the residuals,
\begin{equation}
\widetilde{\epsilon}_i^{(N)} = \epsilon_i^{(N)}
-\frac{1}{M_N}\sum_{j=1}^{M_N}\epsilon_j^{(N)}\,
\label{eq:app_centered_residuals}
\end{equation}
where $M_N$ is the number of anisotropy values entering the fit.
The set of centered residuals $\{\widetilde{\epsilon}_i^{(N)}\}$ is then treated as an empirical pool describing the deviations of the finite-size data from the smooth crossover curve.

For one bootstrap realization $b$, we draw $M_N$ residuals independently with replacement from this pool and construct the
synthetic dataset
\begin{equation}
\langle r\rangle_{i,b}^{(N)} = f_N(x_i) + \widetilde{\epsilon}_{i,b}^{(N)}\,.
\label{eq:app_bootstrap_data}
\end{equation}
The anisotropy values $x_i$ themselves are kept fixed. The synthetic
dataset is then fitted again to exactly the same sigmoid,
Eq.~\eqref{eq:app_sigmoid}, producing a new midpoint
$x_{{\rm cross},b}^{(N)}$ and hence
\begin{equation}
\Delta_{{\rm cross},b}(N) = -x_{{\rm cross},b}^{(N)}\,.
\end{equation}
Repeating this procedure $5000$ times produces an empirical
distribution of extracted crossover locations. We quote the
16th and 84th percentiles of this distribution as the residual-bootstrap
68\% interval shown in the figures and tables.

This interval should therefore be interpreted as a 
{\it fit-extraction uncertainty}: it measures how much the inferred sigmoid midpoint changes when the observed residual structure is redistributed among the sampled anisotropies. It is not a statistical error bar on the deterministic ED data, and it does not include systematic uncertainty associated with the choice of sigmoid itself or with the interval of anisotropies included in the fit. We assess the latter separately below by varying the lower boundary of the fit
window.
\begin{figure*}[t]
\includegraphics[width=1.00\textwidth]{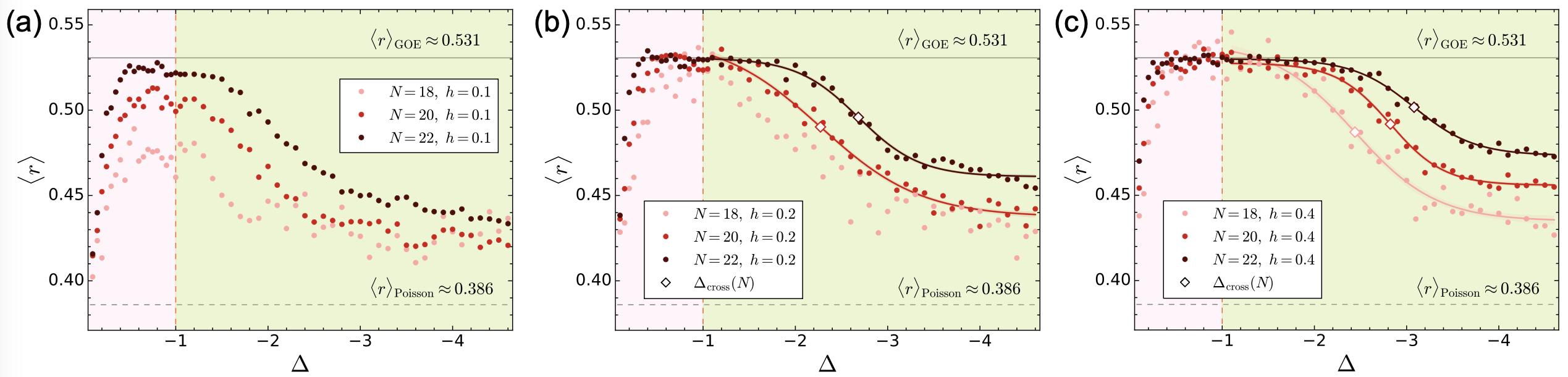}
\caption{
         {\it Field dependence of the mean adjacent-gap ratio for the staggered-field XXZ chain.}
         The quantity $\langle r\rangle$ is shown as a function of anisotropy $\Delta$ for $N=18,20,22$ in the symmetry sector  $(S_{\rm total}^{z},\kappa,\mathcal{I},\mathcal{C}_{\rm flip})
=(0,0,+1,+1)$ at (a) $h=0.1$, (b) $h=0.2$, and (c) $h=0.4$.
         The solid and dashed horizontal lines indicate the large-matrix GOE value $\langle r\rangle_{\rm GOE}\simeq0.5307$ and the Poisson value $\langle r\rangle_{\rm Poisson}\simeq0.3863$, respectively. The vertical dashed line at $\Delta=-1$ and the background shading refer to the zero-field XXZ phase diagram and distinguish the portion of the zero-field critical regime shown here from the antiferromagnetic regime $\Delta<-1$. 
         In panel (a) the dependence is nonmonotonic and no sigmoid midpoint is assigned.
         In panels (b) and (c), the solid curves are sigmoidal fits performed only over $1.1\leq|\Delta|\leq4.6$. The raw points at smaller $|\Delta|$ are displayed but are not included in the fit. Open diamonds mark the accepted values of $\Delta_{\rm cross}(N)$, and the faint envelopes around the fitted curves denote the corresponding 16th--84th percentile residual-bootstrap ranges. For $h=0.2$, a stable midpoint is obtained only for $N=20$ and $22$, whereas for $h=0.4$ stable midpoints are obtained for all three system sizes.
}
\label{figure_level}
\end{figure*}

For the main-text finite-size extrapolation, the bootstrap is
propagated one step further. For every bootstrap index $b$ we retain
the matched triplet
\begin{equation}
\left\{
\Delta_{{\rm cross},b}(18),
\Delta_{{\rm cross},b}(20),
\Delta_{{\rm cross},b}(22)
\right\}
\end{equation}
and perform the same unweighted linear fit used for the central
values,
\begin{equation}\label{eq:app_bootstrap_fss}
\Delta_{{\rm cross},b}(N)
= \Delta_{{\rm cross},b}(\infty) + \frac{a_b}{N}\,. 
\end{equation}
Thus every bootstrap realization generates one complete
finite-size-scaling line. The distribution of the resulting
$\Delta_{{\rm cross},b}(\infty)$ values gives the uncertainty of the
thermodynamic intercept quoted in Sec.~\ref{subsec:results_levelstats},
while the 16th and 84th percentiles of the ensemble of fitted lines
at fixed $1/N$ generate the shaded band in Fig.~\ref{figure3}. Because only three system sizes are available, we use an unweighted linear fit and interpret the extrapolation as an indicative thermodynamic crossover scale rather than a precision determination of a critical anisotropy. In particular, the bootstrap interval does not include possible corrections to the assumed linear $1/N$ form.

\paragraph*{Field dependence and limitations of the crossover construction.} 
To determine how robust the level-statistics evolution is against
changes of the staggered field, we extend the calculation to $h=0.1$, $0.2$, and $0.4$. The complete results are shown in Fig.~\ref{figure_level}. In all three panels the raw ED data are displayed over the full calculated interval extending from the
weak-anisotropy region into the antiferromagnetic Ising regime. The light background to the left of $\Delta=-1$ indicates the part of the critical region of the {\it zero-field} XXZ chain shown in the figure, whereas the green background marks the zero-field antiferromagnetic regime $\Delta<-1$. The vertical dashed line at $\Delta=-1$ is therefore included only as a reference to the $h=0$ phase diagram so that at finite staggered magnetic field it should not be interpreted as an exact phase boundary. 

At the weakest field, $h=0.1$ [Fig.~\ref{figure_level}(a)], the
anisotropy dependence is qualitatively different from the approximately monotonic crossover observed at $h=0.3$. Starting near $\Delta=0$, $\langle r\rangle$ first increases toward a broad maximum in the vicinity of the zero-field boundary and only then decreases upon moving deeper into the antiferromagnetic regime. The maxima of the raw finite-size data occur at approximately
\begin{equation}
\Delta_{\rm max} \simeq -0.55,\,-0.75,\,-0.80
\end{equation}
for $N=18,20$, and $22$, respectively, with
\begin{equation}
\langle r\rangle_{\rm max} \simeq 0.489,\ 0.513,\ 0.528\,.
\end{equation}
Thus the maximum moves closer to the zero-field boundary and toward the GOE benchmark as the system size is increased, but the three curves do not exhibit a common GOE-like upper plateau followed by a single monotonic decline. A fit to the complete curve would therefore violate the basic assumption underlying Eq.~\eqref{eq:app_sigmoid}. Restricting the fit to only part of the descending branch also produces a fit-window-dependent result, particularly for $N=18$. We consequently do not assign $\Delta_{\rm cross}(N)$ at $h=0.1$. This case provides an explicit example of the limitation of the midpoint construction, i.e. a numerical value is quoted only when the data exhibit a sufficiently well-defined single crossover that can be parameterized stably by the sigmoid.

For $h=0.2$ [Fig.~\ref{figure_level}(b)], the evolution in the antiferromagnetic regime becomes much closer to the monotonic form
seen in the main-text $h=0.3$ data. To separate the crossover in the
AF regime from the rise of $\langle r\rangle$ on the $-1<\Delta<0$ side, the sigmoid is fitted only over
\begin{equation}\label{eq:app_fit_window}
1.1 \leq |\Delta| \leq 4.6 
\end{equation}
although all calculated points are retained in the figure. For $N=20$ and $N=22$, the resulting fits are well behaved and yield stable interior midpoints. For $N=18$, in contrast, the optimization drives $x_{\rm cross}$ to the lower allowed boundary $x_{\rm cross}=1.2$ and simultaneously produces an anomalously large extrapolated upper value $r_{\rm high}\simeq0.617$. The midpoint is therefore not determined by a well-resolved crossover contained inside the fitting interval. We regard this fit as unreliable and do not report a value of $\Delta_{\rm cross}(18)$ at $h=0.2$.

At $h=0.4$ [Fig.~\ref{figure_level}(c)], all three sizes show a well-developed weak-anisotropy shoulder followed by a smooth decrease in $\langle r\rangle$ within the antiferromagnetic regime. Application of the same fit interval, Eq.~\eqref{eq:app_fit_window}, gives stable interior midpoints for $N=18,20$, and $22$. The extracted crossover again shifts toward more negative anisotropy with increasing chain length.

The accepted additional-field crossover values are summarized in
Table~\ref{tab:additional_crossovers}. As in Table~\ref{tab:sigmoid_parameters}, the uncertainties are the 16th--84th percentile residual-bootstrap intervals. An entry is reported only when the fitted midpoint lies in the interior of the allowed parameter range and remains stable under modest changes of the fit window. A good value of $R^2$ by itself is not taken as sufficient evidence that the midpoint is meaningful.

\begin{table}[b]
\begin{ruledtabular}
\begin{tabular}{cccc}
$h$ & $N$ & $\Delta_{\rm cross}(N)$ & $R^2$ \\ \hline
0.2 & 18 & --- & --- \\ 
0.2 & 20 & $-2.272^{+0.086}_{-0.073}$ & 0.982 \\
0.2 & 22 & $-2.682^{+0.033}_{-0.032}$ & 0.986 \\
0.4 & 18 & $-2.440^{+0.081}_{-0.070}$ & 0.961 \\
0.4 & 20 & $-2.824^{+0.029}_{-0.027}$ & 0.987 \\
0.4 & 22 & $-3.081^{+0.023}_{-0.023}$ & 0.992
\end{tabular}
\end{ruledtabular}
\caption{
         {\it Accepted finite-size crossover anisotropies at the additional staggered fields $h=0.2$ and $0.4$.} All central values are obtained from fits over $1.1\leq|\Delta|\leq4.6$, and the quoted uncertainties are the 16th--84th percentile residual-bootstrap intervals. The $N=18$, $h=0.2$ fit is omitted because its midpoint is driven to the lower fitting bound and is not stably determined by the data.
}
\label{tab:additional_crossovers}
\end{table}

As a separate check of the dependence on the chosen fitting interval, we keep $|\Delta|_{\rm max}=4.6$ fixed and repeat the fits with
$|\Delta|_{\rm min}=1.0$, $1.1$, $1.2$, and $1.3$. For $h=0.2$, the accepted $N=22$ midpoint changes by less than $0.01$ over this entire range, while the $N=20$ result varies by approximately $0.10$, comparable to its residual-bootstrap uncertainty. At $h=0.4$, the total variation is approximately $0.052$, $0.020$, and $0.004$ for $N=18$, $20$, and $22$, respectively. These variations are small compared with the overall finite-size drift of the crossover and confirm that the accepted midpoints are not artifacts of choosing precisely $|\Delta|_{\rm min}=1.1$. We therefore use $1.1\leq|\Delta|\leq4.6$ as the common fitting interval in Fig.~\ref{figure_level} and in Table~\ref{tab:additional_crossovers}.

Taken together, the additional-field results show both the robustness and the limitations of the finite-size crossover construction. At $h=0.2$, $0.3$, and $0.4$, whenever a stable midpoint can be identified, $\Delta_{\rm cross}(N)$ moves systematically toward more negative anisotropy with increasing $N$. For a fixed size, the accepted midpoints also move to more negative $\Delta$ as the staggered field is increased. At the weakest field $h=0.1$, however, the spectral evolution becomes nonmonotonic and cannot be reduced reliably to the same single-midpoint description. We therefore use the $h=0.3$ data as the primary quantitative finite-size extrapolation in the main text and regard the additional fields in Fig.~\ref{figure_level} as tests of the field dependence and applicability of the crossover construction, rather than performing separate thermodynamic extrapolations for each $h$.

\subsection{Additional domain-wall-number diagnostics}
\label{appC_domain_walls}
The level-statistics crossover discussed in
Sec.~\ref{subsec:results_levelstats} and Appendix~\ref{appC_levelstats}
is accompanied by the emergence of an approximate domain-wall-number decomposition of the many-body spectrum. We quantify the extent of this reorganization by supplementing the energy-resolved results of Fig.~\ref{fig:direct_W_diagnostics} with a spectrum-wide summary of the domain-wall-number variance. 

\begin{figure}[t]
\includegraphics[width=1.00\columnwidth]{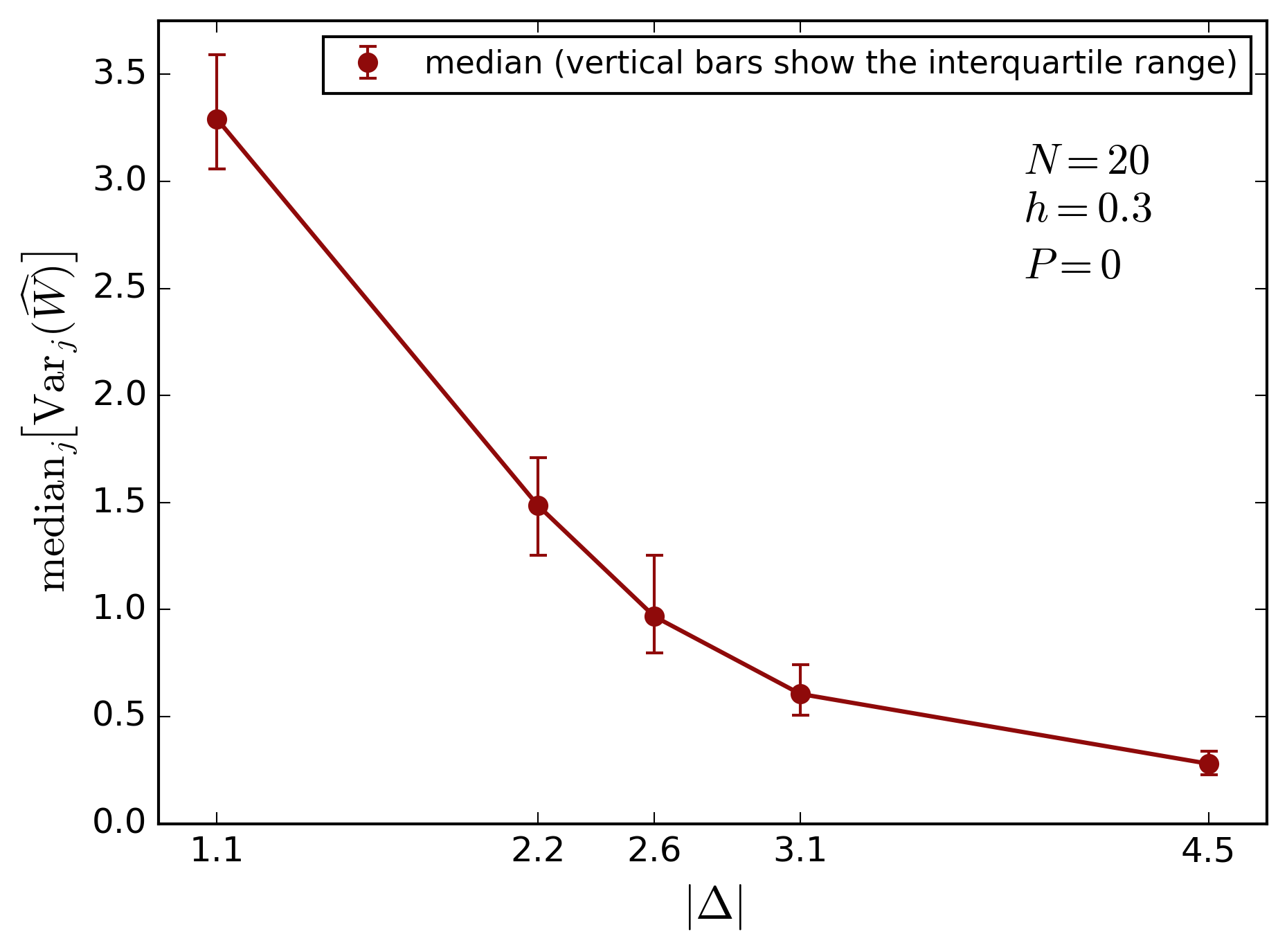}
     \caption{
             {\it Spectrum-averaged domain-wall-number fluctuations for $N=20$, $h=0.3$, and $(S_{\mathrm{tot}}^{z},\kappa,I,C_{\mathrm{flip}}) =(0,0,+1,+1)$.}
             The points show $\operatorname{median}_{j} [\operatorname{Var}_{j}(\widehat W)]$ over all many-body eigenstates in the symmetry-resolved block. The vertical bars extend from the 25th to the 75th percentile and therefore represent the interquartile range, rather than a statistical error estimate. The thin line is a guide to the eye. The monotonic decrease with increasing $|\Delta|$ demonstrates the broad suppression of mixing between different domain-wall sectors deep in the antiferromagnetic regime, while the remaining finite variance is consistent with approximate domain-wall sectorization rather than exact fragmentation. 
}
\label{fig:W_variance_summary}
\end{figure}

The energy-resolved results in
Fig.~\ref{fig:direct_W_diagnostics} show that the eigenstate
fluctuations of the domain-wall-number operator are progressively
suppressed as the system is taken deeper into the antiferromagnetic
Ising regime. To quantify this trend across the full many-body
spectrum, we consider the median of $\operatorname{Var}_{j}(\widehat W)$ over all eigenstates in the symmetry sector
$(S_{\mathrm{tot}}^{z},\kappa,I,C_{\mathrm{flip}})
=(0,0,+1,+1)$.
The resulting spectrum-averaged diagnostic is shown in
Fig.~\ref{fig:W_variance_summary} for $N=20$ and $h=0.3$.
In addition to the four anisotropies displayed in the main-text
energy-resolved figure, we include the intermediate value
$\Delta=-2.2$.

The median domain-wall-number variance decreases monotonically as $|\Delta|$ is increased. Specifically, we obtain
\begin{align}
\begin{split}
\operatorname{median}_{j}
&\!\left[\operatorname{Var}_{j}(\widehat W)\right] = \\
&3.291,\ 1.486,\ 0.967,\ 0.606,\ 0.279
\end{split}
\label{eq:median_W_variance_values}
\end{align}
for
\begin{equation}
|\Delta|=1.1,\ 2.2,\ 2.6,\ 3.1,\ 4.5,
\end{equation}
respectively. Between $\Delta=-1.1$ and $\Delta=-4.5$, the
median variance is therefore reduced by approximately a factor of $11.8$. The interquartile range, indicated by the vertical bars in Fig.~\ref{fig:W_variance_summary}, narrows simultaneously. Thus, the suppression of domain-wall-number fluctuations is not confined to a small number of specially selected eigenstates, but occurs broadly throughout the spectrum.

The variance nevertheless remains finite at all anisotropies studied. This is consistent with the fact that $\widehat W$ does not commute exactly with the full Hamiltonian, i.e. the transverse exchange terms continue to hybridize states with different domain-wall numbers. The monotonic reduction of both the median variance and its spectral spread therefore supports the interpretation of \(W\) as an emergent approximate quantum number and of the resulting spectral organization as fragmentation-like rather than exactly fragmented. 

As an independent numerical consistency check, we also compared the directly evaluated correlation measure $C_{j}^{zz}$ with the exact operator relation
\begin{equation}
C_{j}^{zz} = 2\left(\langle\widehat W\rangle_{j} - \langle\widehat W\rangle_{0}\right)\,.
\end{equation}
For all eigenstates and anisotropies entering the analysis, the
maximum discrepancy is below $10^{-12}$, confirming the
domain-wall interpretation of the correlation bands to numerical
precision. 

\subsection{Momentum-sector robustness of the level statistics}
\label{app:momentum_robustness}

The level-statistics analysis in the main text is performed in the
symmetry sector
\begin{equation}
(S^z_{\rm total},\kappa,I,C_{\rm flip})=(0,0,+1,+1) 
\label{eq:C_kappa0_sector}
\end{equation}
for which the two-site translation phase is $\kappa=0$. This sector is particularly convenient because $\kappa=0$ is invariant under spatial inversion and therefore permits an additional resolution by the parity quantum number $I=\pm1$. A generic nonzero translation sector has a different residual symmetry structure.  It is therefore useful to verify explicitly that the crossover in the level statistics identified in Sec.~\ref{subsec:results_levelstats} is not a peculiarity of the special $\kappa=0$ block. 
\begin{figure}[t]
\centering
\includegraphics[width=1.0\columnwidth]{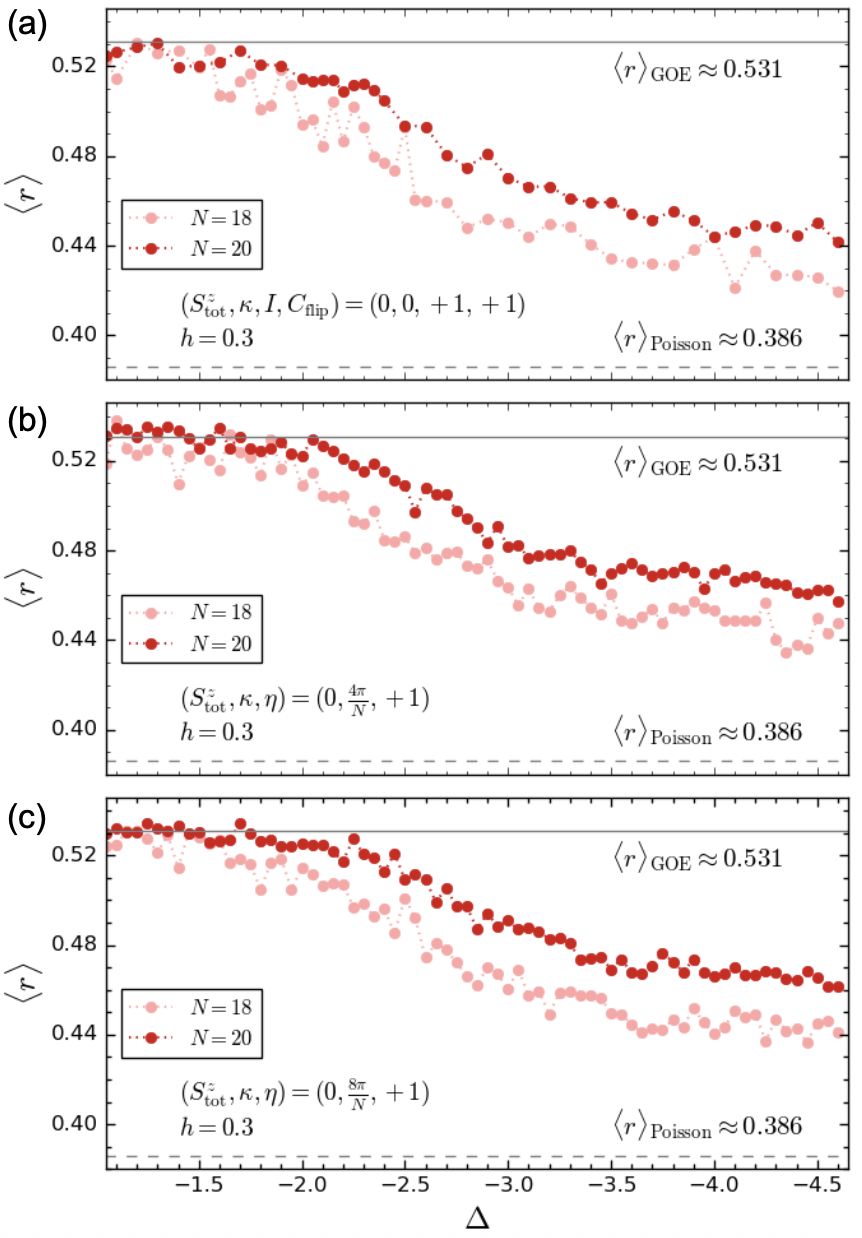}
\caption{
        {\it Momentum-sector robustness of the mean adjacent-gap ratio for the staggered-field XXZ chain at $h=0.3$.} 
       (a) Results for the inversion-invariant sector $(S^z_{\rm total},\kappa,I,C_{\rm flip})=(0,0,+1,+1)$.
       (b) Results in the first generic nonzero two-site translation sector, $(S^z_{\rm total},\kappa,\zeta)=(0,4\pi/N,+1)$ ($m=1$). 
       (c) Results in the second generic nonzero sector, $(S^z_{\rm total},\kappa,\zeta)=(0,8\pi/N,+1)$ ($m=2$). In all panels the data are shown for $N=18$ and $20$ over $1.1\leq|\Delta|\leq4.6$. The solid and dashed horizontal lines indicate the GOE value $\langle r\rangle_{\rm GOE}\simeq0.5307$ and the Poisson value $\langle r\rangle_{\rm Poisson}=2\ln2 - 1\simeq0.3863$, respectively. The same crossover from GOE-like statistics at weak anisotropy to reduced level repulsion at strong anisotropy is visible in all three sectors, demonstrating that the spectral evolution is not peculiar to the special $\kappa=0$ block.
    }
\label{fig:kappa_levelstats}
\end{figure}

As discussed in Sec.~\ref{section_spinchain} and Appendix~\ref{app:Z2}, translation by two lattice sites obeys
\begin{equation}\label{eq:C_T2_kappa}
\mathcal{T}_2|\psi_\kappa\rangle
= e^{i\kappa}|\psi_\kappa\rangle 
\end{equation}
with the finite-size translation phases
\begin{equation}\label{eq:C_kappa_grid}
\kappa_m=\frac{4\pi m}{N},
\qquad
m=0,\ldots,\frac{N}{2}-1\,. 
\end{equation}
The sublattice-preserving inversion reverses the translation phase,
\begin{equation}
\mathcal{I}\mathcal{T}_2\mathcal{I}^{-1}
= \mathcal{T}_2^{-1}
\label{eq:C_I_T2}
\end{equation}
and hence maps $\kappa \rightarrow -\kappa$. Consequently, at a generic $\kappa\neq0,\pi$, inversion does not act within a single $\kappa$ sector and an independent parity quantum number cannot be assigned. The combined translation--spin-flip symmetry remains available and satisfies
\begin{equation}
\mathcal{C}_{\rm flip}^{\,2}=\mathcal{T}_{2}\,.
\label{eq:C_Cflip_square}
\end{equation}
Thus, within a fixed $\kappa$ sector, its two possible eigenvalues are
\begin{equation}
c_{\rm flip}
= \zeta\, e^{i\kappa/2},
\qquad
\zeta=\pm1 \,.
\label{eq:C_Cflip_eta}
\end{equation}
The generic nonzero-$\kappa$ blocks used below are therefore resolved
by $(S^z_{\rm total},\kappa,\zeta)$ rather than by
$(S^z_{\rm total},\kappa,I,C_{\rm flip})$. 
\begin{table}[b]
\centering
\begin{ruledtabular}
\begin{tabular}{ccccc}
$N$ & $\dim\mathcal{H}_{S^z_{\rm total}=0}$ & $m=0$ & $m=1$ & $m=2$ \\[3pt] \hline
18 & 48620  & 1530 & 2672 & 2728 \\
20 & 184756 & 4947 & 9225 & 9247
\end{tabular}
\end{ruledtabular}
\caption{
        {\it Dimensions of the symmetry-resolved Hilbert-space blocks used in the momentum-sector robustness analysis at $S^z_{\rm total}=0$.} For $m=0$, the two-site translation phase is $\kappa=0$ and the additional inversion symmetry is resolved, so the sector is $(S^z_{\rm total},\kappa,I,C_{\rm flip})=(0,0,+1,+1)$. For the generic nonzero translation sectors $m=1,2$, inversion does not act within a fixed $\kappa$ block and the sectors are labeled by $(S^z_{\rm total},\kappa,\zeta)$ with $\zeta=+1$. The full $S^z_{\rm total}=0$ dimensions are shown for comparison.
}
\label{tab:kappa_block_dimensions}
\end{table}

Although the Hamiltonian matrix in a generic momentum basis is complex Hermitian, the corresponding fully resolved block remains in the orthogonal random-matrix symmetry class.  To see this, let $\mathcal{K}_0$ denote complex conjugation in the computational $\sigma^z$ basis. The Hamiltonian is real in this basis, so $\mathcal{K}_0$ is an antiunitary symmetry of the full Hamiltonian. Both $\mathcal{K}_0$ and $\mathcal{I}$ reverse the two-site translation phase separately, whereas their product
\begin{equation}
\mathcal{A}\equiv\mathcal{I}\mathcal{K}_0
\label{eq:C_A_def}
\end{equation}
acts within a fixed generic-$\kappa$ sector. Moreover,
$\mathcal{I}\,\mathcal{C}_{\rm flip}\,\mathcal{I}^{-1}
=\mathcal{C}_{\rm flip}^{-1}$, so the antiunitary operation $\mathcal{A}$ also preserves the fixed $\mathcal{C}_{\rm flip}$ eigenvalue. Since the inversion operator is real and satisfies $\mathcal{I}^2=1$, one has
\begin{equation}
\mathcal{A}^2=+1 \,.
\label{eq:C_A_square}
\end{equation}
The appropriate chaotic reference for each fully resolved generic $(\kappa,\zeta)$ block is therefore the GOE value
$\langle r\rangle_{\rm GOE}\simeq0.5307$, rather than the GUE value. 
\begin{figure*}[t]
\centering
\includegraphics[width=\textwidth]{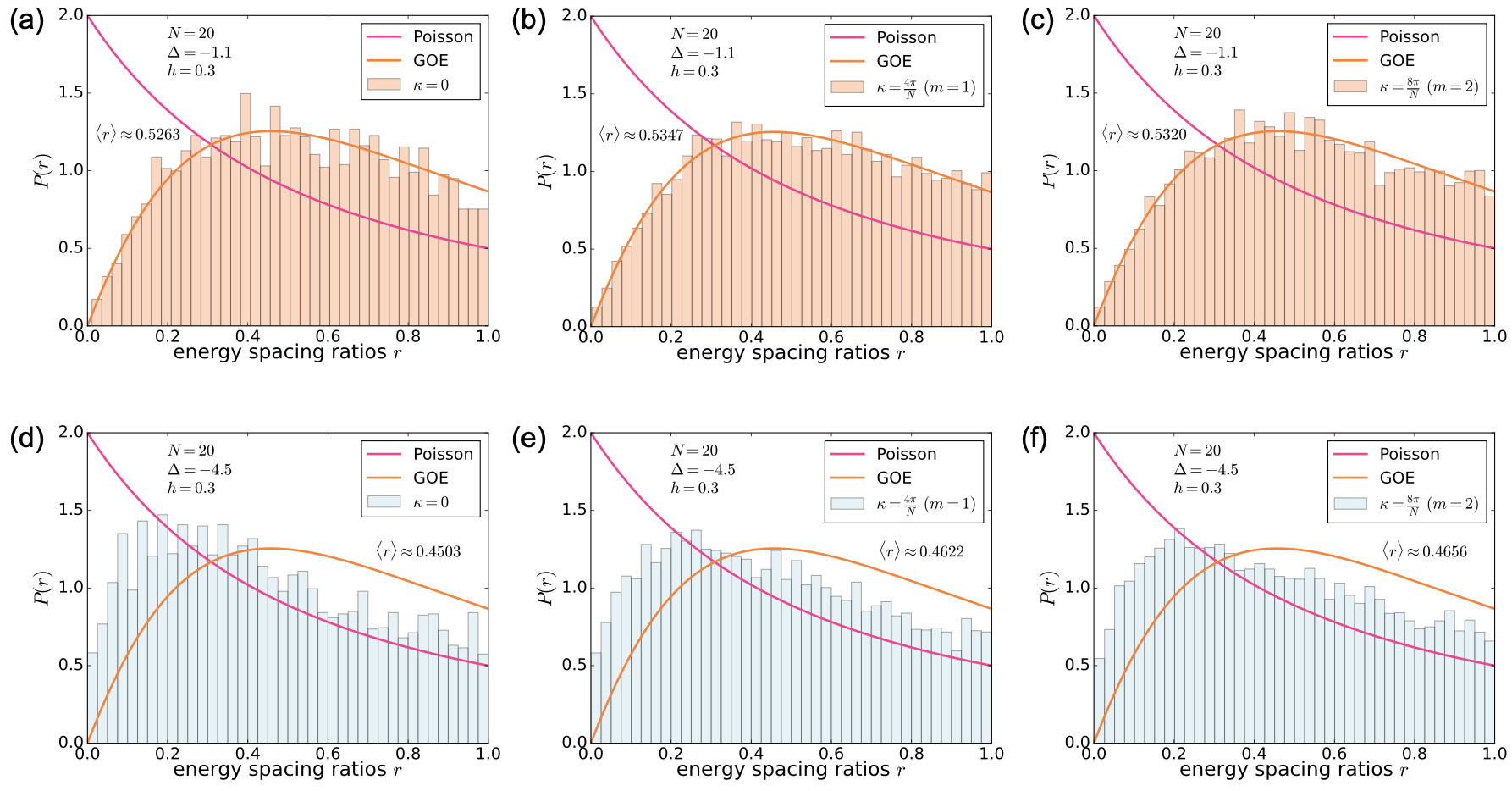}
\caption{
        {\it Momentum-sector comparison of the full adjacent-gap-ratio distribution $P(r)$ for $N=20$ and $h=0.3$.} 
        Panels~(a)--(c) show $\Delta=-1.1$ for
       (a) $\kappa=0$, $I=+1$, $C_{\rm flip}=+1$,
       (b) $\kappa=4\pi/N$ ($m=1$), $\zeta=+1$, and
       (c) $\kappa=8\pi/N$ ($m=2$), $\zeta=+1$, respectively. The corresponding means are $\langle r\rangle\simeq0.5263$, $0.5347$, and $0.5320$. Panels~(d)--(f) show the same three momentum sectors at $\Delta=-4.5$, with $\langle r\rangle\simeq0.4503$, $0.4622$, and $0.4656$, respectively. The histograms are compared with the Poisson distribution (magenta) and the GOE Wigner--Dyson surmise (orange). At $\Delta=-1.1$, all three momentum sectors closely follow the GOE form. At $\Delta=-4.5$, all three exhibit a pronounced loss of GOE level repulsion and a redistribution of weight toward smaller $r$, while remaining intermediate between the GOE and Poisson limits.
}
\label{fig:kappa_histograms}
\end{figure*}

We test the momentum-sector dependence at the same field used for the main-text analysis, namely $h=0.3$, and for chain lengths $N=18$ and $20$. In addition to the $\kappa=0$ sector of Eq.~(\ref{eq:C_kappa0_sector}), we consider the first two nonzero two-site translation sectors: 
\begin{align}
m=1:\qquad&
(S^z_{\rm total},\kappa,\zeta)
=\left(0,\frac{4\pi}{N},+1\right)
\label{eq:C_m1_sector}\\
m=2:\qquad&
(S^z_{\rm total},\kappa,\zeta)
=\left(0,\frac{8\pi}{N},+1\right)\,.
\label{eq:C_m2_sector}
\end{align}
These are generic momentum sectors for both system sizes considered here and therefore do not admit an additional inversion resolution. Their Hilbert-space dimensions consequently differ from those of the $\kappa=0$ blocks. 

The dimensions of the symmetry-resolved blocks used in this comparison are summarized in Table~\ref{tab:kappa_block_dimensions}. For $N=18$ ($20$), the special $\kappa=0$, $I=+1$, $C_{\rm flip}=+1$ block contains 1530 (4947) states, whereas the generic $m=1$ and $m=2$, $\zeta=+1$ blocks contain 2672 and 2728 (9225 and 9247) states, respectively. The finite-momentum comparison therefore probes substantially larger symmetry blocks with a different residual symmetry structure from the one used in the main-text $\kappa=0$ analysis. 

Figure~\ref{fig:kappa_levelstats} compares the mean adjacent-gap ratio over the antiferromagnetic interval $1.1\leq|\Delta|\leq4.6$.  Panel~(a) reproduces the raw $\kappa=0$ data for $N=18$ and $20$, while panels~(b) and (c) show the corresponding results in the $m=1$ and $m=2$ sectors, respectively. 
No sigmoidal fits are performed for the additional momentum sectors, i.e. their purpose here is to test the robustness of the spectral evolution, rather than to extract separate momentum-dependent crossover scales.

All three momentum sectors display the same qualitative evolution. On the weak-anisotropy side the adjacent-gap ratio lies close to the GOE reference value. Upon increasing $|\Delta|$, $\langle r\rangle$ decreases systematically and moves away from GOE toward the Poisson benchmark. The precise finite-size values are momentum dependent, as expected for distinct symmetry blocks, but the loss of GOE-like level
repulsion occurs in both generic nonzero-$\kappa$ sectors in the same anisotropy regime as in the $\kappa=0$ sector. In particular, the closely similar behavior of the $m=1$ and $m=2$ data shows that the result is not restricted to the lowest nonzero translation sector.

To test this conclusion beyond the single spectrum-averaged number $\langle r\rangle$, we also compare the full adjacent-gap-ratio distributions $P(r)$ at two representative anisotropies. The results for $N=20$ are shown in Fig.~\ref{fig:kappa_histograms}. The upper row, panels~(a)--(c), corresponds to $\Delta=-1.1$ in the $\kappa=0$, $m=1$, and $m=2$ sectors, respectively. For the case $\Delta=-1.1$ the  numerical means are
\begin{equation}
\langle r\rangle
= 0.5263,\quad 0.5347,\quad 0.5320
\label{eq:C_rmeans_weak}
\end{equation}
for $m=0,1,2$.  In all three cases the distributions closely follow the GOE form and exhibit the characteristic suppression of small spacing ratios associated with level repulsion.

The lower row, panels~(d)--(f), shows the same three sectors at
$\Delta=-4.5$, where
\begin{equation}
\langle r\rangle
= 0.4503,\quad 0.4622,\quad 0.4656
\label{eq:C_rmeans_strong}
\end{equation}
for $m=0,1,2$. 
Here all three distributions deviate strongly from the GOE form and develop enhanced weight at small $r$. They have shifted toward the Poisson distribution but remain intermediate between the GOE and Poisson limits, in agreement with the interpretation of the main-text $\kappa=0$ data as a finite-size crossover toward reduced level repulsion rather than a complete Poisson limit. 

The agreement between the momentum sectors at the level of both
$\langle r\rangle(\Delta)$ and the full distributions $P(r)$ provides a direct check that the level-statistics crossover identified in
Sec.~\ref{subsec:results_levelstats} is not generated by the special symmetry structure of the $\kappa=0$ sector. Generic nonzero-$\kappa$ blocks, in which inversion cannot be separately resolved and whose Hilbert-space dimensions differ substantially from those at $\kappa=0$, exhibit the same qualitative evolution from GOE-like statistics to reduced level repulsion as the system is taken deeper into the antiferromagnetic Ising regime. The momentum-sector dependence visible in the precise finite-size values of $\langle r\rangle$ therefore does not alter the central conclusion that increasing confinement is accompanied by a robust finite-size departure from GOE-like spectral statistics.

\subsection{Supporting diagnostics for the main-text meson comparison at
$\boldsymbol{h=0.1}$}\label{appC_mesons_h01}

In this subsection we provide three complementary diagnostics for the main-text meson comparison at $N=22$, $\Delta=-4.5$, and $h=0.1$, focusing again on the $W=2$ one-meson band in the $(S^z_{\rm total},\kappa,I,C_{\rm flip})=(0,0,+1,+1)$ sector. Whereas Sec.~\ref{sec:numerical_comparison_mesons} compares the ED spectrum to the three analytic descriptions using absolute energies, absolute level spacings, and theory-specific diagnostics, we supplement those results here by considering dimensionless relative level spacings, normalized spacing ratios, and compact scaled representations of the same meson ladder. 

As a dimensionless complement to the absolute level spacings
$\Delta E_n^\gamma=E_{n+1}^\gamma-E_n^\gamma$ considered in
Sec.~\ref{sec:numerical_comparison_mesons}, we define the relative
spacing
\begin{equation}
s_{n}^{\gamma}
\equiv
\frac{\Delta E_{n}^{\gamma}}{E_{n}^{\gamma}},
\qquad
\gamma=\mathrm{ED},\mathrm{th},\mathrm{WKB},\mathrm{SAE}\,.
\label{eq:relative_spacing}
\end{equation}
This normalization measures the separation of successive meson levels relative to the energy of the corresponding state and therefore provides a complementary view of the accuracy of the analytic ladders.

\begin{figure}[t]
\centering
\includegraphics[width=1.00\columnwidth]{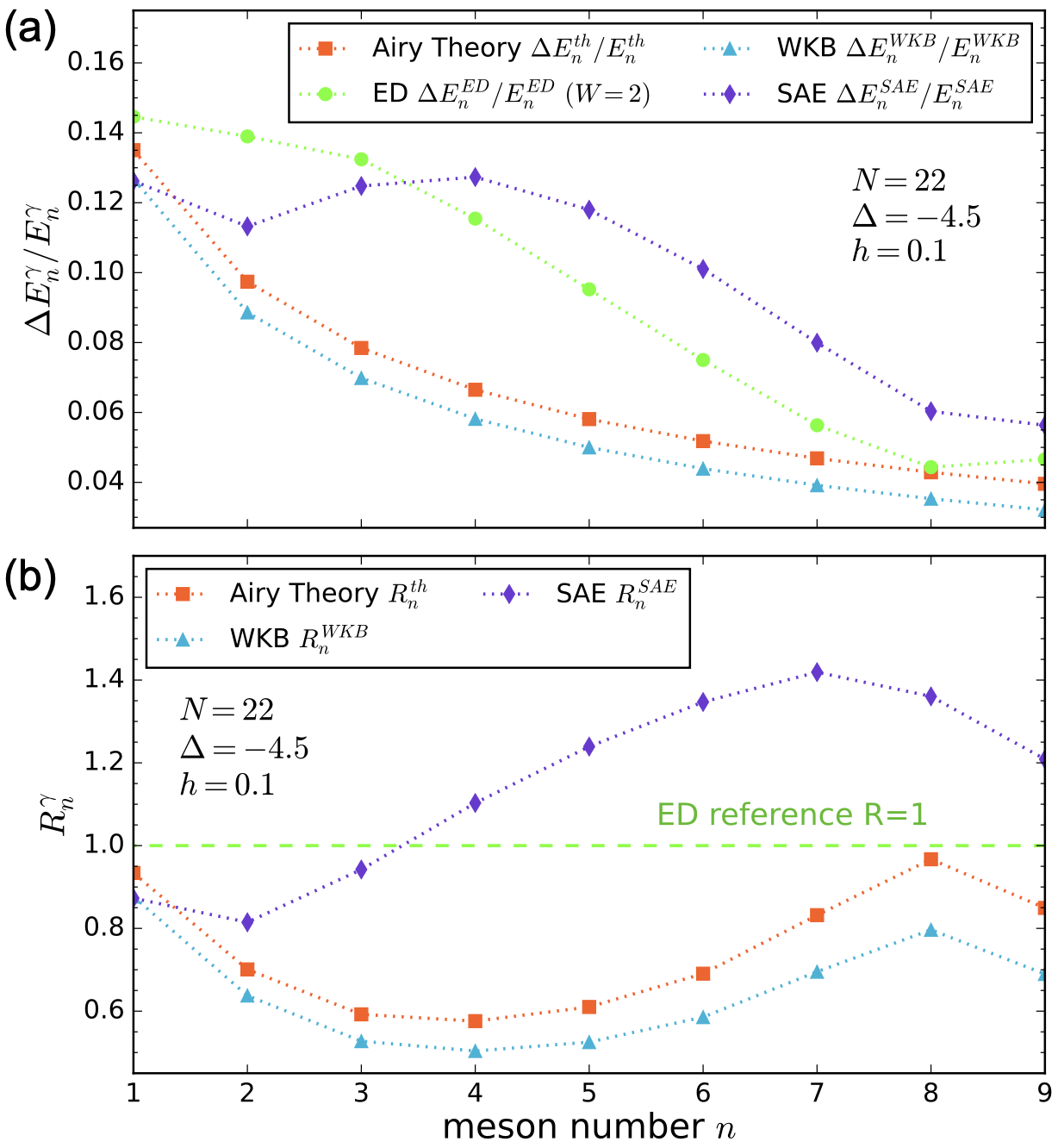}
\caption{
        {\it Relative level-spacing diagnostics for the $W=2$ meson spectrum at $P=0$ for the periodic chain with $N=22$, $\Delta=-4.5$, and $h=0.1$.}
        (a) Dimensionless relative spacings $s_n^\gamma=\Delta   E_n^\gamma/E_n^\gamma$ for the ED spectrum and the three analytic descriptions $\gamma=\mathrm{ED},\mathrm{th}, \mathrm{WKB},\mathrm{SAE}$. All descriptions reproduce the overall decrease of the relative spacing with meson number, while systematic quantitative deviations remain.
        (b) Corresponding ratios $R_n^\gamma=s_n^\gamma/s_n^{\mathrm{ED}}$ for the three analytic descriptions. The dashed horizontal line at $R=1$ denotes exact agreement with the ED relative spacing.  The Airy and WKB results generally underestimate the ED relative spacings through the interior of the ladder, whereas the finite-$N$ strong-anisotropy result crosses from an underestimate at low meson number to an overestimate over part of the higher-$n$ sequence.  The normalized comparison therefore provides a complementary measure of the theory-dependent deviations seen in the absolute-spacing analysis of the main text.
}
\label{fig:relative_spacing_h01}
\end{figure}

Figure~\ref{fig:relative_spacing_h01}(a) compares $s_{n}^\gamma$ for the ED spectrum and the three analytic descriptions. All four curves show an overall decrease of the relative spacing with increasing meson number, although systematic quantitative differences remain. The Airy and WKB results lie below the ED values through most of the displayed sequence. The finite-$N$ strong-anisotropy result shows a different behavior, i.e. it is below the ED relative spacing for the lowest mesons, becomes comparable to it at intermediate meson number, and exceeds it over part of the upper sequence.

To display these deviations directly, we further introduce
\begin{equation}
R_n^\gamma \equiv
\frac{s_n^\gamma}{s_n^{\mathrm{ED}}}
= \frac{\Delta E_n^\gamma/E_n^\gamma}
     {\Delta E_n^{\mathrm{ED}}/E_n^{\mathrm{ED}}},
\qquad
\gamma=\mathrm{th},\mathrm{WKB},\mathrm{SAE}\,.
\label{eq:relative_spacing_ratio}
\end{equation}
Perfect agreement with the ED relative spacing corresponds to
$R_n^\gamma=1$. As shown in Fig.~\ref{fig:relative_spacing_h01}(b), the normalized ratios make the theory-dependent deviations particularly transparent. The Airy and WKB descriptions generally underestimate the ED relative spacings in the interior of the ladder, while the finite-$N$ strong-anisotropy result crosses through the ED reference and subsequently overestimates the relative spacings over part of the higher-$n$ sequence. Thus the relative-spacing analysis confirms that the analytic descriptions reproduce the overall scale and evolution of the meson ladder, while none of them reproduces the complete finite-size spacing sequence uniformly. 

\begin{figure}[t]
\includegraphics[width=1.00\columnwidth]{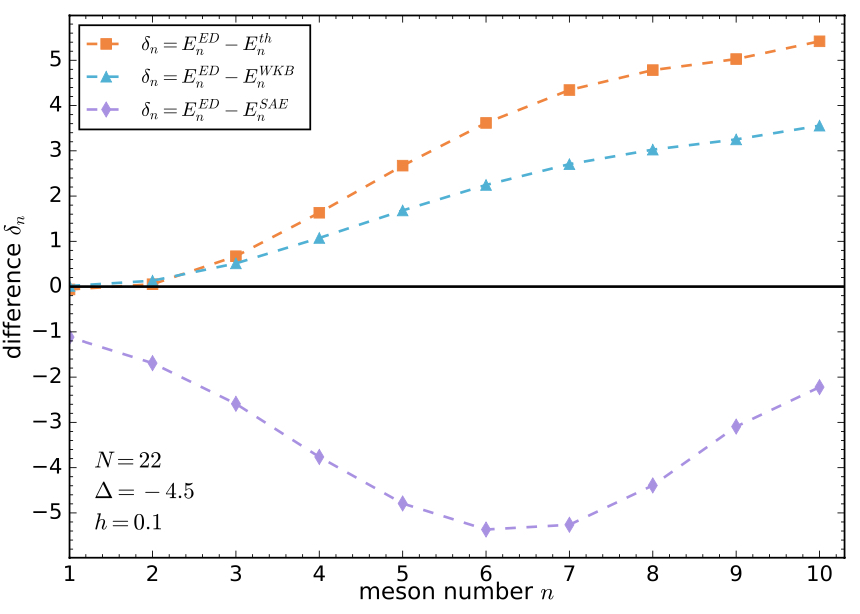}
\caption{
         {\it Residual comparison for the $W=2$ meson spectrum at $P=0$ for $N=22$, $\Delta=-4.5$, and $h=0.1$.} 
         The plot shows the differences $\delta_n^\gamma=E_n^{\mathrm{ED}}-E_n^\gamma$ for the three analytic descriptions $\gamma=\mathrm{th},\mathrm{WKB},\mathrm{SAE}$ as a function of meson number $n$. Presented in this compact form, the figure highlights the same hierarchy discussed in the main text, i.e.  the Airy-ladder theory is most accurate for the lowest mesons,  
         the WKB approximation provides the relevant semiclassical benchmark for the interior of the ladder but remains systematically offset from the finite-size ED spectrum and the finite-$N$ strong-anisotropy expansion gives the smallest residuals at higher meson number $n$. 
}
\label{figure11}
\end{figure}

Figure~\ref{figure11} provides a complementary view in terms of the absolute energy residuals
\begin{equation}
\delta_n^\gamma=E_n^{\rm ED}-E_n^\gamma
\end{equation}
with $\gamma=\mathrm{th},\mathrm{WKB},\mathrm{SAE}$ for the same
parameter set. Figure~\ref{figure11} reinforces the hierarchy discussed in Sec.~\ref{sec:numerical_comparison_mesons}. Presented on a common set of axes, these residuals make the systematic deviations of the three analytic descriptions directly visible. The low-energy Airy theory gives the smallest deviations for the lowest few mesons nearest the deconfined threshold, the WKB result remains systematically offset from the ED spectrum, and the finite-$N$ strong-anisotropy expansion gives the smallest residuals toward the upper part of the $W=2$ sequence for $N=22$ and $h=0.1$. 

\begin{figure}[t]
\includegraphics[width=1.00\columnwidth]{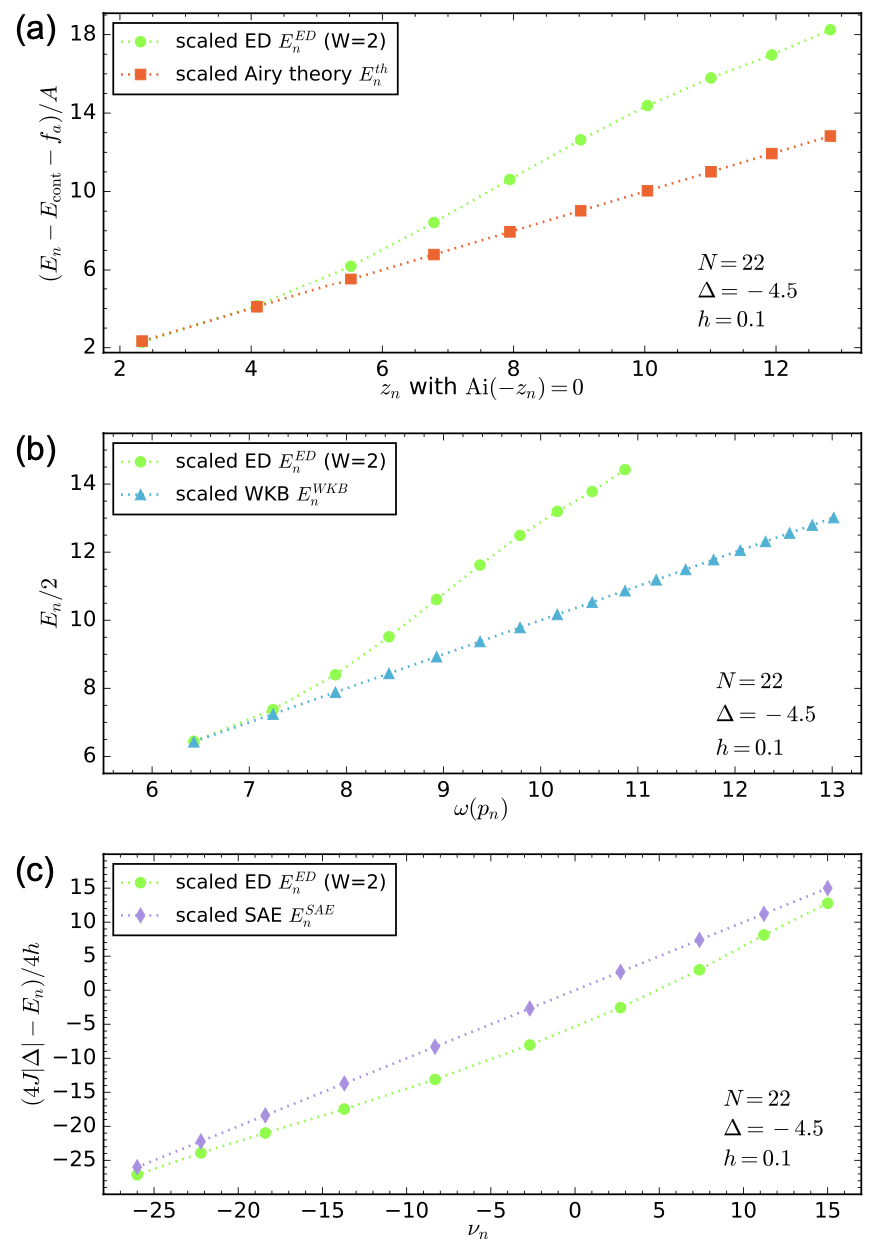}
\caption{
         {\it Scaled comparisons for the $W=2$ meson spectrum at $P=0$ for the periodic chain with $N=22$, $\Delta=-4.5$, and $h=0.1$.} 
         (a) Continuum-relative binding energies plotted as $(E_n-E_{\rm cont}(0)-f\,a_+(0|\eta))/A(0)$ versus the Airy magnitudes $z_n$, comparing the ED data to the low-energy Airy-ladder theory. 
         (b) ED energies compared to the semiclassical/WKB relation $E_n^{\mathrm{WKB}}=2\omega(p_n)$, shown as a function of $\omega(p_n)$. 
         (c) ED energies compared to the finite-$N$ strong-anisotropy form $E_n^{\mathrm{SAE}}=4J|\Delta|-4h\nu_n$, shown as a function of $\nu_n$. The three panels provide a compact visual summary of how the Airy, WKB, and SAE descriptions organize the same $W=2$ ED ladder in their natural variables. 
}
\label{figure10}
\end{figure}

Finally, Figure~\ref{figure10} reorganizes the same ED ladder in the natural variables of each of the three analytic descriptions. 
The top panel shows the continuum-relative binding energies in the Airy representation, where the natural scaling variable is the Airy magnitude $z_{n}$. In this form, the low-energy theory predicts an approximately linear dependence with slope $A(0)$ once the continuum threshold and the channel-dependent offset are removed. The middle panel reorganizes the same ED data in the WKB variables, displaying the energies as a function of $\omega(p_{n})$, which is the natural quantity entering the semiclassical relation $E_{n}^{\mathrm{WKB}}=2\omega(p_{n})$. 
The bottom panel shows the strong-anisotropy representation, in which the finite-$N$ theory predicts a linear dependence on the roots $\nu_{n}$ through $E_{n}^{\mathrm{SAE}} = 4J|\Delta| - 4h\nu_{n}$. Taken together, the three panels provide a compact summary of how the same ED ladder is organized by the Airy, WKB, and SAE descriptions when each is viewed in its own natural coordinates.
 
Taken together, these Appendix diagnostics complement the main-text comparison in three different ways. The relative-spacing analysis quantifies the level separations after normalization by the overall meson-energy scale, the residual comparison exposes the remaining absolute-energy deviations directly, and the scaled representations reorganize the ED spectrum in the natural variables of each analytic description. These complementary views reinforce the conclusion that the Airy, WKB, and finite-$N$ strong-anisotropy descriptions probe
different regimes of the finite-size meson ladder, while none provides a uniformly accurate description over the entire sequence.

\subsection{Additional meson-spectroscopy data at
$\boldsymbol{h=0.05}$ and $\boldsymbol{h=0.2}$} \label{appC_mesons_additional}

In this subsection we extend the meson-spectroscopy comparison of Sec.~\ref{sec:numerical_comparison_mesons} to the additional field values $h=0.05$ and $h=0.2$, keeping the system size and anisotropy fixed at $N=22$ and $\Delta=-4.5$. As in the main text, we focus on the $W=2$ one-meson band in the
$(S^z_{\rm total},\kappa,I,C_{\rm flip})=(0,0,+1,+1)$ sector and compare the ED spectrum to the low-energy Airy-ladder theory, the semiclassical/WKB approximation, and the finite-$N$ strong-anisotropy expansion.

\begin{figure*}[t]
\includegraphics[width=1.00\textwidth]{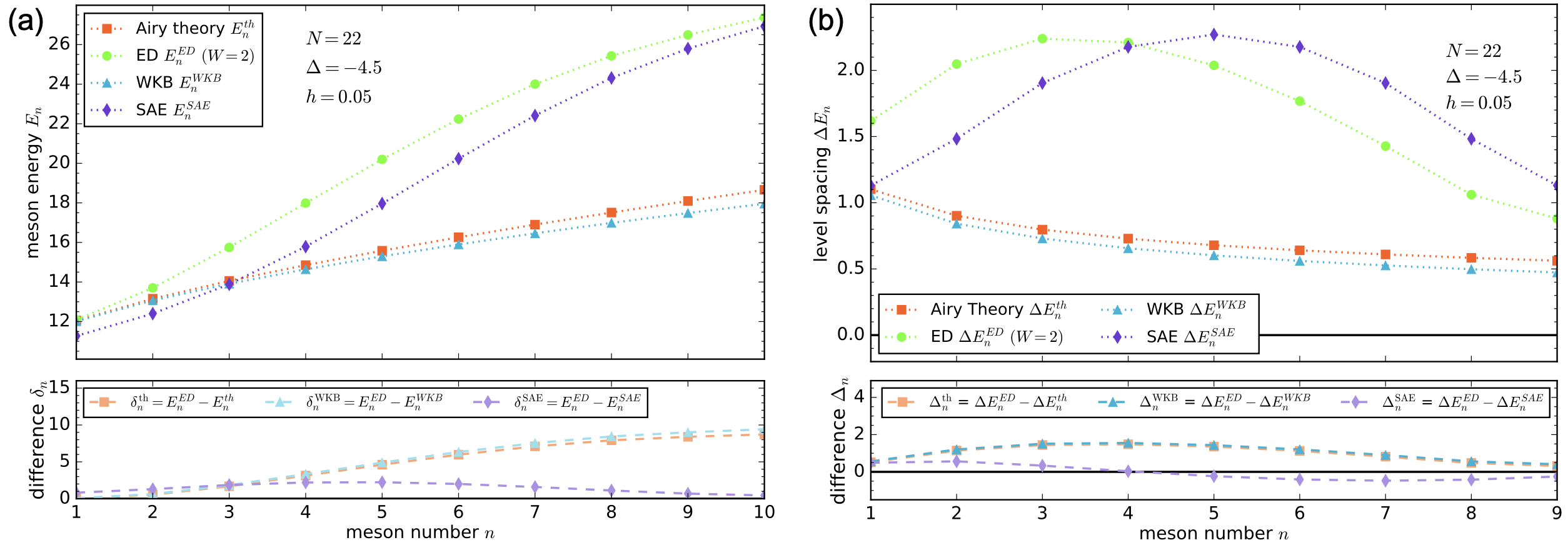}
\caption{
        {\it Additional meson-spectroscopy comparison for the periodic chain with $N=22$, $\Delta=-4.5$, and $h=0.05$ at $P=0$.}
        (a) Absolute $W=2$ meson energies $E_{n}$ and residuals $\delta_{n}^{\gamma}$ with $\gamma=\mathrm{th},\mathrm{WKB},\mathrm{SAE}$ comparing the ED ladder to the three analytic descriptions: Airy theory, semiclassical/WKB, and the finite-$N$ strong-anisotropy expansion.
        (b) Level spacings $\Delta E_{n}$ and spacing residuals $\Delta_{n}^{\gamma}$, $\gamma=\mathrm{th},\mathrm{WKB},\mathrm{SAE}$, comparing ED to the three analytic descriptions: Airy theory, semiclassical/WKB, and finite-$N$ strong-anisotropy expansion. At this weaker confinement strength, the disagreement between ED and the threshold Airy theory grows strongly with meson number, while the finite-$N$ strong-anisotropy expansion provides the closest description of the upper part of the ladder. 
}\label{fig:app_h005_overview}
\end{figure*}

\begin{figure*}[t]
\includegraphics[width=1.00\textwidth]{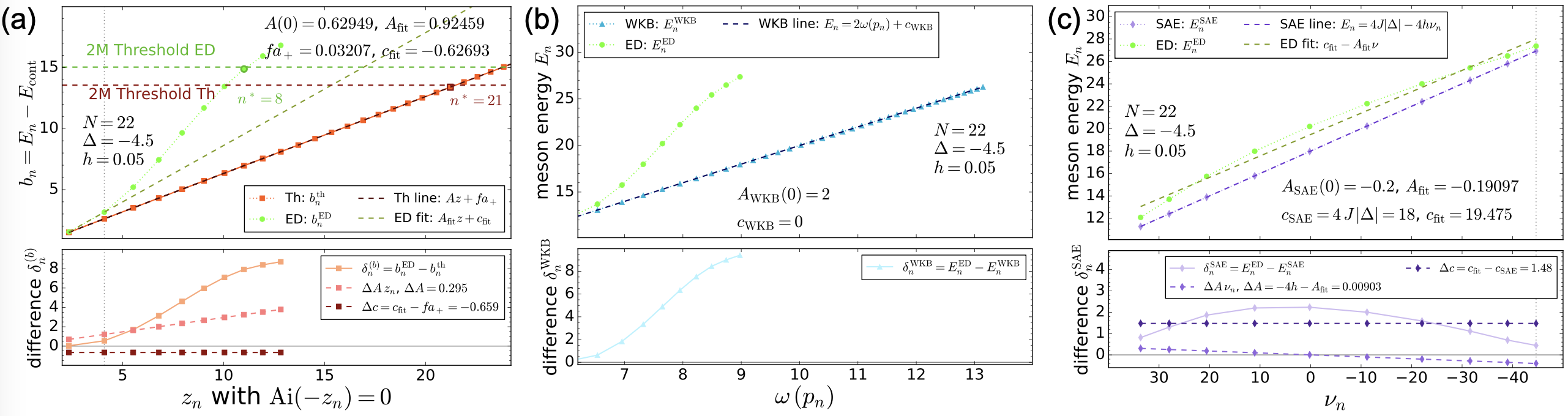}
\caption{
        {\it Theory-specific diagnostics for the $W=2$ meson spectrum at   $P=0$ for $N=22$, $\Delta=-4.5$, and $h=0.05$.}
        (a) Airy-ladder comparison: continuum-relative binding energies $b_{n}=E_{n}-E_{\rm cont}(0)$ plotted versus the Airy magnitudes $z_{n}$. The two horizontal lines mark the analytic two-meson threshold
        $b_{\rm th}^{(2M)}(0)$ and the ED-anchored threshold
        $b_{\rm ED}^{(2M)}(0)$ in the same binding reference, from which the stability counts $n_{\rm th}^{\star}=21$ and $n_{\rm ED}^{\star}=8$ are read off.
        (b) WKB comparison: ED energies plotted against the corresponding $\omega(p_n)$ values entering the semiclassical relation $E_n^{\mathrm{WKB}}=2\omega(p_n)$.
        (c) Strong-anisotropy comparison: ED energies plotted against the finite-$N$ strong-anisotropy variables $\nu_{n}$ entering $E_n^{\mathrm{SAE}}=4J|\Delta|-4h\nu_n$. At $h=0.05$, the threshold Airy description is restricted to the very lowest states, the WKB description remains systematically offset, and the finite-$N$ strong-anisotropy expansion gives the best account of the upper part of the finite-size ladder.  
}\label{fig:app_h005_diagnostics}
\end{figure*}

\begin{figure*}[t]
\includegraphics[width=1.00\textwidth]{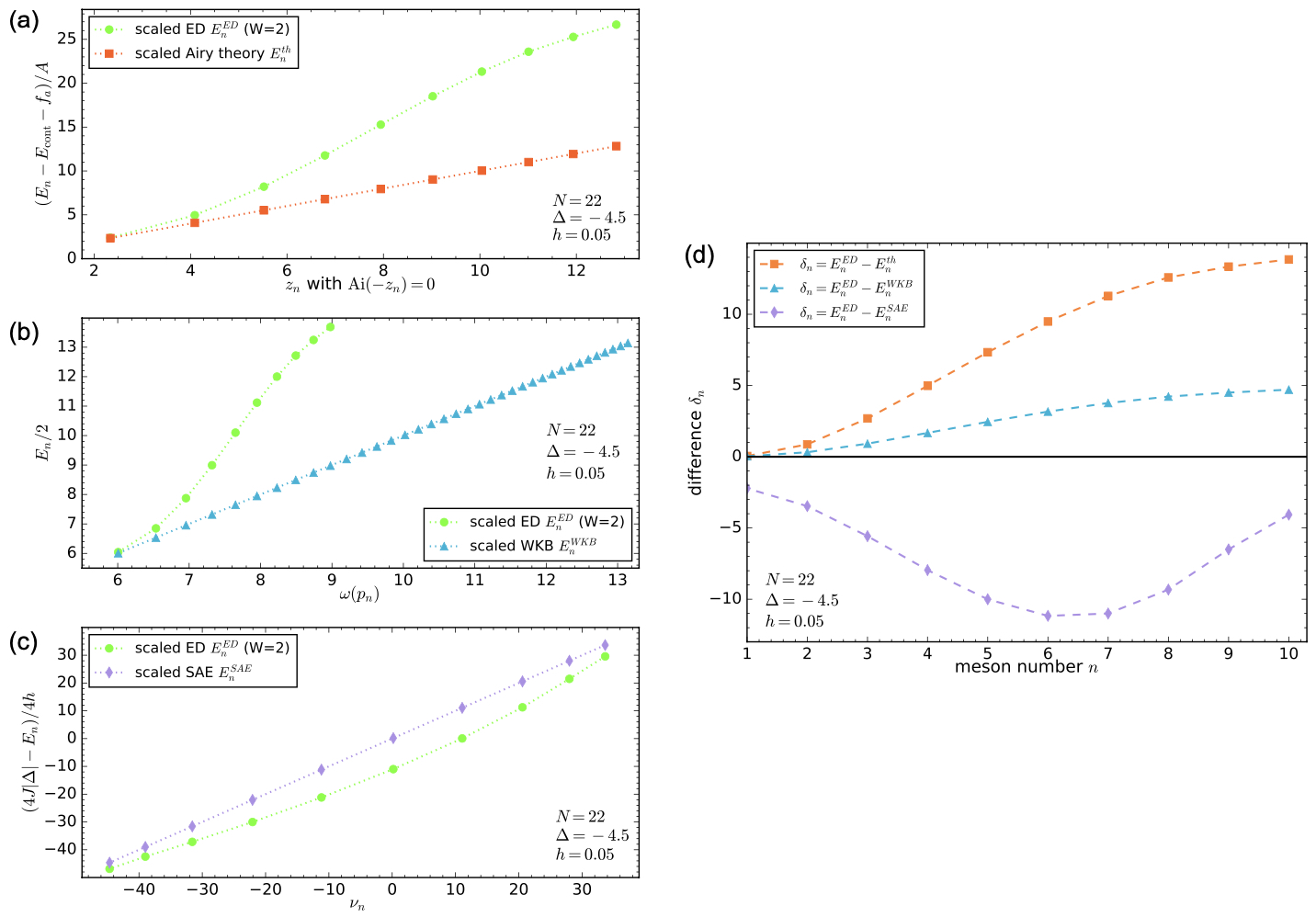}
\caption{
         {\it Compact supporting diagnostics for the $W=2$ meson spectrum at $P=0$ for $N=22$, $\Delta=-4.5$, and $h=0.05$.}
         (a)--(c) Scaled comparisons in the natural variables of the three analytic descriptions: (a) Airy magnitudes $z_{n}$, (b) semiclassical variables $\omega(p_{n})$, and (c) finite-$N$ strong-anisotropy variables $\nu_n$.
         (d) Residual comparison showing $\delta_{n}^{\gamma}=E_{n}^{\mathrm{ED}}-E_{n}^{\gamma}$ for $\gamma=\mathrm{th},\mathrm{WKB},\mathrm{SAE}$ on a common set of axes. These compact diagnostics reinforce the conclusion that at weaker confinement the Airy and WKB descriptions drift strongly away from the ED ladder, whereas the finite-$N$ strong-anisotropy expansion remains the closest analytic description for the higher mesons. 
}\label{fig:app_h005_support}
\end{figure*}

\paragraph*{Weaker confinement: $h=0.05$.}
The comparison for the weaker confinement case $h=0.05$ is shown in
Figs.~\ref{fig:app_h005_overview}--\ref{fig:app_h005_support}. 
Figure~\ref{fig:app_h005_overview}(a) shows that the Airy-ladder theory still reproduces the lowest mesons reasonably well, but its deviation from ED grows strongly with increasing meson number. The corresponding spacing comparison in 
Fig.~\ref{fig:app_h005_overview}(b) leads to the same conclusion, i.e. the threshold theory captures the low-lying spacing pattern, but becomes progressively less accurate deeper in the ladder. This behavior is consistent with the fact that weaker confinement produces a denser ladder of shallow states and therefore enhances the distinction between the threshold regime and the rest of the spectrum. 

The theory-specific diagnostics in
Fig.~\ref{fig:app_h005_diagnostics}(a)--(c) clarify this point further. In the Airy representation, Fig.~\ref{fig:app_h005_diagnostics}(a), the analytic two-meson threshold predicts a much broader one-meson stability window than is observed directly in the finite-size ED spectrum, i.e. 
the threshold counts extracted from the figure are 
$n_{\rm th}^{\star}=21$ and $n_{\rm ED}^{\star}=8$. 
Thus, while the low-energy theory captures the lowest portion of the spectrum, it substantially overestimates the number of cleanly resolved stable one-meson levels at this weaker field. 
The WKB comparison in Fig.~\ref{fig:app_h005_diagnostics}(b)  provides the appropriate semiclassical benchmark for interior states, but the residuals remain large and systematic at this weaker field. By contrast, the finite-$N$ strong-anisotropy comparison in Fig.~\ref{fig:app_h005_diagnostics}(c) provides the best description of the upper part of the ladder, although its residuals still exhibit a pronounced curvature as a function of meson number.
 
The compact supporting diagnostics in
Fig.~\ref{fig:app_h005_support} summarize 
the same picture in a more condensed form.
At $h=0.05$, the Airy and WKB descriptions both drift markedly away from ED as one moves upward through the $W=2$ sequence, whereas the finite-$N$ SAE remains closest to the upper levels.
Overall, the weaker-confinement case emphasizes most clearly that the three analytic descriptions apply in distinct regimes. 
The Airy ladder is restricted to the immediate threshold region, the WKB construction supplies the semiclassical interior benchmark but remains quantitatively offset at this finite size, and the finite-$N$ strong-anisotropy expansion best tracks the upper part of the finite-size ladder.

\begin{figure*}[t]
\includegraphics[width=1.00\textwidth]{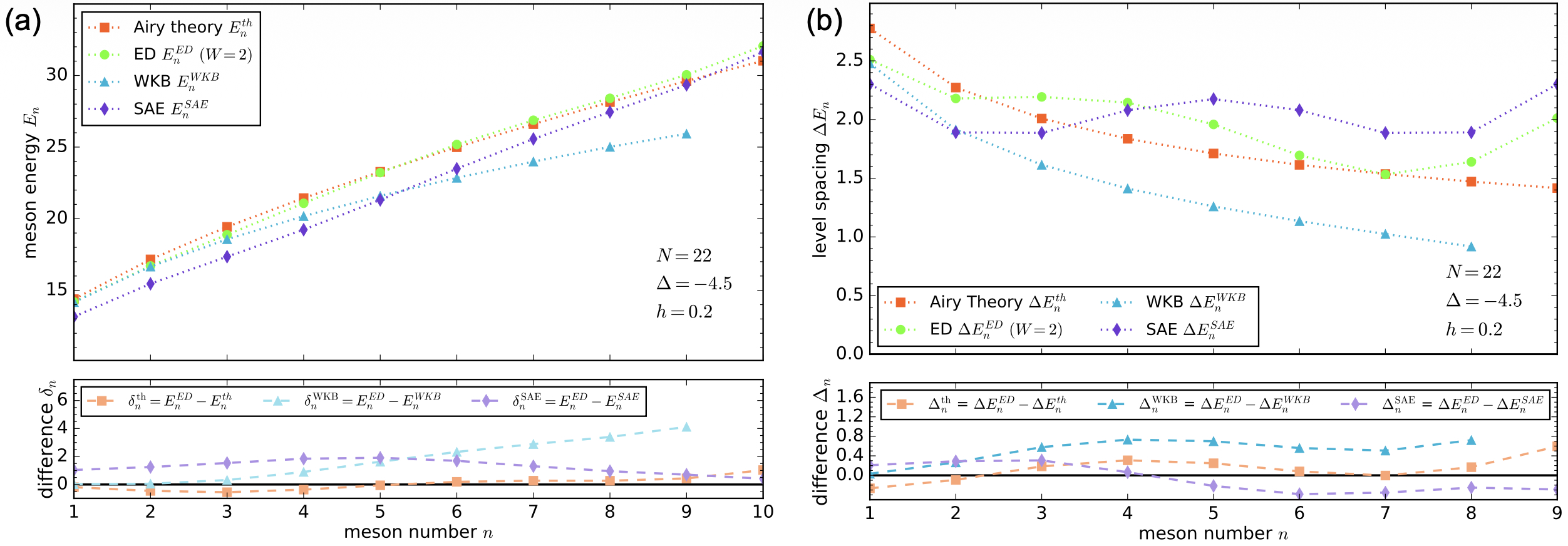}
\caption{
         {\it Additional meson-spectroscopy comparison for the periodic chain with $N=22$, $\Delta=-4.5$, and $h=0.2$ at $P=0$.}
         (a) Absolute $W=2$ meson energies $E_{n}$ and residuals $\delta_{n}^{\gamma}$ with $\gamma=\mathrm{th},\mathrm{WKB},\mathrm{SAE}$  comparing the ED ladder to the three analytic descriptions: Airy theory, semiclassical/WKB, and the finite-$N$ strong-anisotropy expansion.  
         (b) Level spacings $\Delta E_n$ and spacing residuals $\Delta_{n}^{\gamma}$, $\gamma=\mathrm{th},\mathrm{WKB},\mathrm{SAE}$,  comparing ED to the three analytic descriptions: Airy theory, semiclassical/WKB, and the finite-$N$ strong-anisotropy expansion.  
         At this stronger confinement strength, the Airy-ladder description remains close to the ED data over a substantially broader range of meson number than at $h=0.05$, both for the energies and for the spacings. The WKB comparison is improved relative to the $h=0.05$ case but remains displaced from the ED spectrum, whereas the finite-$N$ strong-anisotropy expansion continues to describe the upper part of the ladder well.
}\label{fig:app_h02_overview}
\end{figure*}

\begin{figure*}[t]
\includegraphics[width=1.00\textwidth]{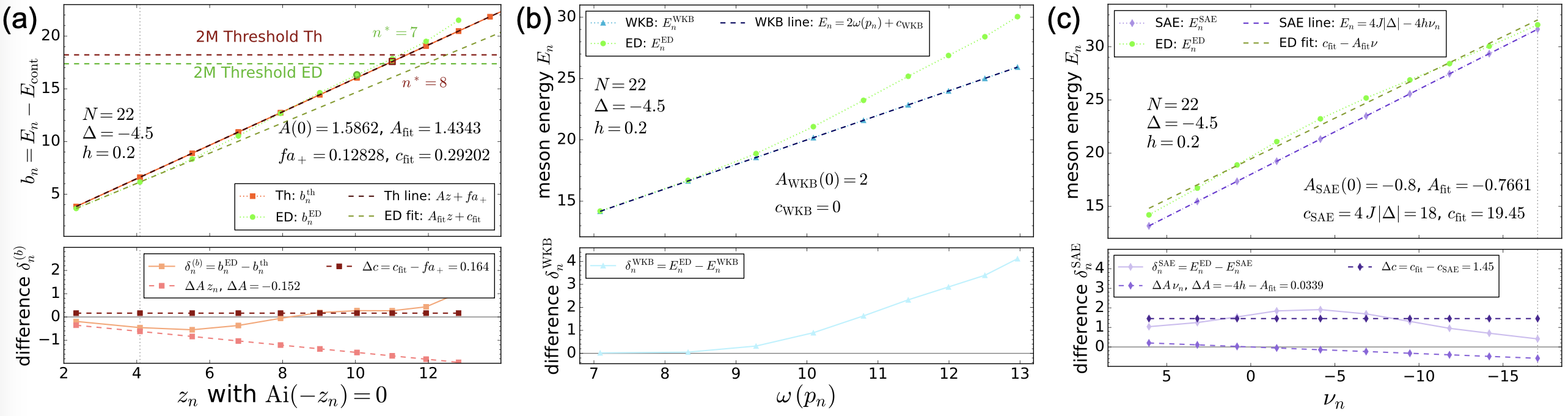}
\caption{
         {\it Theory-specific diagnostics for the $W=2$ meson spectrum at  $P=0$ for $N=22$, $\Delta=-4.5$, and $h=0.2$.}
         (a) Airy-ladder comparison: continuum-relative binding energies $b_{n}=E_{n}-E_{\rm cont}(0)$ plotted versus the Airy magnitudes $z_{n}$. The analytic and ED-anchored two-meson thresholds are close, corresponding to stability counts $n_{\rm th}^{\star}=8$ and $n_{\rm ED}^{\star}=7$.
         (b) WKB comparison: ED energies plotted against the corresponding $\omega(p_{n})$ values entering the semiclassical relation $E_{n}^{\mathrm{WKB}}=2\omega(p_{n})$.
         (c) Strong-anisotropy comparison: ED energies plotted against the finite-$N$ strong-anisotropy variables $\nu_{n}$ entering $E_{n}^{\mathrm{SAE}}=4J|\Delta|-4h\nu_{n}$.  
         At $h=0.2$, the Airy comparison is substantially improved relative to $h=0.05$, while the ED points in the WKB representation lie closer to the semiclassical reference and exhibit reduced residuals. The finite-$N$ strong-anisotropy expansion continues to provide an accurate description of the upper part of the ladder. 
}\label{fig:app_h02_diagnostics}
\end{figure*}

\begin{figure*}[t]
\includegraphics[width=1.00\textwidth]{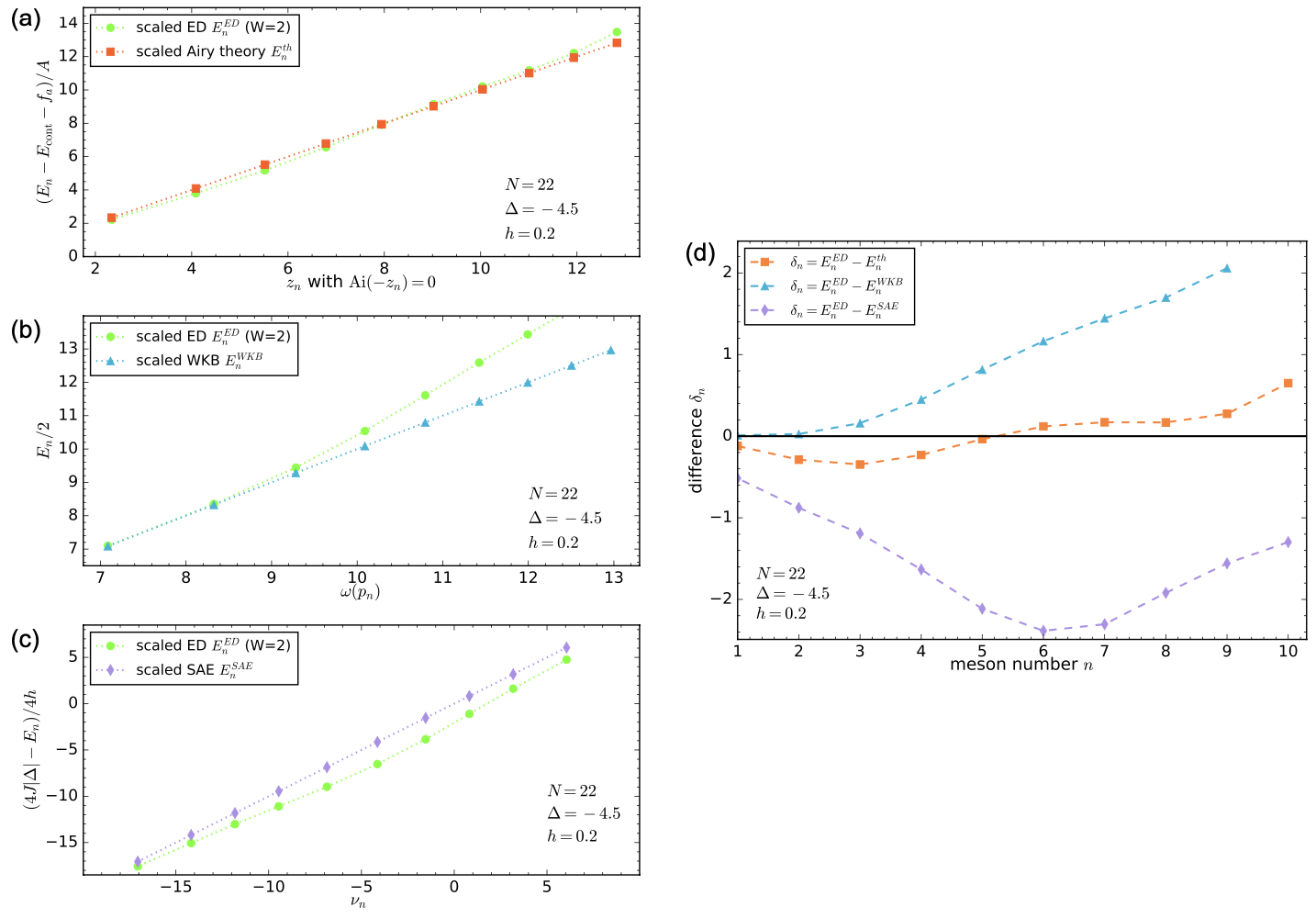}
\caption{
         {\it Compact supporting diagnostics for the $W=2$ meson spectrum at $P=0$ for $N=22$, $\Delta=-4.5$, and $h=0.2$.}
         (a)--(c) Scaled comparisons in the natural variables of the Airy, WKB, and finite-$N$ strong-anisotropy descriptions.
        (d) Residual comparison showing $\delta_{n}^{\gamma}=E_{n}^{\mathrm{ED}}-E_{n}^{\gamma}$ for $\gamma=\mathrm{th},\mathrm{WKB},\mathrm{SAE}$ on a common set of axes. Compared to the $h=0.05$ case, the stronger-confinement data show an improved collapse onto the Airy and WKB variables, while the finite-$N$ strong-anisotropy expansion remains a robust description of the upper levels. 
}\label{fig:app_h02_support}
\end{figure*}

\paragraph*{Stronger confinement: $h=0.2$.}
The corresponding comparison for the stronger confinement case $h=0.2$ is
shown in Figs.~\ref{fig:app_h02_overview}--\ref{fig:app_h02_support}. 
Here the overall agreement between ED and the analytic descriptions is significantly improved. In Fig.~\ref{fig:app_h02_overview}(a), the Airy-ladder energies remain close to the ED data over a much broader range of meson number than at $h=0.05$, and the residuals are correspondingly smaller. The spacing comparison in Fig.~\ref{fig:app_h02_overview}(b) shows that the same is true for the level spacings, i.e. the Airy description now captures not only the lowest levels but a substantial fraction of the entire $W=2$ ladder.

This improvement is particularly transparent in the Airy
diagnostic of Fig.~\ref{fig:app_h02_diagnostics}(a). The analytic and ED two-meson thresholds are now much closer, corresponding to the stability counts $n^\star_{\rm th}=8$ and $n^\star_{\rm ED}=7$. Thus, at stronger confinement the threshold theory provides a substantially more accurate estimate of the finite-size one-meson stability window.

The WKB comparison is also visibly improved relative to the
$h=0.05$ case: the ED points lie closer to the semiclassical
reference line and the corresponding residuals are reduced.
This improved agreement should nevertheless be understood as
a relative improvement over the weaker-confinement case rather
than as uniformly quantitative agreement across the full
finite-size meson ladder. 
The finite-$N$ SAE comparison in Fig.~\ref{fig:app_h02_diagnostics}(c) continues to describe the upper part of the ladder well, with residuals that remain small and slowly varying over the displayed range.

The compact supporting plots in Fig.~\ref{fig:app_h02_support} reinforce these conclusions. 
Compared to $h=0.05$, the stronger-confinement case shows an improved overall collapse of the ED data onto the Airy and WKB variables, while the finite-$N$ SAE remains a robust description of the upper part of the spectrum. Taken together, the $h=0.05$ and $h=0.2$ results illustrate clearly how the relative quality of the three analytic descriptions evolves with the confinement strength, i.e. as $h$ increases, the near-threshold Airy description remains useful over a broader fraction of the finite-size ladder, while the WKB and finite-$N$ SAE descriptions continue to provide the relevant analytic benchmarks for the interior and upper parts of the
spectrum. 

We emphasize that this finite-size trend should not be interpreted as a statement that the weak-confinement expansion becomes asymptotically better as $h$ is increased. Formally, the Airy description is controlled in the small-$h$ limit. At fixed $N=22$, however, decreasing $h$ also produces an increasingly dense ladder of shallow states, so that finite-size effects and the identification of a clean one-meson sequence become more restrictive. The improved agreement observed at $h=0.2$ therefore reflects the interplay between the weak-confinement expansion and finite-size resolvability of the meson ladder, rather than a reversal of the formal small-$h$ ordering.

\clearpage
\bibliography{bibfile2}
\end{document}